\documentclass[a4paper,11pt]{article}
\pdfoutput=1

\usepackage{jcappub} 

\usepackage[T1]{fontenc} 
\usepackage{gensymb}
\usepackage{graphicx}

\usepackage{caption}
\usepackage{subcaption}

\title{A Big Ring on the Sky}

\author[a]{A.M. Lopez,}
\author[a]{R.G. Clowes}
\author[b]{and G.M. Williger}

\affiliation[a]{Jeremiah Horrocks Institute,\\University of Central Lancashire,
Preston, PR1 2HE, United Kingdom}
\affiliation[b]{Department of Physics and Astronomy, University of Louisville,
Louisville, KY 40292, USA}

\emailAdd{amlopez@uclan.ac.uk}
\emailAdd{rgclowes@uclan.ac.uk}
\emailAdd{gmwill06@louisville.edu}

\abstract{We present the discovery of `A Big Ring on the Sky' (BR), the
  second ultra-large large-scale structure (uLSS) found in Mg~{\sc
    II}-absorber catalogues, following the previously reported Giant Arc
  (GA).  In cosmological terms the BR is close to the GA --- at the same
  redshift $z \sim 0.8$ and with a separation on the sky of only $\sim
  12^\circ$. Two extraordinary uLSSs in such close configuration raises the
  possibility that together they form an even more extraordinary cosmological
  system.  The BR is a striking circular, annulus-like, structure of diameter
  $\sim 400$~Mpc (proper size, present epoch).  The method of discovery is as
  described in the GA paper, but here using the new Mg~{\sc II}-absorber
  catalogues restricted to DR16Q quasars.  Using the Convex Hull of Member
  Spheres (CHMS) algorithm, we estimate that the annulus and inner absorbers
  of the BR have departures from random expectations, at the density of the
  control field, of up to $5.2\sigma$.  We present the discovery of the BR,
  assess its significance using the CHMS, Minimal Spanning Tree (MST),
  FilFinder and Cuzick \& Edwards (CE) methods, discuss it in the context of the
  GA+BR system, and suggest some implications for the origins of uLSS and for
  our understanding of cosmology.  For example, it may be that unusual
  geometric patterns, such as these uLSSs, have an origin in cosmic strings.
}

\begin{document}
\maketitle
\flushbottom

\section{Introduction}
\label{sec:intro}
We continue to make use of the method of intervening Mg~{\sc II} absorbers in
the spectra of quasars to trace faint matter at intermediate redshifts
\cite{Lopez2019, Lopez2022}.  The Mg~{\sc II} method relies on both the
spectroscopic measurement of luminous, high redshift quasars from the Sloan
Digital Sky Survey (SDSS), and the highly accurate, spectroscopic redshifts
of the intervening Mg~{\sc II} absorption doublets present in the quasar
spectra, documented by independent authors \citep{Zhu2013, Anand2021}.  The
intervening Mg~{\sc II} absorption doublet indicates the presence of galaxies
and galaxy clusters \cite{Bergeron1988, Churchill2005, Steidel1995}.
Together, with the quasars and Mg~{\sc II} absorbers, we have the information
on the on-sky position of intervening matter and the redshift of the
intervening matter, so in mapping the 3D distribution of the Mg~{\sc II}
absorption features in the spectra of quasars, we can infer the LSS of
intermediate-to-high redshift, faint matter.

The multiple discoveries of LSSs made throughout the past few decades are
well known to challenge our understanding of the Standard Cosmological Model
($\Lambda$CDM) \cite{Lopez2022, Bagchi2017, Balazs2015, Horvath2014,
  Clowes2013, Clowes2012}, in particular due to a possible violation of a
fundamental assumption, the Cosmological Principle (CP), which states that
our Universe is both homogeneous and isotropic on large scales.  In addition,
there are numerous results in cosmology posing similar challenges and
tensions for our current standard model (see \cite{Aluri2023} for a recent
review).  Large structures in the Universe are interesting for several
reasons, such as: how did the structures form so early on in the evolution of
the Universe, given the current understanding of CDM structure formation; how
might the structures evolve to the present day; do the seeds of such large
LSSs lie in the density perturbations that are amplified by inflation
\cite{Huterer2023, Bari2022}.  Answers to such questions may lie outside
concordance cosmology, with either the inclusion of extensions to the
standard model, e.g., cosmic strings (CS) \cite{Ahmed2023, Peebles2023,
  Balazs2015, Shlaer2012}, or alternative theories to the $\Lambda$CDM model,
e.g., Conformal Cyclic Cosmology (CCC) \cite{Penrose2014} and modified gravity
theories \cite{Mazurenko2024, Llinares2008, Milgrom1983}.

During the discovery of the Giant Arc (GA) \citep{Lopez2022} (hereafter
Lopez22), the SDSS Data Release 16 quasar database (DR16Q) became available
\cite{Lyke2020}.  Then, independent authors created the most up-to-date
Mg~{\sc II} database from the DR16Q quasars \cite{Anand2021} (hereafter
Anand21).  Following the discovery of the GA, made using the older Mg~{\sc
  II} database from \citep{Zhu2013} (hereafter Z\&M, also the corresponding
DR7QSO and DR12Q quasar catalogues from \cite{Schneider2010, Paris2017}), we
are now in a position to continue LSS investigations with the new DR16Q
database and corresponding DR16Q Mg~{\sc II} database.  We have found an
interesting ring shape in the Mg~{\sc II} absorbers, indicating a LSS of
galaxies and galaxy clusters, that spans a diameter of $\sim 400$~Mpc scaled
to the present epoch.  Incidentally, the estimated size of the BR is close to
that which could be expected in a detection of an individual Baryon Acoustic
Oscillation (BAO), $r \sim 150$~Mpc \cite{Tully2023, Einasto2016, Planck2015,
  Anderson2014, Eisenstein2005}, but we later suggest that the BR is unlikely
to have its origins in BAOs.  The BR shape and size are both hard to
understand in our current theoretical framework.  Additionally, the BR is in
the same redshift slice as the GA and to the north of the GA by $\sim 12
\degree$ which raises further questions about their origin both together and
independently.

For the work that led to the original discovery of the GA, using the Z\&M
database, we had looked at only a few small areas of sky and redshift slices
\cite{Lopez2019}, essentially to test the viability of the Mg~{\sc II}
approach itself.  Following the discovery of the GA, and now using the
Anand21 database, we have so far concentrated on the GA field and redshift
slice because we immediately made the further discovery of the BR there; the
only exceptions to this statement are (i) the use of adjacent redshift slices
to test that the BR was not arising from artefacts and to test for extensions
of the BR into adjacent redshift slices, and (ii) the use of neighbouring
fields at the same redshift as the GA/BR field for comparing the spatial
clustering results of the Cuzick and Edwards test (see
section~\ref{subsec:CE}).  Consequently, a `look-elsewhere' effect on the
statistical assessments should not be a factor. In future, of course, we
intend to explore both databases in their entireties.

The Mg~{\sc II} data we use here are complicated and quite difficult to
manage. The advantage, of course, is the precise redshifts, and the
concomitant possibility of discovering intriguing structures such as the Big
Ring and the Giant Arc. In future, the Mg~{\sc II} approach to LSS should be
enhanced by DESI spectra \cite{Napolitano2023}, taken with the KPNO 4m
telescope, allowing detection of Mg~{\sc II} to lower equivalent widths, and
hence allowing the exploration of finer detail in the morphology of
structures.

In this paper we assess the reality of the BR and its statistical
significance with respect to the assumed, homogeneous large-scale
distribution of matter.

Note that we have introduced the term `ultra-large LSS' (uLSS) to distinguish
those structures that exceed the estimated $\sim 370$~Mpc upper limit to the
scale of homogeneity \cite{Yadav2010}. This limit is often adopted in
discussions of homogeneity and the CP.

\subsection{Cosmological model}

The concordance model is adopted for cosmological calculations, with
$\Omega_{T0} = 1$, $\Omega_{M0} = 0.27$, $\Omega_{\Lambda 0} = 0.73$, and
$H_0 = 70$~km\,s$^{-1}$Mpc$^{-1}$. All sizes given are proper sizes at the
present epoch. (For consistency, we are using the same values for the
cosmological parameters that were used in Lopez22.)

\section{The Big Ring}
\label{sec:BR}
The existence of a large, circular structure of Mg~{\sc II} absorbers, the
Big Ring (BR), became apparent when investigating the new SDSS DR16Q
catalogues, and corresponding Mg~{\sc II} databases, at the same redshift and
position as the previously documented Giant Arc (GA).  
\begin{figure}[h]
\centering 
\includegraphics[scale=0.7]{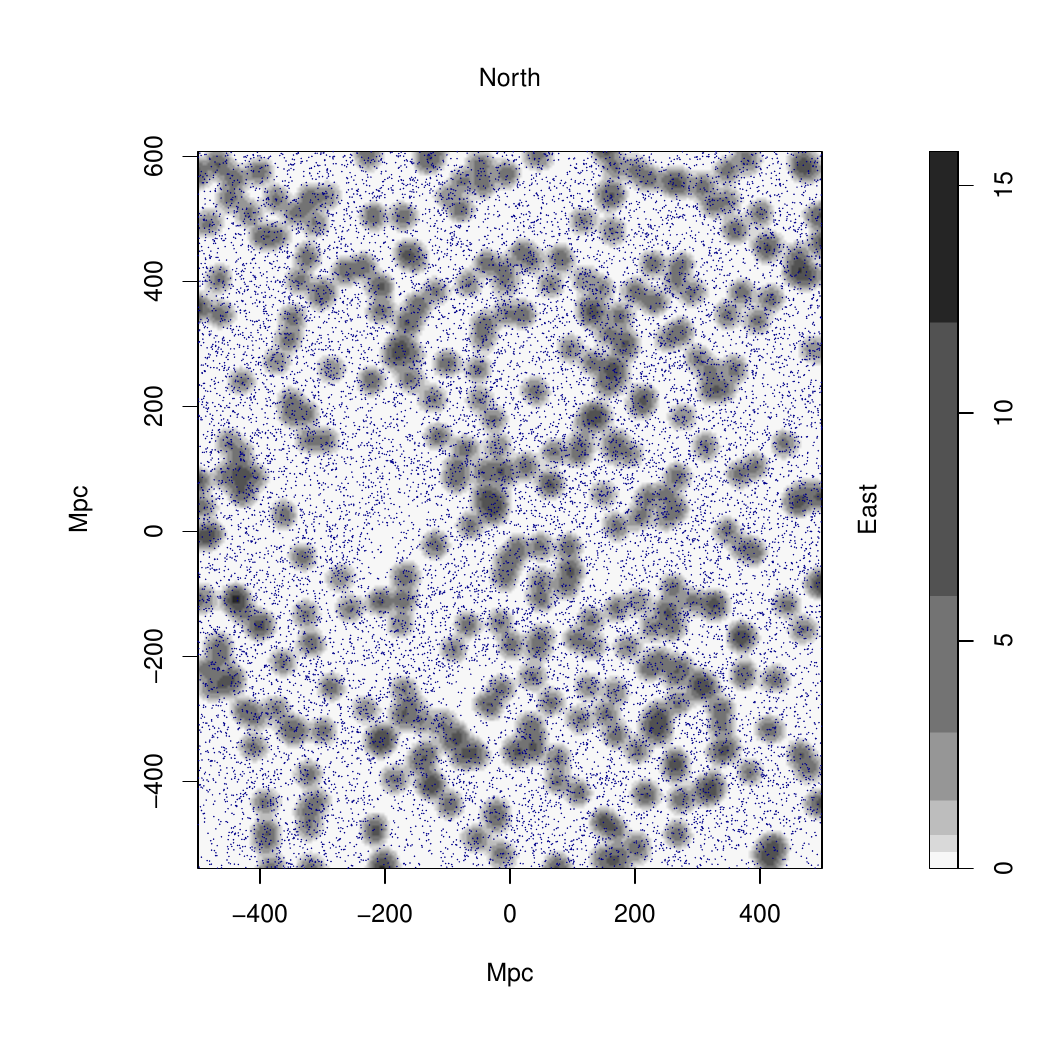}
\caption{\label{fig:BR} The tangent-plane distribution of Mg~{\sc II}
  absorbers in the redshift slice $z=0.802 \pm 0.060$. The grey contours,
  increasing by a factor of two, represent the density distribution of the
  absorbers which have been smoothed using a Gaussian kernel of $\sigma =
  11$~Mpc, and flat-fielded with respect to the distribution of background
  probes (quasars). The dark blue dots represent the background probes. S/N
  limits of: $4, 2$ and $4$ were applied to the $\lambda_{2796},
  \lambda_{2803}$ Mg~{\sc II} lines and quasar continuum, respectively
  (details of S/N cuts are discussed in Section~\ref{subsec:SLHC}). The BR
  can be seen to the north of the centre point spanning $\sim 400$~Mpc in
  diameter. The field-of-view corresponds to the small, pink area seen in
  Figure~\ref{fig:SDSS_control_fields}.}
\end{figure}
In Figure~\ref{fig:BR}, we are seeing the BR, which is the visually overdense
ring shape of Mg~{\sc II} absorbers centred at approximately $x=0$~Mpc and
$y=240$~Mpc.  (The large `void' to the south-west of the BR is also
particularly striking.)  In this figure, and others, the grey contours,
increasing by a factor of two, represent the density distribution of Mg~{\sc
  II} absorbers in the specified redshift slice and field-of-view (FOV).  The
Mg~{\sc II} contours have been smoothed using a Gaussian kernel of $\sigma =
11$~Mpc, and flat-fielded with respect to the distribution of background
probes (quasars).  The smoothing gives a useful impression of the
connectivity.  The background probes (quasars) are represented by the small,
blue points.  The axes are labelled in Mpc, scaled to the present epoch; for
details on obtaining the Mpc scaling, refer to the GA paper (Lopez22).  East
is towards the right and north is towards the top.  From Figure~\ref{fig:BR}
we can estimate the BR diameter is $\sim 300 - 400$~Mpc, which would make its
circumference comparable to the extent of the GA.

\subsection{Data sources}
\label{subsec:data}
The use of Mg~{\sc II} absorbers for analysing LSS has the particular
advantage of providing very precise redshifts. A disadvantage of course,
which can require very careful handling, is that one must take the background
probes --- the quasars --- where they are given by the catalogues. The
catalogues now have suitably dense coverage on the sky, but generally they
are affected by variations in selection criteria.

At $z_{abs} \sim 0.8$, the parameter in the Anand21 catalogues for the
redshift error (\scalebox{0.7}{{\sc Z\_ABS\_ERR}}) indicates a median error
of $\sigma_{z_{abs}} \approx 4.2 \times 10^{-5}$. (The emission-redshift
error for a quasar catalogue might be about two orders of magnitude larger at
$\sim 0.004$.) This $\sigma_{z_{abs}}$ corresponds to a velocity difference
of $\sim 7 $~km\,s$^{-1}$.  A comparison of repeated observations in the
basic Anand21 database suggests that a practical estimate of the redshift
error at $z_{abs} \sim 0.8$ is a little larger at $\sigma_{z_{abs}} \approx
1.7 \times 10^{-4}$, corresponding to a velocity difference of $\sim
28$~km\,s$^{-1}$.

When considering the finer details of the morphology of individual LSSs, any
blurring will then be due to peculiar velocities, for which plausible values
might be $\sim 400$~km\,s$^{-1}$, corresponding to $\sigma_{z_{pec}} \sim
0.0024$ at $z \sim 0.8$, or $\sim 7$~Mpc in proper distance for the present
epoch. We therefore expect that any blurring effects should be minor.

The BR is detected in the new Anand21 Mg~{\sc II} database, whereas the
previously documented GA was detected in the older Z\&M Mg~{\sc II} database,
so we investigate the differences in the databases (e.g. the Mg~{\sc II}
absorbers detected from probes that were in common to both databases) in the
following manner.  First, we choose the standard BR FOV at $z=0.802 \pm
0.060$ as in Figure~\ref{fig:BR}.  Then, we select the absorbers and probes
arising in the chosen field from both Anand21 and Z\&M (making no additional
cuts to the S/N or $i$-magnitude).  Using \scalebox{0.7}{{\sc
    TOPCAT}}\footnote{https://www.star.bris.ac.uk/mbt/topcat/} \cite{Topcat},
the probe (quasar) and Mg~{\sc II} files are paired in various ways for
comparison.

A summary of the results is as follows.
(1) There are over three times as many background probes in the field for
Anand21 than Z\&M. (These are the probes that were initially searched by the
authors for Mg~{\sc II} absorbers.) Almost all of the additional probes
searched by Anand21 were new observations between SDSS DR12 to DR16. However,
a small fraction (just $37$ out of $7257$ probes) were not included in the
Mg~{\sc II} search by Anand21.
(2) There are just over twice as many Mg~{\sc II} absorbers detected by
Anand21 than by Z\&M: $852$ Mg~{\sc II} absorbers in Anand21 and $359$
Mg~{\sc II} absorbers in Z\&M. Of those absorbers, $597$ were unique to
Anand21 and $104$ were unique to Z\&M.
(3) Of the probes that both authors had in common and searched, Anand21 found
$67$ additional absorbers that Z\&M missed, and Z\&M found $103$ additional
absorbers that Anand21 missed.  In total, there were $255$ Mg~{\sc II}
absorbers in common to both authors ($\sim 60 \% $ total agreement of the
probes in common to both authors).  We found that, despite the greater number
of absorbers in Anand21, they missed around $1.5$~times as many absorbers
found by Z\&M, compared with the number of absorbers missed by Z\&M that were
found by Anand21, for shared probes.  Without access to their software it
will probably remain unclear why Anand21 and Z\&M have this disagreement.
One possibility might be that Anand21 have more refined search criteria,
given that their percentage Mg~{\sc II} detection rate per quasar is much
lower than for Z\&M.  Or, perhaps Z\&M are detecting to lower thresholds,
without increasing spurious detections.

\subsection{Initial checks of the data}
\label{subsec:initial_checks}
\begin{figure}[tbp]
\centering 
\includegraphics[scale=0.6]{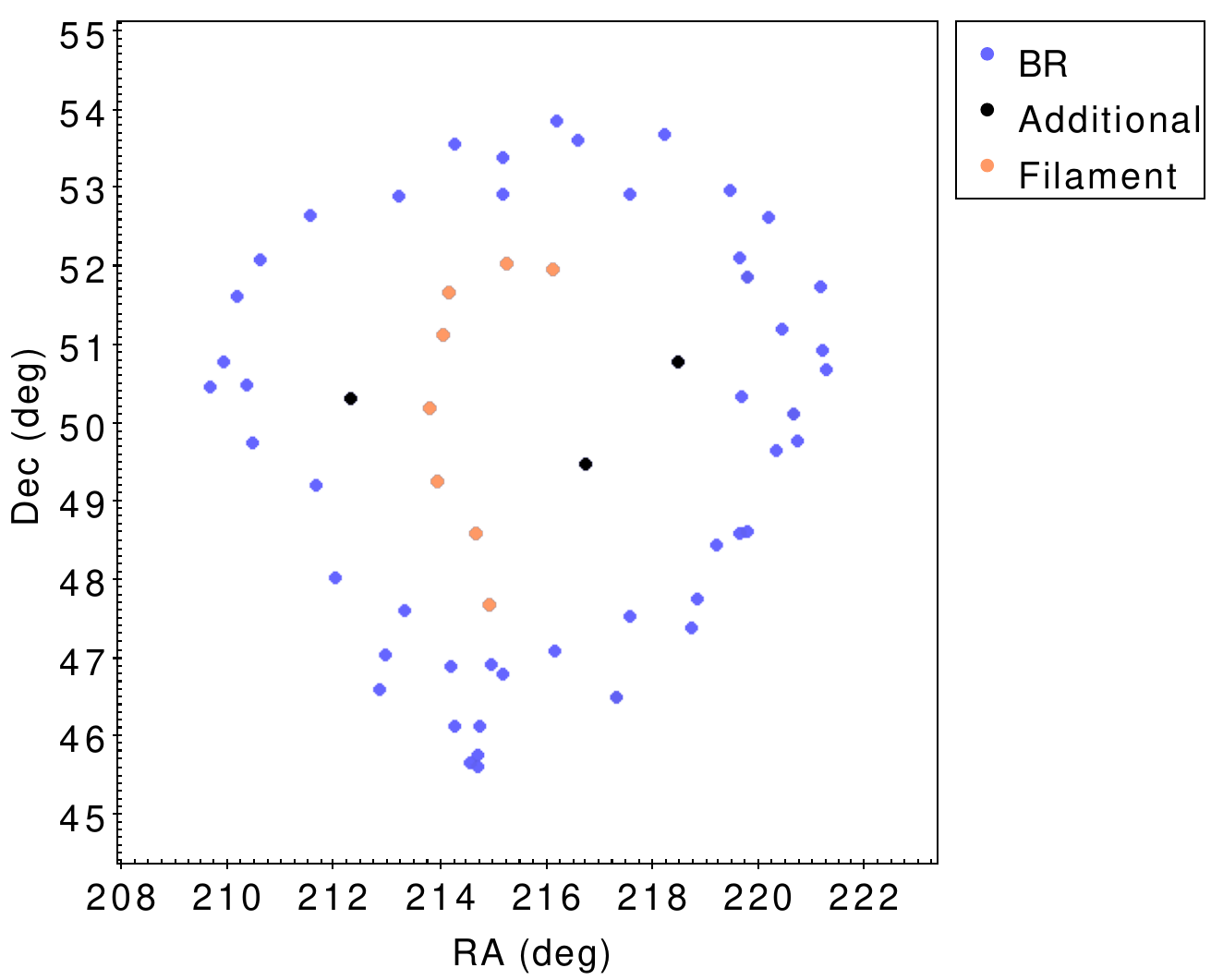}
\caption{\label{fig:BR_and_filament_topcat} The Mg~{\sc II} absorbers in the
  visually-identified BR (blue), the inner filament (orange), and the
  additional absorbers inside the BR (black) from the on-sky
  perspective. Three of the blue points (BR) correspond to two absorbers
  occurring in one spectrum (see Figure~\ref{fig:BR_and_SLHC_groups}).}
\end{figure}

First, we check that the Mg~{\sc II} absorbers belonging to the
visually-identified BR are real (not false positive detections).  We visually
inspected spectra of $56$ DR16Q quasars that are the probes that correspond
to the visually-identified BR and inner filament absorbers (see
Figure~\ref{fig:BR_and_filament_topcat}).
Since $6$ of the Mg~{\sc II} absorbers are multiples per probe occurring in
three quasar spectra, we were checking for a total of $59$ Mg~{\sc II}
doublet systems.  Each Mg~{\sc II} absorption doublet that we searched for
was visually confirmed and in agreement with the documented redshifts from
Anand21, so we can confirm that $100\%$ of the Mg~{\sc II} absorber members
of the BR and inner filament are real, physical Mg~{\sc II} absorbers
indicating the presence of intervening matter.

There are two unusual Mg~{\sc II} systems, occurring in the same spectrum,
for which the Anand21-documented redshifts suggest that the $\lambda_{2796}$
of the lower-$z$ absorption doublet appears at the same wavelength as the
$\lambda_{2803}$ of the higher-$z$ absorption doublet.  This is indeed the
case: the two Mg~{\sc II} absorption doublets appear as $3$ absorption lines
in the spectrum, which is a rare oddity.  Although Anand21 recognise the $3$
absorption lines as two systems, they appear not to have disentangled the EWs
of the centre absorption line of the triplet (the higher-$z$ $\lambda_{2796}$
and the lower-$z$ $\lambda_{2803}$ EWs being the same).

\begin{figure}[h]
\centering 
\includegraphics[scale=0.7]{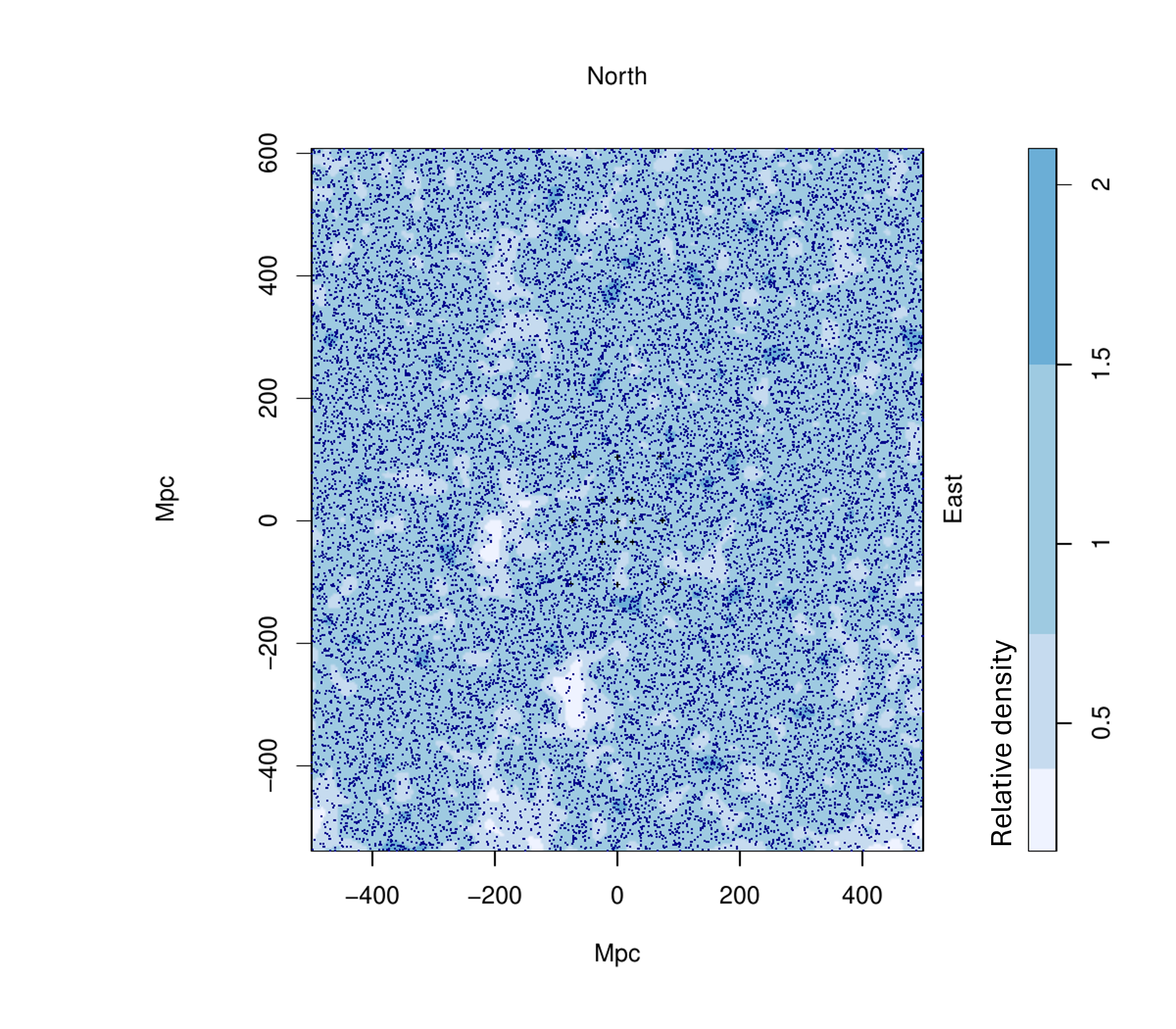}
\caption{\label{fig:Rplot_probes} The tangent-plane distribution of probes
  (background quasars) in the BR FOV with the redshift condition $z > 0.862$
  --- i.e., the probes that are responsible for the Mg~{\sc II} absorbers
  arising in the BR field. The blue contours, increasing by a factor of two,
  represent the density distribution of the probes which have been smoothed
  using a Gaussian kernel of $\sigma = 11$~Mpc. S/N limits were applied to
  the quasar continuum such that S/N$_{con} > 4$ (details of S/N are
  discussed in Section~\ref{subsec:SLHC}). The field-of-view corresponds to
  the small, pink area seen in Figure~\ref{fig:SDSS_control_fields}. The
  figure shows many areas of overdensities and underdensities. In particular,
  there are a few overdense regions (small, dark clumps), centred at $0$~Mpc
  on the $x$-axis and between roughly $100$~Mpc to $400$~Mpc on the $y$-axis,
  a few of which appear to coincide with the inner filament of BR. There is
  also a much larger region of underdense probes spanning $-100$~Mpc to
  $400$~Mpc in the $y$-axis and centred at $-200$~Mpc in the $x$-axis which
  coincides with the LHS of the BR.}
\end{figure} 

Secondly, we investigate if the visually obvious BR is an artefact of the
probes.  This can be done in two ways: simply checking the density
distribution of background probes and checking for obvious artefacts; and
looking at the next redshift slice down from the BR field (on the near side)
and checking for repeating Mg~{\sc II} features that correspond to any
obvious artefacts in the probes.  For the former, see
Figure~\ref{fig:Rplot_probes}.

\begin{figure}[h]
\centering 
\includegraphics[scale=0.5]{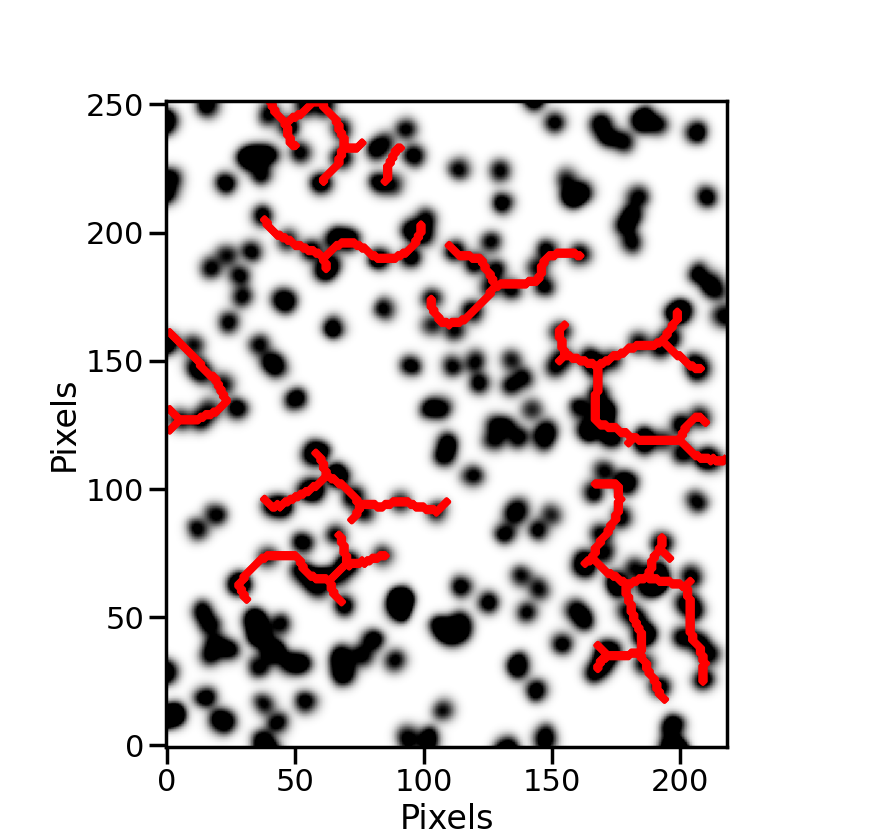}
\caption{\label{fig:FilFinder_BR_0_682} The FilFinder algorithm applied to
  the tangent-plane distribution of Mg~{\sc II} absorbers in the closest,
  non-overlapping redshift slice to the BR field --- i.e., the field centred
  at $z=0.682 \pm 0.060$ and corresponding to the usual BR
  field-of-view. Axes are labelled in pixels, where $1$~pixel~$=
  4^2$~Mpc$^2$. S/N limits of: $4, 2$ and $4$ were applied to the
  $\lambda_{2796}, \lambda_{2803}$ Mg~{\sc II} lines and quasar continuum,
  respectively (details of S/N cuts are discussed in
  Section~\ref{subsec:SLHC}). The probes (quasars) responsible for the
  Mg~{\sc II} arising here have redshifts $z > 0.862$ so that they are the
  same probes responsible for the Mg~{\sc II} absorbers in the BR field
  centred on the usual redshift slice ($z=0.802 \pm 0.060$). The figure shows
  that there are no filaments correlating to the BR indicating that the BR is
  not a result of artefacts in the probes.}
\end{figure}
Many overdensities and underdensities can be seen clearly in Figure~\ref{fig:Rplot_probes}.  In
particular, there are a few overdense regions (small, dark clumps) centred at
$0$~Mpc on the $x$-axis and between roughly $100$~Mpc to $400$~Mpc on the
$y$-axis, a few of which appear to coincide with the inner filament of BR.
This could imply that the inner filament of the BR is suspect, so we will
need to be sure that the filament is not an artefact of the probes.  We can
visually check the rate of occurrence of Mg~{\sc II} absorbers at the
position of the overdense artefacts (dense blobs) by blinking the image of
the probes (Figure~\ref{fig:Rplot_probes}) with the image of the absorbers
(Figure~\ref{fig:BR}).  Doing this shows that $8$ out of $20$ randomly
selected overdense artefacts had Mg~{\sc II} absorbers present, so less than
half of the artefacts.  Incidentally, we could also confirm that most of the
artefacts that appeared to coincide with parts of the BR (including the inner
filament) were in fact offset, so not responsible for the Mg~{\sc II}
absorbers arising there.  Checking the artefacts shows that there is no
particular association of the artefacts with the Mg~{\sc II} absorbers
present in the BR field.
There is also a much larger region of underdense probes spanning $-100$~Mpc
to $400$~Mpc in the $y$-axis and centred at $-200$~Mpc in the $x$-axis which
partially coincides with the LHS of the BR.  The fact that part of the BR is
located in an underdense region of probes is noteworthy.

For the latter way to test if the Mg{\sc II} absorbers are artefacts of the
probes, we can check the absorbers arising from the same set of probes
corresponding to the BR field in the next, non-overlapping redshift slice on
the near side of the BR.  To do this we keep the probes of the BR field the
same (having $z>0.862$ --- i.e., a redshift greater than the far edge of the
BR Mg~{\sc II} redshift slice) and map the Mg~{\sc II} absorption in the
nearest, non-overlapping redshift slice (i.e., $z=0.682 \pm 0.060$).  In this
way we are able to search for any obvious artefacts of the probes that could
be responsible for the specific distribution of Mg~{\sc II} absorbers in the
BR field by comparing the Mg~{\sc II} image in the usual BR field with the
neighbouring redshift slice.  We apply the SLHC / CHMS and FilFinder
algorithms to the field centred at $z=0.682 \pm 0.060$ and corresponding to
the usual BR FOV (see Sections~\ref{subsec:SLHC} and \ref{subsec:FilFinder}
for details on the SLHC / CHMS and FilFinder methods).  In the redshift slice
$z=0.682 \pm 0.060$ we find with the SLHC / CHMS method that there are no
structures detected by the SLHC algorithm corresponding to the BR, and in
addition, no structures at all that are statistically significant.  We also
find with the FilFinder method, no filaments detected in the field that
correspond to the BR (Figure~\ref{fig:FilFinder_BR_0_682}).
Therefore, we conclude that the BR is not an artefact of the probes.

It is worth highlighting that the BR appears in the same redshift slice and
FOV as the previously documented GA, but we are now using the new Anand21
databases and not the previously used Z\&M databases.  As mentioned earlier,
the overall agreement between the two datasets is $\sim 60\%$ for the probes
in common in the BR field, so the GA appears somewhat different in this new
dataset.  The GA is still the most significant, most numerous and overdense
structure detected in the field despite the slight change in appearance, so
quantitatively, there is very little difference in the GA in the new dataset.
However, qualitatively, there are two main reasons the GA appears different:
(1) the field overall is much more dense with the new dataset due to many
more quasar observations; (2) Anand21 miss several GA absorbers.  We
investigate the second point by manually checking each of the quasar spectra
that are probes to the Mg~{\sc II} absorbers in the GA that Anand21 missed.
There were $16$ from $51$ absorbers that Anand21 missed, and none was due to
the small fraction of removed quasars in DR16Q that was mentioned earlier.
In each of the quasar spectra (corresponding to the GA) the Mg~{\sc II}
doublets were visually confirmed, but the $16$ absorbers missed by Anand21
had profiles that were generally complex, broad or weak.  This again suggests
that the Anand21 Mg~{\sc II} detection algorithm has a narrower detection
window at the cost of losing some real absorbers.  Conversely, the Z\&M
Mg~{\sc II} detection algorithm could conceivably contain more spurious
absorbers while likely managing to detect a higher percentage of the real
absorbers.

\section{Statistical analysis}
\label{sec:stats}
The discovery of the BR was made serendipitously when looking at the
previously documented GA field with the Anand21 data.  We now present the
statistical analysis of the BR, taking a similar approach to the previous
work in Lopez22.  Given the nature of the discovery, the analysis is
necessarily post-hoc.  However, given the previous work presented in Lopez22
we are able to follow the guidelines set there, consequently alleviating many
of the problems associated with post-hoc analysis.

Simulations are often advocated in contemporary astrophysics and cosmology,
but we do not consider them likely to be effective or efficient here. Their
complexity would be too great and would have too many unknowns and
uncertainties.  Consider, for example, that the simulations would have to
incorporate: simulating the universe in general; the occurrence of quasars in
that simulated universe; the observational parameters of the imaging and
spectroscopic surveys and their on-sky variations; and the detection of the
Mg~{\sc II} by software.  Instead, we have taken the more practical approach
of (i) using the data to correct the data, and (ii) seeking independent
corroboration of features using independent tracers.

This section is divided in the following manner.
(\ref{subsec:SLHC}) We assess the BR from a `first look' perspective by using
a heuristic process of stepping through redshift slices and determining the
optimum redshift for the BR.
(\ref{subsec:CHMS_MST}) We use the Convex Hull of Member Spheres (CHMS) and
the Minimal Spanning Tree (MST) significance calculations for assessing the
significance of the BR.  We apply these two methods of significance
calculations to four sets of BR absorber-member estimates: the SLHC groups;
the visually-identified BR absorbers (both including and excluding the inner
absorbers); and the FilFinder-identified absorbers.
(\ref{subsec:FilFinder}) The 2D FilFinder algorithm is applied to the pixel
image containing the BR to objectively identify filaments in the field.
(\ref{subsec:CE}) Finally, we apply the 2D Cuzick and Edwards test to the
BR field to determine the significance of clustering in the field (not the
candidate structure itself).

\subsection{Single-Linkage Hierarchical Clustering algorithm}
\label{subsec:SLHC}
The Single-Linkage Hierarchical Clustering (SLHC) algorithm is equivalent to
a Minimal Spanning Tree (MST in a generic sense, not to be confused with the
MST \emph{significance} calculation in Section~\ref{subsec:CHMS_MST}) when
separated at a specified linkage scale.  Our particular application of the
algorithm was first described in \cite{Clowes2012}, in combination with the
Convex Hull of Member Spheres (CHMS) algorithm which assesses the
significance of a specified structure.  The SLHC / CHMS method has been used
to locate and assess LSS in both quasars and Mg~{\sc II} absorbers
\cite{Clowes2012, Clowes2013, Lopez2022}.
\begin{figure}[h]
\centering 
\includegraphics[scale=0.5]{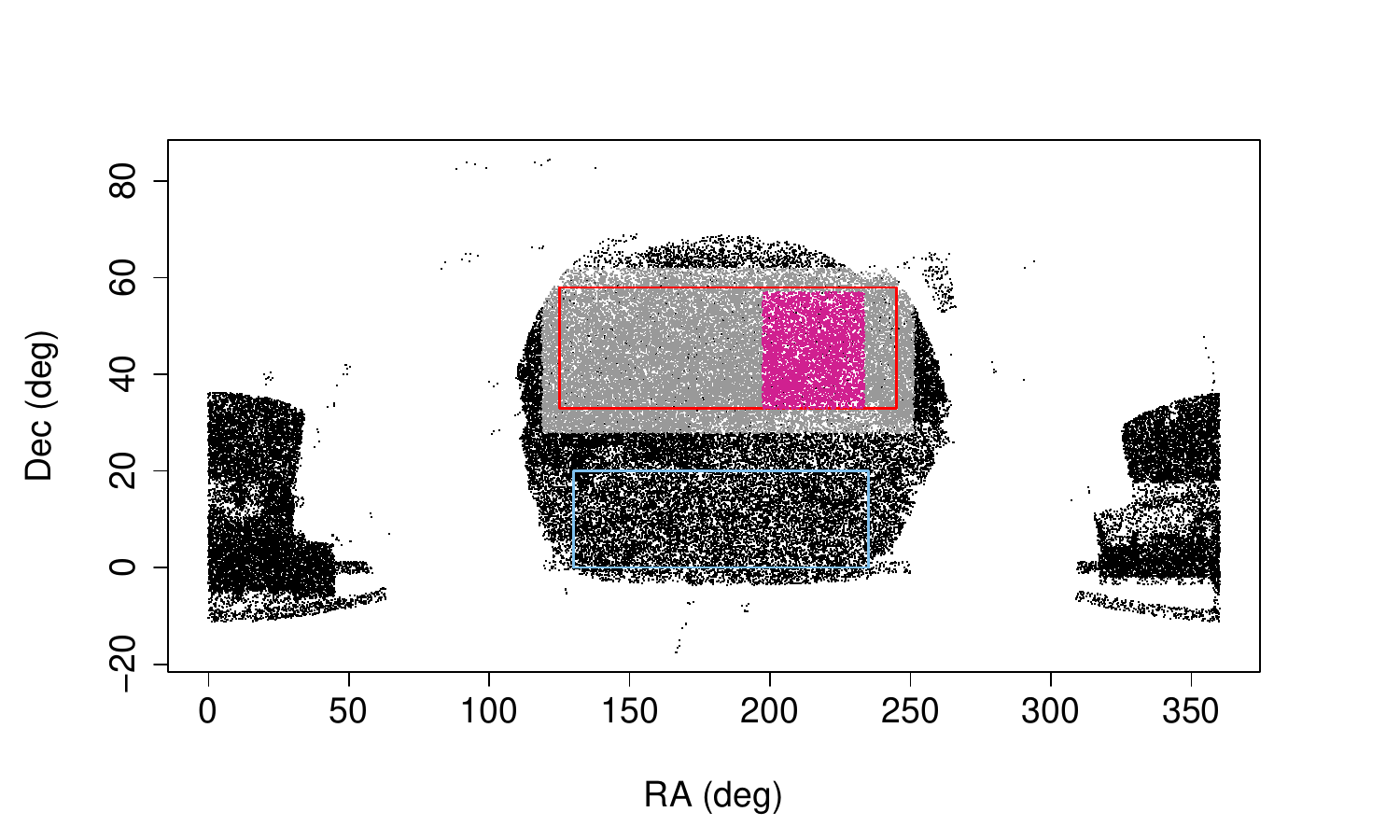}
\caption{\label{fig:SDSS_control_fields} The SDSS DR16Q footprint. The grey
  points are the input quasars for the control field and the pink points
  correspond to the Mg~{\sc II} absorbers in the field of interest. The
  control fields are outlined by lines in red (overdense region, version-1)
  and blue (underdense region, version-2). }
\end{figure}

Previously, the inherent difficulties of the Mg~{\sc II} method have been
discussed, especially in relation to applying statistics on an intrinsically
inhomogeneous dataset.  The quasars act as probes of the intervening matter,
so the varying availability of background quasars leads to an incomplete
image or map of the intervening matter, in this case, the Mg~{\sc II}
absorbers.  The inhomogeneity of the quasars, both intrinsic and of the
survey, contribute in a complicated way to the inhomogeneity of the Mg~{\sc
  II} absorbers.  Of course, with large data and large survey areas,
approximations can be made.  Before applying the SLHC algorithm we define the
field-of-view (FOV) for the analysis presented here.

First, the BR appears in the north of the usual GA field, so we re-centre the
Mg~{\sc II} images accordingly.  Second, the field containing the BR is close
to the northern SDSS footprint border as well as a southern border arising
from much lower quasar coverage.  We choose to shrink slightly the typical
size of a Mg~{\sc II} image to avoid these areas.  Third, given the generally
patchy DR16Q quasars, and thus the corresponding Mg~{\sc II} absorbers in the
Anand21 databases, we apply signal-to-noise (S/N) limits to the Mg~{\sc II}
absorption lines and the continuum.  We calculate the S/N of the Mg~{\sc II}
lines by:
\begin{equation*}
S/N = W_{r} / W_{err}
\end{equation*}
\noindent where $W{r}$ is the equivalent width of the line, and $W_{err}$ is
the corresponding error in the Mg~{\sc II} line.
For the continuum, Anand21 provide the median quasar S/N.  Applying S/N
limits has the effect of removing spurious and potentially false positive
absorbers as well as generally reducing the patchiness in the data.  Setting
additional\footnote{A base-line magnitude limit of $i <= 20.5$ is applied to
the DR16Q quasars as part of the read-in process since the completeness
declines steeply for fainter quasars.} limits to the magnitude of the quasars
would also have a similar, desirable effect.  However, not all quasars will
have the same integration time, so faint quasars could have good S/N due to
long exposures.  Following the example by Z\&M, we apply a S/N limit of $4$
and $2$ for the $\lambda_{2796}$ and $\lambda_{2803}$ lines respectively.
Since the S/N of the quasar continuum necessarily has equal or higher S/N
than the $\lambda_{2796}$ line we apply a S/N limit to the quasar continuum
of $6$.  Applying the condition of S/N~$\geq 6$ to the quasar continuum could
be too restrictive, and later we will apply a less conservative condition to
the continuum for comparison.
    
The CHMS significance is calculated by the rate of occurrence of volumes
smaller than the CHMS volume of the structure by randomly distributing the
absorbers belonging to the structure at a density equal to the control field
density --- the simulations are repeated 1000 times.  Previously, the control
field was chosen to be that of the field being assessed.  However, this would
mean that a percentage of the field absorbers are those belonging to the
structures of interest, e.g., the GA and BR in the GA/BR FOV.  In addition,
the FOV containing the BR is small, so small-scale inhomogeneities have a
much larger effect on the average density.  Conversely, choosing a control
field that is too large will lead to problems involving the large-scale
inhomogeneities of the SDSS survey, explained above.  Subsequently, we have
designed two versions of control fields accounting for the northern portion
of the SDSS footprint (overdense region) and the southern portion of the SDSS
footprint (underdense).  Each version can be chosen depending on the location
of the field of interest.  So here, we will be using version-1 of the control
field --- Figure~\ref{fig:SDSS_control_fields}.

The SLHC is equivalent to an MST when separated at a specified linkage scale:
thus the choice of linkage scale will determine the maximum distance between
points that would be considered `joined', or a candidate structure.  The term
`structure'
  \footnote{For a working definition, in investigating LSS, we often consider
  a candidate structure to be a set of $N$ connected tracers, the containing
  volume of which is a $n\sigma$ departure from the containing volume
  expected for a uniform, random distribution. We might choose to consider
  further only those candidates for which the amplitude $n\sigma$ exceeds
  some threshold. `Connected' and `containing' volume will often be
  determined {\em algorithmically}; for both, there is an implicit assumption
  of a uniform host survey, which might be approximately true only in
  restricted areas.}  here is not be confused with a gravitationally-bound
  system, but is instead referring to a grouping of more than $10$ (a
  specified minimum) members (in this case, Mg~{\sc II} absorbers) that have
  an MST with distances smaller than the linkage scale.  LSSs are not
  expected to be gravitationally bound, as, indeed, superclusters and great
  walls are not expected to be gravitationally bound.  Similar usages and
  definitions of `structure' are common in LSS studies \cite{Pomarede2020,
    Balazs2015, Horvath2014, Gott2005}.  The candidate structures are then
  assessed by the CHMS to determine whether their volumes are statistically
  significant.  In Lopez22 we discuss the effects of varying the linkage
  scale, so here we choose to scale the linkage scale according to the
  Mg~{\sc II} number density of the control field and the number density of
  the GA field --- the GA field being the candidate field from which to scale
  all other fields, i.e.,
\begin{equation*}
s = (\rho_0 / \rho)^{1/3}s_0
\end{equation*}
\noindent where $s$ and $s_0$ are the linkage scale for the control field and
the GA field respectively, and similarly $\rho$ and $\rho_0$ are the
densities of the corresponding fields.  The linkage scale is to be taken as a
guideline; if the linkage scale is too small then potentially interesting
candidate structures will be missed, and if the linkage scale is too high
then too many points will be grouped as one seemingly coherent structure, but
that would of course reflect in the CHMS significance.  We find that the
linkage scale set for the GA field (the field which we now use as the
base-line) was an appropriate choice for the specific field density.
However, we saw that even with the chosen linkage scale the SLHC algorithm
identified the GA as two, individual, overlapping candidate structures.  We
reason that, when using a smaller linkage scale, candidate structures that
are overlapping or adjacent could reasonably belong to the same structure.
It is important to recognise that multiple candidate structures overlapping
and adjacent to each other will still need to be assessed with the CHMS or
MST significance, which would then objectively determine whether their
agglomeration is statistically significant (remember, if the whole field was
joined as one structure then this would of course not be statistically
significant).
  
As with the GA analysis, we step through overlapping redshift slices and use
the SLHC / CHMS algorithms to determine the redshift of the peak signal of
the BR.  Given the much larger Mg~{\sc II} database from Anand21, even after
S/N limits are applied, it is expected that the field density containing the
BR is much higher than the GA field in the Z\&M databases, and therefore the
chosen linkage scale will be correspondingly lower.  The five redshift slices
assessed are centred at: 0.682, 0.742, 0.802, 0.862 and 0.922, each with a
redshift thickness of $\Delta z = 0.060$; the results are shown in
Table~\ref{tab:5_z_slices_SN_4_2_6}.
  
\begin{table}[h]
\begin{tabular}{|c|c|c|c|c|c|}
\hline
\multicolumn{1}{|l|}{\textbf{\begin{tabular}[c]{@{}l@{}}Central \\ redshift\end{tabular}}} & \multicolumn{1}{l|}{\textbf{\begin{tabular}[c]{@{}l@{}}Linkage \\ scale \\ (Mpc) \end{tabular}}} & \multicolumn{1}{l|}{\textbf{\begin{tabular}[c]{@{}l@{}} No. candidate\\ structures\end{tabular}}} & \multicolumn{1}{l|}{\textbf{\begin{tabular}[c]{@{}l@{}} No. candidate \\structures with\\ $\sigma_{CHMS} \geq 3.5 \sigma$\end{tabular}}} & \multicolumn{1}{l|}{\textbf{\begin{tabular}[c]{@{}l@{}}Maximum \\ $\sigma_{CHMS}$ ($\sigma$) \end{tabular}}} & \multicolumn{1}{l|}{\textbf{\begin{tabular}[c]{@{}l@{}}Maximum \\Mg~{\sc II} absorber \\membership\end{tabular}}} \\ \hline
\textbf{0.682} & 79.3 & 4 & 0 & 3.3 & 17  \\ \hline
\textbf{0.742} & 79.3 & 5 & 1 & 3.5 & 16  \\ \hline
\textbf{0.802} & 79.5 & 8 & 1 & 4.5 & 28  \\ \hline
\textbf{0.862} & 79.6 & 6 & 2 & 4.7 & 42  \\ \hline
\textbf{0.922} & 81.1 & 4 & 1 & 4.1 & 26  \\ \hline
\end{tabular}
\caption{\label{tab:5_z_slices_SN_4_2_6} Results from the SLHC / CHMS on
  five, overlapping redshift slices to determine the optimum redshift slice
  for the BR signal. The FOV of each redshift slice corresponds to the small,
  pink area seen in Figure~\ref{fig:SDSS_control_fields}. S/N limits of $4,
  2$ and $6$ were applied to the $\lambda_{2796}, \lambda_{2803}$ Mg~{\sc II}
  lines and quasar continuum, respectively. All redshift slices have a
  thickness of $\Delta z = \pm 0.060$. The columns from left to right are:
  the central redshift of the field being assessed; the linkage scale used
  for the field being assessed, calculated as $s = (\rho_0 / \rho)^{1/3}s_0$
  (see the main text); the number of candidate structures identified in the
  field; the number of candidate structures identified in the field with a
  CHMS significance equal to or exceeding $3.5 \sigma$; the maximum CHMS
  significance calculated from the candidate structures; the maximum Mg~{\sc
    II} absorber membership identified from the candidate structures. }
\end{table}
\begin{table}[h]
\begin{tabular}{|c|c|c|c|c|c|}
\hline
\multicolumn{1}{|l|}{\textbf{\begin{tabular}[c]{@{}l@{}}Central \\ redshift\end{tabular}}} & \multicolumn{1}{l|}{\textbf{\begin{tabular}[c]{@{}l@{}}Linkage \\ scale \\ (Mpc) \end{tabular}}} & \multicolumn{1}{l|}{\textbf{\begin{tabular}[c]{@{}l@{}}No. candidate\\ structures\end{tabular}}} & \multicolumn{1}{l|}{\textbf{\begin{tabular}[c]{@{}l@{}}No. candidate\\ structures with \\$\sigma_{CHMS} \geq 3.5 \sigma$\end{tabular}}} & \multicolumn{1}{l|}{\textbf{\begin{tabular}[c]{@{}l@{}}Maximum\\ $\sigma_{CHMS}$ ($\sigma$) \end{tabular}}} & \multicolumn{1}{l|}{\textbf{\begin{tabular}[c]{@{}l@{}}Maximum \\ Mg~{\sc II} absorber \\ membership \end{tabular}}} \\ \hline
\textbf{0.682} & 76.7 & 6 & 0 & 3.2 & 16  \\ \hline
\textbf{0.742} & 76.2 & 7 & 0 & 3.3 & 19  \\ \hline
\textbf{0.802} & 75.8 & 10 & 1 & 4.1 & 30 \\ \hline
\textbf{0.862} & 75.7 & 7 & 1 & 3.7 & 27  \\ \hline
\textbf{0.922} & 77.1 & 6 & 0 & 3.4 & 15  \\ \hline
\end{tabular}
\caption{\label{tab:5_z_slices_SN_4_2_4} As with Table~\ref{tab:5_z_slices_SN_4_2_6}. Slightly relaxed S/N limits of $4, 2, 4$ were applied to the $\lambda_{2796}, \lambda_{2803}$ Mg~{\sc II} lines and quasar continuum, respectively. All redshift slices have a thickness of $\Delta z = \pm 0.060$. The columns from left to right are: the central redshift of the field being assessed; the linkage scale used for the field being assessed, calculated as $s = (\rho_0 / \rho)^{1/3}s_0$ (see the main text); the number of candidate structures identified in the field; the number of candidate structures identified in the field with a CHMS significance equal to or exceeding $3.5 \sigma$; the maximum CHMS significance calculated from the candidate structures; the maximum Mg~{\sc II} absorber membership identified from the candidate structures.  }
\end{table}
  
The BR that was originally identified visually appears almost fully (in a
partly-open ring) in only the central redshift slice $z = 0.802$ indicating
that this is the optimum redshift slice for the BR, as it was for the
GA. (Note, the GA is also identified, and is statistically significant in the
central redshift slice).  The four structures contributing to the
visually-identified BR are adjacent or overlapping on the sky, indicating
that the separate structures plausibly belong to the same structure.  The
apparent splitting of a seemingly coherent structure was also seen with the
GA, which was made up of two overlapping SLHC groups; the splitting of
structures is an example of the limitations of applying the SLHC algorithm to
an essentially incomplete dataset.  Interestingly, the SLHC group
corresponding to the bottom portion of the BR appears also to extend into the
two higher redshift slices, $z=0.862$ and $z=0.922$, since there are
similarly-shaped arcs (corresponding to the bottom portion of the BR)
appearing at the same on-sky position in all three redshift slices ($z=0.802,
0.862, 0.922$). (Note, the probes are not here restricted to be identical.)
  
As mentioned earlier, the condition of S/N~$\geq 6$ for the quasar continuum
could be too restrictive.  Anand21 calculate the S/N over the whole quasar
continuum, rather than the local continuum at the point of an absorber.
Therefore, the whole quasar continuum could have lower S/N overall compared
with local S/N at the position of an absorber.  Accordingly, we slightly
relax the S/N conditions to $4, 2, 4$ for the $\lambda_{2796},
\lambda_{2803}$ Mg~{\sc II} lines and quasar continuum, respectively and
repeat the above described analysis (see
Table~\ref{tab:5_z_slices_SN_4_2_4}).

\begin{figure}[h]
\centering
\includegraphics[scale=0.5]{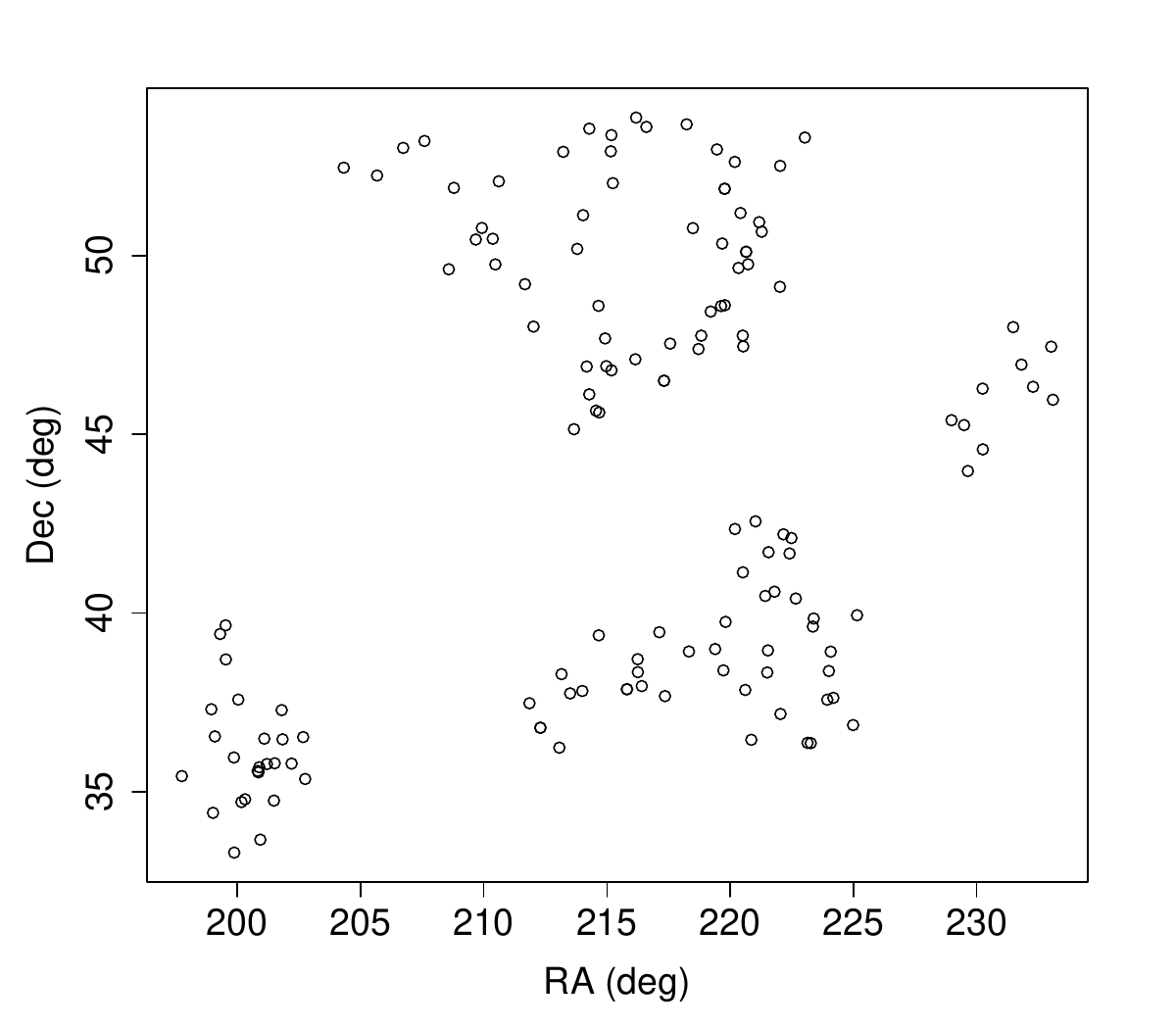}
\caption{\label{fig:SLHC_all_SN_4_2_4} All the Mg~{\sc II} absorbers
  belonging to a candidate structure identified by the SLHC algorithm from
  results-2 in the redshift slice centred at $z=0.802 \pm 0.060$. The
  field-of-view seen here corresponds to the pink points in
  Figure~\ref{fig:SDSS_control_fields}. S/N limits of: $4, 2, 4$ were applied
  to the $\lambda_{2796}, \lambda_{2803}$ Mg~{\sc II} lines and quasar
  continuum, respectively. The BR can been seen to the north of the centre
  point. The visually-identified BR is seen here with an additional extension
  heading towards to the north-west direction, and the visually-identified
  inner filament is the central line cutting through the BR here. The BR
  spans $\sim 10$~degrees in RA and Dec coordinates. The large structure
  south of the BR belongs to the previously identified GA.}
\end{figure}

We see again that the full BR is detected --- this time a full, closed ring,
and with the inclusion of the inner filament --- in the central redshift
slice $z=0.802 \pm 0.060$, Figure~\ref{fig:SLHC_all_SN_4_2_4}.  The BR is
located north of the centre point, spanning $\sim 10$~degrees in the RA and
Dec axes ($x$ and $y$ axes respectively), and is made up of a collection of
five structures identified by the SLHC algorithm.

\begin{figure}[h]
\centering 
\includegraphics[scale=0.5]{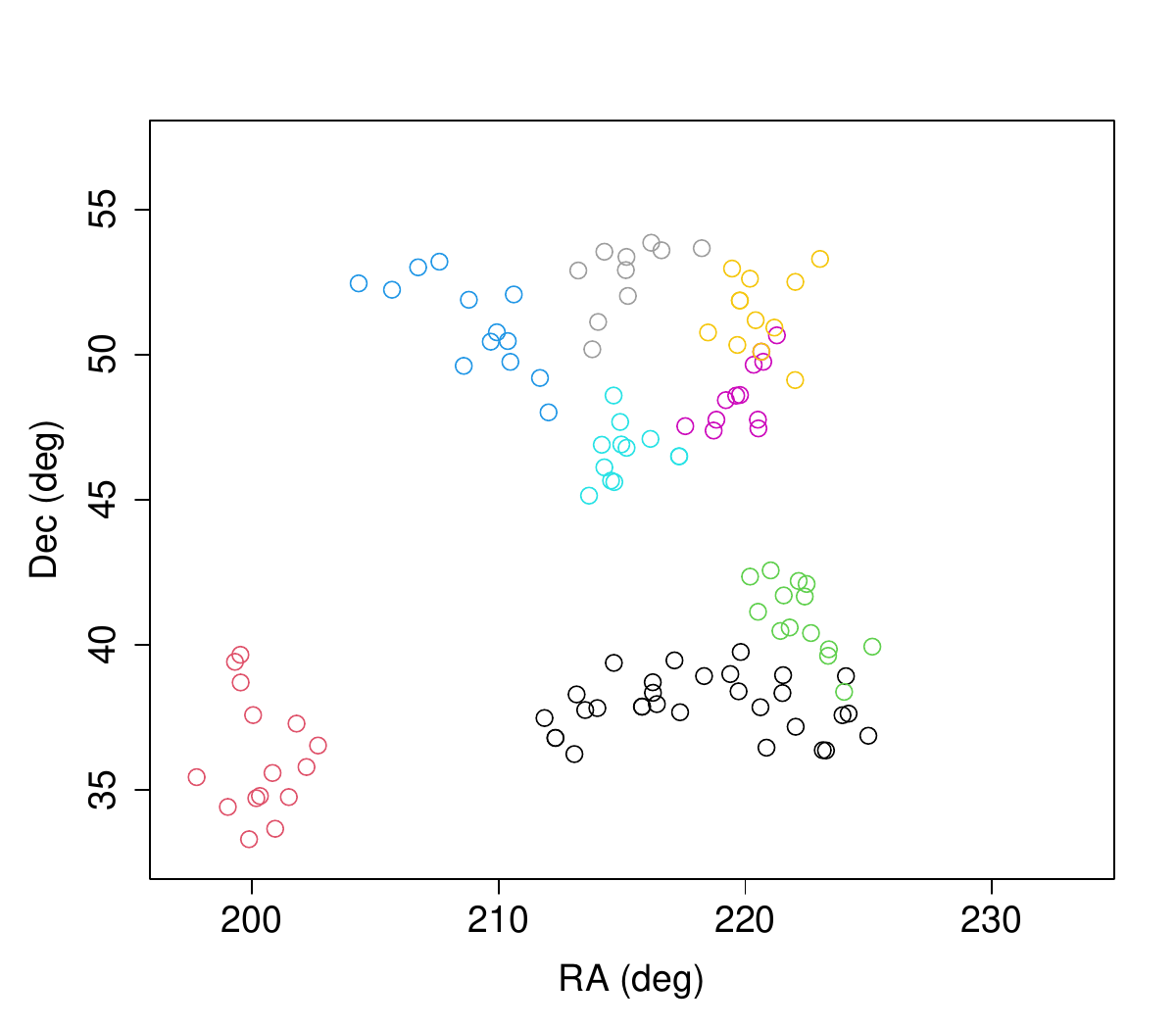}
\caption{\label{fig:SLHC_CHMS_BR_SN_4_2_4} Eight of the $10$ highest
  membership candidate structures identified by the SLHC / CHMS algorithms
  from results-2 in the redshift slice centred at $z=0.802 \pm 0.060$. The
  colours represent the memberships which are ordered from high to low in the
  following way: black, red, green, blue, turquoise, pink, yellow, grey. The
  field-of-view here corresponds to the pink points in
  Figure~\ref{fig:SDSS_control_fields}. The BR and inner filament are
  detected, but separated into five structures, that can visually be seen
  adjacent to each other or overlapping. In this figure, only the black
  points, representing absorbers belonging to the GA, are statistically
  significant. }
\end{figure}

Note the very high similarity of the two tables of SLHC / CHMS results
(Tables~\ref{tab:5_z_slices_SN_4_2_6} and \ref{tab:5_z_slices_SN_4_2_4}),
where the only change is reducing the S/N limit of the quasar continuum from
$6$ to $4$.  For convenience, the first set of results with the more
restrictive S/N limits will be referred to by MST results-1, and the second
set of results with the less restrictive S/N limits will be referred to by
MST results-2.  We deduce that setting restrictive S/N limits ($>4$) to the
quasar continuum, after already applying S/N limits to the Mg~{\sc II} lines,
is not absolutely essential, and possibly adds to the incompleteness of data
in a detrimental way.  An overview of the results is as follows.  (1) In the
two lowest redshift slices, in MST results-1 and results-2, there is only one
statistically-significant ($>3.5 \sigma$) structure detected in total.  The
statistically-significant structure belongs to the redshift slice centred at
$z=0.742 \pm 0.060$ from results-1; it is a small group of absorbers located
at the lower LHS of the Mg~{\sc II} image, and of no relevance to the BR.  In
addition, other than a possibility of a thin filament forming in the Mg~{\sc
  II} image in the redshift slice centred at $z=0.742$, there is no strong
indication or detection of the BR in the two lowest redshift slices.  (2) In
the central redshift slice, $z=0.802$, both results-1 and results-2 find one
significant structure corresponding to the GA. The BR is separated into four
and five (respective to results-1 and results-2) individual, adjacent or
overlapping, structures, that are statistically insignificant on their own
(see Figure~\ref{fig:SLHC_CHMS_BR_SN_4_2_4}).
\begin{figure}[h]
\centering 
\includegraphics[scale=0.5]{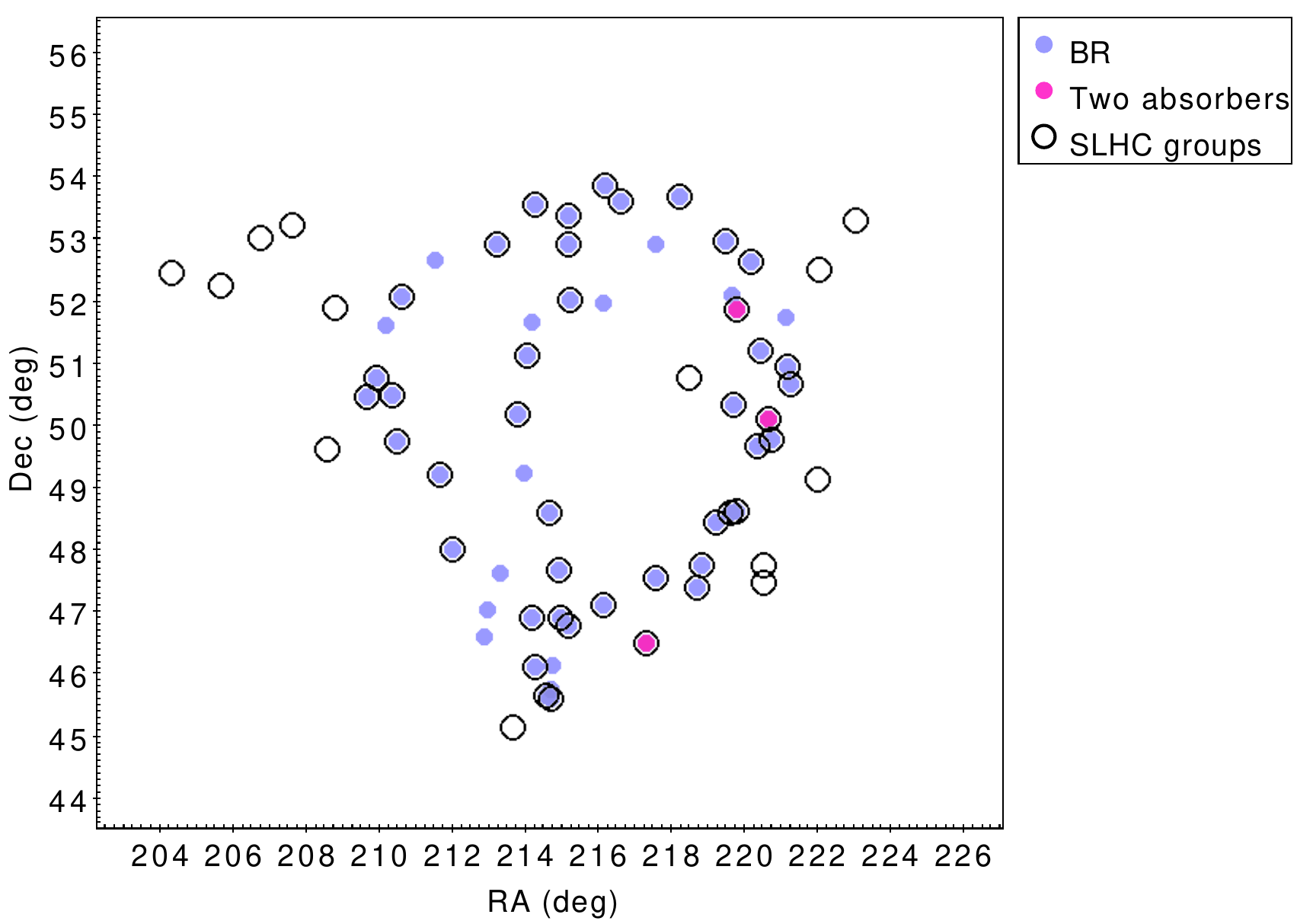}
\caption{\label{fig:BR_and_SLHC_groups} The visually-identified BR absorbers
  (lilac) and the SLHC-identified absorbers belonging to the $5$ candidate
  structures from results-2 at $z=0.802 \pm 0.060$ (black circles) that
  correspond to the visually-identified BR. The pink points indicate the
  positions where there are two absorbers occurring in a single
  spectrum. There are $46$ out of $59$ absorbers in common to the
  SLHC-identified and visually-identified BR and inner filament absorbers. Of
  the $13$ absorbers in the visually-identified BR and inner filament
  absorbers that were not connected by the SLHC algorithm, $9$ of these occur
  at the most extreme edges of the redshift range, possibly explaining their
  exclusion. }
\end{figure}
\begin{figure}[h]
\centering
\includegraphics[scale=0.5]{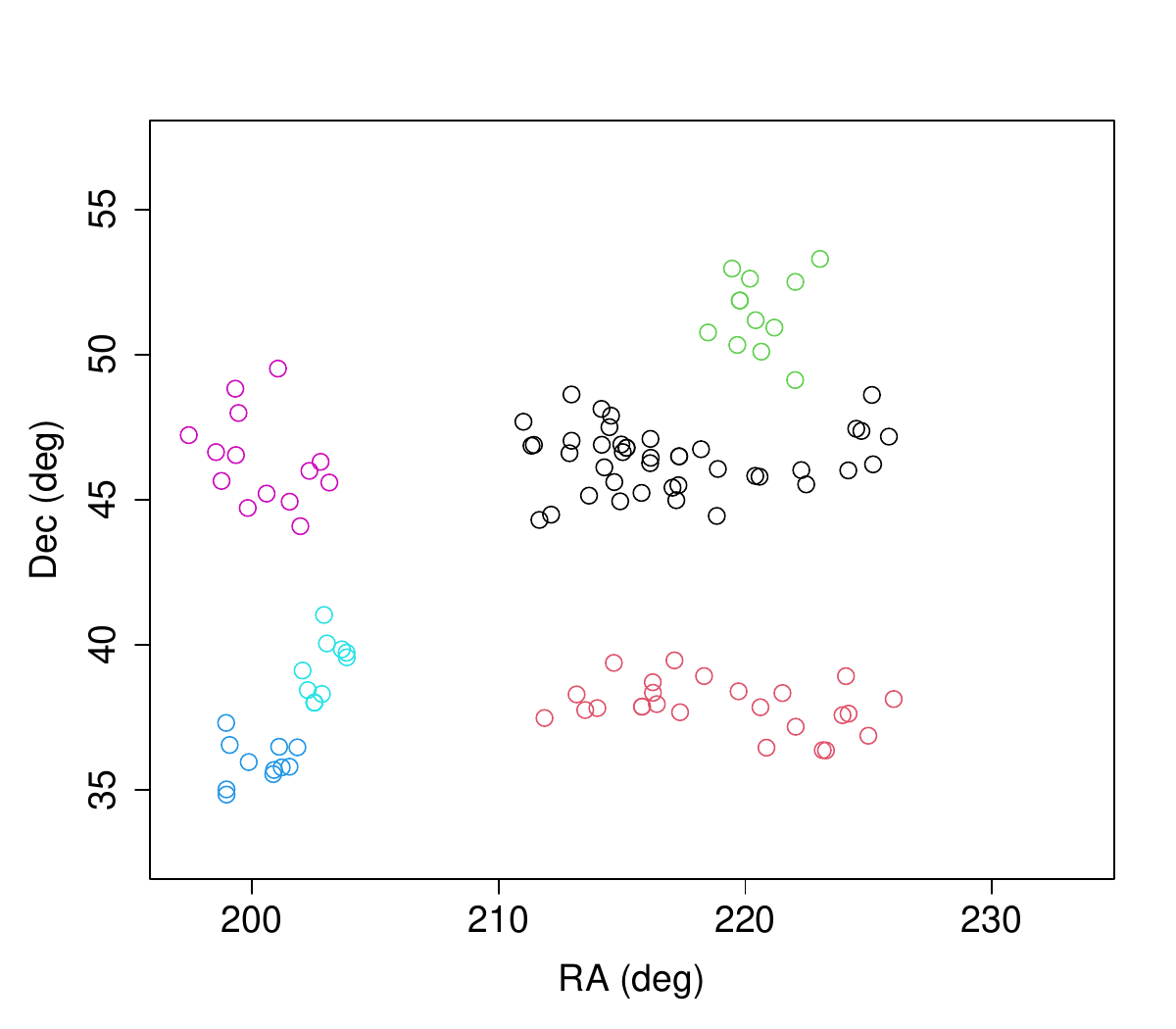}
\caption{\label{fig:SLHC_0_862_results_1} The $6$ candidate structures
  identified by the SLHC algorithm from results-1 in the redshift slice
  centred at $z=0.862 \pm 0.060$. The different colours represent the
  significances, and are ordered from high to low in the following way:
  black, red, green, blue, turquoise, pink. The two most significant
  structures belong to the BR and GA respectively, both which are also
  statistically significant.}
\end{figure}
\begin{figure}[h]
\centering
\includegraphics[scale=0.5]{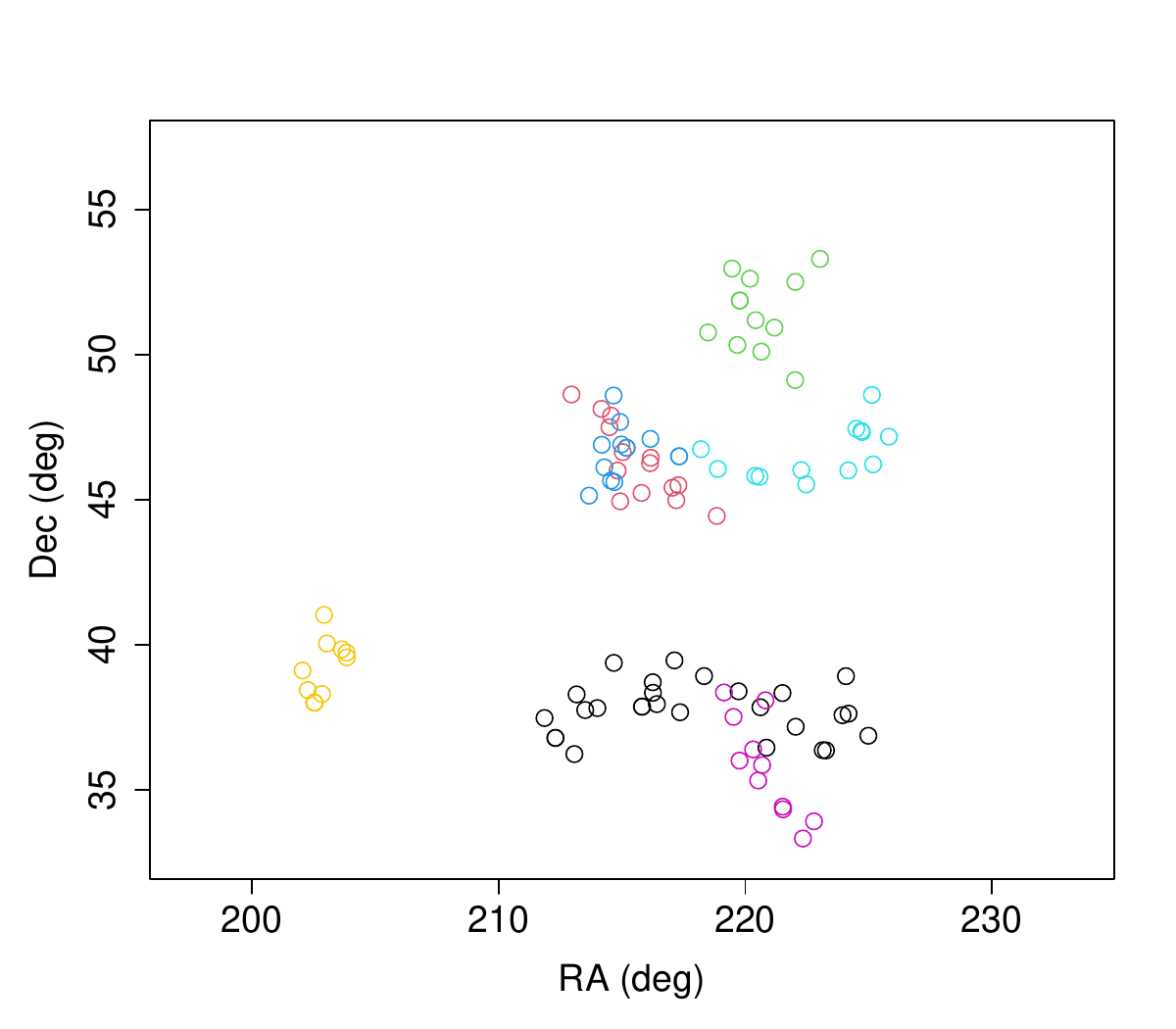}
\caption{\label{fig:SLHC_0_862_results_2} The $7$ candidate structures
  identified by the SLHC algorithm from results-2 in the redshift slice
  centred at $z=0.862 \pm 0.060$. The different colours represent the
  significances, and are ordered from high to low in the following way:
  black, red, green, blue, turquoise, pink, yellow. The absorbers belonging
  to the GA are statistically significant, shown in black. By comparing this
  figure with Figure~\ref{fig:SLHC_0_862_results_1} we can see that the
  absorbers identified by the $3$ individual, overlapping or adjacent
  structures, coloured red, blue and turquoise, are clearly the same
  absorbers identified by \emph{one} full, statistically-significant
  structure that was found in results-1, highlighting the complications of
  applying the SLHC algorithm to an essentially incomplete dataset.}
\end{figure}

We could again be seeing here limitations of the Mg~{\sc II} method, as was
seen with the GA, since the individual structures are overlapping or
adjacent.  In results-2, the visually-identified BR and inner filament
absorbers are mostly detected ($46$ out of $59$ absorbers, $78 \%$) and form
the full BR shape (see Figure~\ref{fig:BR_and_SLHC_groups}).
(3) In the two highest redshift slices there are totals of $3$ and $1$
significant structures from MST results-1 and results-2 respectively.  For
results-1 the most significant structure in both redshift slices is an arc
corresponding to the lower portion of the BR (see
Figure~\ref{fig:SLHC_0_862_results_1}).

\begin{figure}[tbp]
\centering
\includegraphics[scale=0.4]{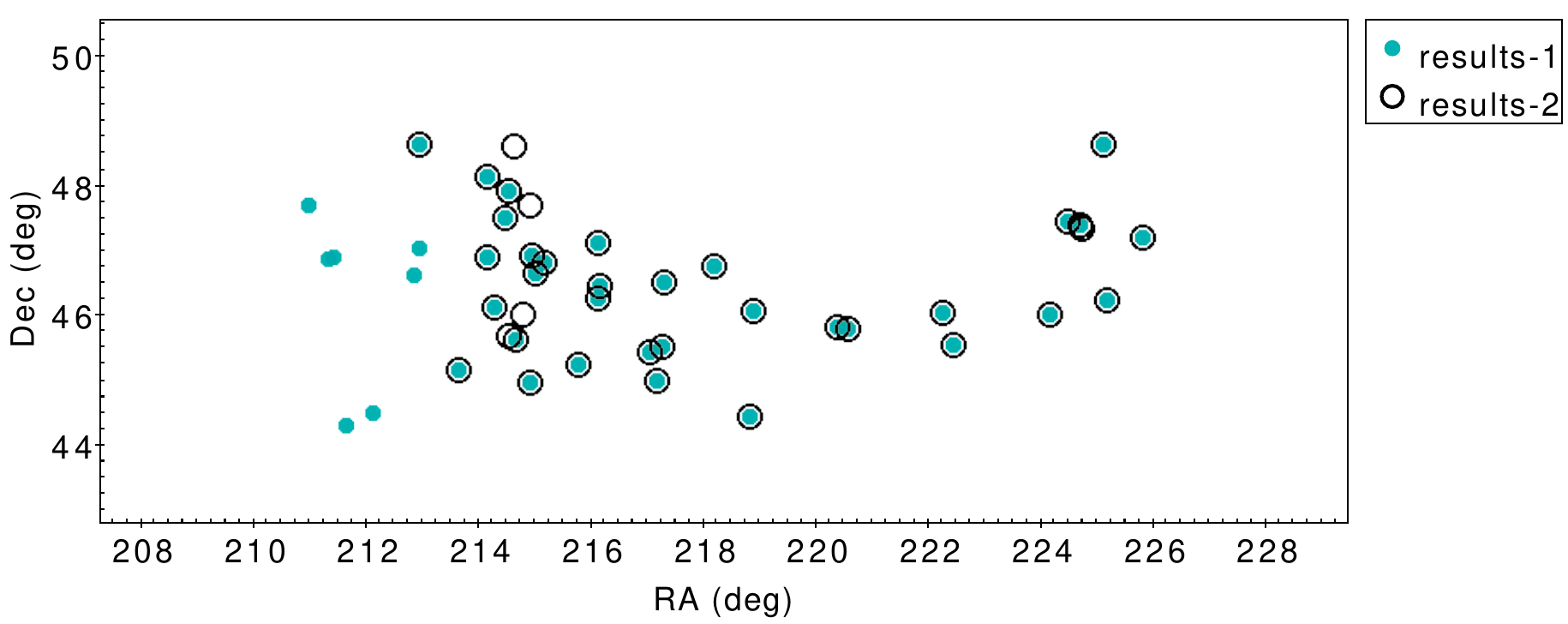}
\caption{\label{fig:BR_lower_arc_results-1_with_results-2} Mg~{\sc II}
  absorbers in the redshift slice $z = 0.862 \pm 0.060$ belonging to the arc
  identified by the SLHC algorithm as one, full, candidate structure from
  results-1 (turquoise) and as three, overlapping or adjacent, candidate
  structures from results-2 (black circles). There are $42$ absorbers belonging
  to results-1, four of which appear as multiples per probe in two quasars,
  and $40$ absorbers belonging to results-2, the same four of which appear as
  multiples per probe in two quasars. Both results connect many of the same
  absorbers, but in results-1, where the linkage scale is higher, there are
  additional absorbers connected by the SLHC algorithm all appearing on the
  LHS of the arc. The additional absorbers in results-2 occur in the middle
  of the arc, and there are much fewer additional absorbers than for
  results-1, suggesting that these absorbers were identified due to the
  lowered restriction on the S/N on the continuum.}
\end{figure}

However, in results-2, we find the same arc that was detected in results-1,
but over multiple structures, highlighting again the nuances of applying the
SLHC algorithm to an essentially incomplete dataset.  To clarify this, reducing
the S/N limits in the quasar continuum from $6$ to $4$ initially appears
inconsequential --- i.e., their Mg~{\sc II} images appear on the whole
unchanged and there is only a $15 \%$ increase of absorbers in the whole
field from results-1 to results-2.  But, when the SLHC algorithm is applied,
then this small increase in absorbers increases the density-scaled linkage
scale thus creating more broken structures.  We scale the linkage scale of
the SLHC algorithm to the density for a general approach to the wide-varying
densities in the dataset (mostly due to survey bias).  We can see that the
absorbers identified as candidate structures corresponding to the arc in the
redshift slice $z=0.862 \pm 0.060$ in results-2 are very similar to the
absorbers identified as one candidate structure in results-1.  To further
clarify this point, compare Figure~\ref{fig:SLHC_0_862_results_1} with
Figure~\ref{fig:SLHC_0_862_results_2}, and see also
Figure~\ref{fig:BR_lower_arc_results-1_with_results-2}.
There are $42$ absorbers in the SLHC-identified arc in the redshift slice
$z = 0.862 \pm 0.060$ from results-1 represented by the black points in
Figure~\ref{fig:SLHC_0_862_results_1}, and a combined total of $40$ absorbers
in the three SLHC-identified candidate structures corresponding to the arc in
the redshift slice $z = 0.862 \pm 0.060$ from results-2 represented by the
red, blue and turquoise points in Figure~\ref{fig:SLHC_0_862_results_2}.
$35$ of these absorbers from results-1 and results-2 are in common to both
results, confirming that the SLHC identified a high fraction ($67 \%$
overlap) of the same absorbers between the two results.  In fact, all $7$ of
the additional absorbers connected by results-1 all occur on the LHS of the
arc, indicating that the increased linkage scale in results-1 (due to lower
field density) is only \emph{extending} the arc, and not responsible for the
central components of the arc.  Due to the above, we reason that individual,
SLHC-identified candidate structures that are overlapping or adjacent could
reasonably be connected as one structure if given a more complete dataset.

In conclusion, the BR appears most obvious in the original, central redshift
slice, $z=0.802 \pm 0.060$, both visually and by the presence of individual
SLHC structures overlapping or adjacent on the sky that comprise the
visually-identified BR.  The strongly, detected arc corresponding to the
lower portion of the BR detected in a slightly higher redshift slice
($z=0.862 \pm 0.060$) is also of particular interest in relation to the BR+GA
system.

\subsection{Significance: CHMS and MST}
\label{subsec:CHMS_MST}
The CHMS and MST significance calculations are applied to: the
SLHC-identified absorbers; the visually-identified absorbers; and the
FilFinder-identified absorbers.  In addition, the Alpha Hull algorithm is
applied to the visually-identified absorbers for an estimate of the volume,
overdensity and significance using simple Poisson statistics.

\subsubsection{SLHC-identified Mg~{\sc II} absorbers}
\label{sec:SLHC_identified}
To determine the significance of the BR in its entirety, we take the
SLHC-identified absorbers from results-2 that make up the BR (i.e., the BR in
Figure~\ref{fig:SLHC_all_SN_4_2_4}), and apply the CHMS algorithm.  Remember
that the CHMS has the ability to assess the significance of a structure by
comparing the observed convex-hull volume with the volumes that would be
expected for a set of random distributions of those same absorbers at the
control density of absorbers for the same redshift interval.  Based on the
definition of the CHMS volume and significance calculation, the algorithm is
optimal when applied to clumpy structures, with no obvious gaps, holes or
curvature that would lead to an overestimation of the volume.  Clearly, this
is not the case with the BR, having a large volume mostly unoccupied by
absorbers in its centre.  Instead, the MST-significance calculation
introduced by \cite{Pilipenko2007} could be more appropriate, which uses the
mean MST edge-length between neighbouring data points.  Using both methods we
then find that the SLHC-identified BR in its entirety has a CHMS significance
of $3.6 \sigma$, and an MST significance of $4.7 \sigma$.  Both tests
indicate statistical significance, but we can see that CHMS has likely
overestimated the BR volume leading to a much lower significance compared
with the MST-significance test.

\subsubsection{Visually-identified Mg~{\sc II} absorbers}
\label{sec:visually_identified}
We then take the visually selected absorbers of the BR, and everything within
the BR, for which there are a total of $62$ absorbers (see
Figure~\ref{fig:BR_and_filament_topcat}), and apply the CHMS to these
absorbers only.  The CHMS algorithm then calculates a significance of $5.2
\sigma$.  The significance calculated here is likely the upper limit estimate
for the BR, as the algorithm was applied to those absorbers that were
visually-selected.  We can similarly repeat this work for the BR-only
absorbers, as well as compare the MST significance with the CHMS
significance.  Clearly, removing the BR innards will reduce the significance
calculations for the CHMS, since the volume will remain the same but the
number of absorbers will be reduced.  In contrast, for the MST significance,
the mean MST edge length may not be greatly affected by removing the BR
innards, as seen in Table~\ref{tab:CHMS_MST_results}.
\begin{table}[]
\centering
\begin{tabular}{|c|c|c|c|}
\hline
& \begin{tabular}[c]{@{}l@{}}No. Mg~{\sc II} absorber \\ members\end{tabular} & \begin{tabular}[c]{@{}l@{}}CHMS signif.\end{tabular} & \begin{tabular}[c]{@{}l@{}}MST signif.\end{tabular} \\ \hline
BR only & 51 & 3.3 & 4.0 \\ \hline
BR all & 62 & 5.2 & 4.1 \\ \hline
\end{tabular}
\caption{\label{tab:CHMS_MST_results} The CHMS and MST significances for the
  BR-all and BR-only absorbers. The CHMS significance is dependent on the
  volume and number of absorbers so, clearly, removing the inner absorber
  members and keeping the volume the same will reduce the CHMS
  significance. However, for the MST significance, the volume is not directly
  related, but instead it is related to the MST mean edge lengths. This is
  clearly shown in the results as the MST significance stays mostly the same,
  at $\sim 4 \sigma$, and the CHMS significance decreases from $5.2 \sigma$
  to $3.3 \sigma$ by excluding the absorbers enveloped by the BR.}
\end{table}
We find that on both occasions, using the BR-all and BR-only absorbers, the
MST significance is roughly the same, at $\sim 4 \sigma$.  In contrast, for
the CHMS calculation, the significance drops from $5.2 \sigma$ to $3.3
\sigma$ after removing the absorbers contained within the BR.  However, even
after removing all of the BR inner absorbers the CHMS significance is still
greater than $3.0 \sigma$, but not quite reaching $3.5 \sigma$ which is the
usual standard we apply for comparing structures.  Of course, the CHMS
calculation on the BR-\emph{only} absorbers gives a drastic
under-representation of the true significance since we have forcefully
removed absorbers contained within the BR while keeping the unique volume the
same.

For comparison, we can estimate the volume of the BR using the 2D Alpha
Hull algorithm.  First, the Alpha Hull area of the BR is repeatedly
calculated $500$ times from the BR by drawing a cloud of points in a circle
with radius equal to half the mean MST edge length around each of the Mg~{\sc
  II} absorbers in the BR.  Then, the area is multiplied by the physical size
of the redshift range of the BR absorber members.  In this manner, we are
calculating the volume of a somewhat cylindrical, tube shape.  The benefits
of this method versus the CHMS is that we can eliminate the central region of
the BR where there are very few absorbers.  The downside of this method is
that, although the absorbers on the sky make up a ring shape, we later see
that the 3D distribution of the absorbers is more of a coil shape, so we
are again overestimating the volume of the BR (and underestimating the
overdensity and significance of the structure).  Nevertheless, using this
method we obtain a volume of $21.8 \times 10^{6}$~Mpc$^3$, an overdensity of
$0.75$ and a significance of $4.0 \sigma$ for the number of absorbers in this
volume based on Poisson statistics.  The significance calculated from the
Alpha Hull here, although simple, agrees with the MST-significances of the
SLHC-identified absorbers and the visually-identified absorbers, which adds
further confidence to the statistical assessment of this structure.

\subsubsection{FilFinder-identified Mg~{\sc II} absorbers}
\label{sec:FilFinder_identified}
\begin{figure}[tbp]
\centering
\includegraphics[scale=0.5]{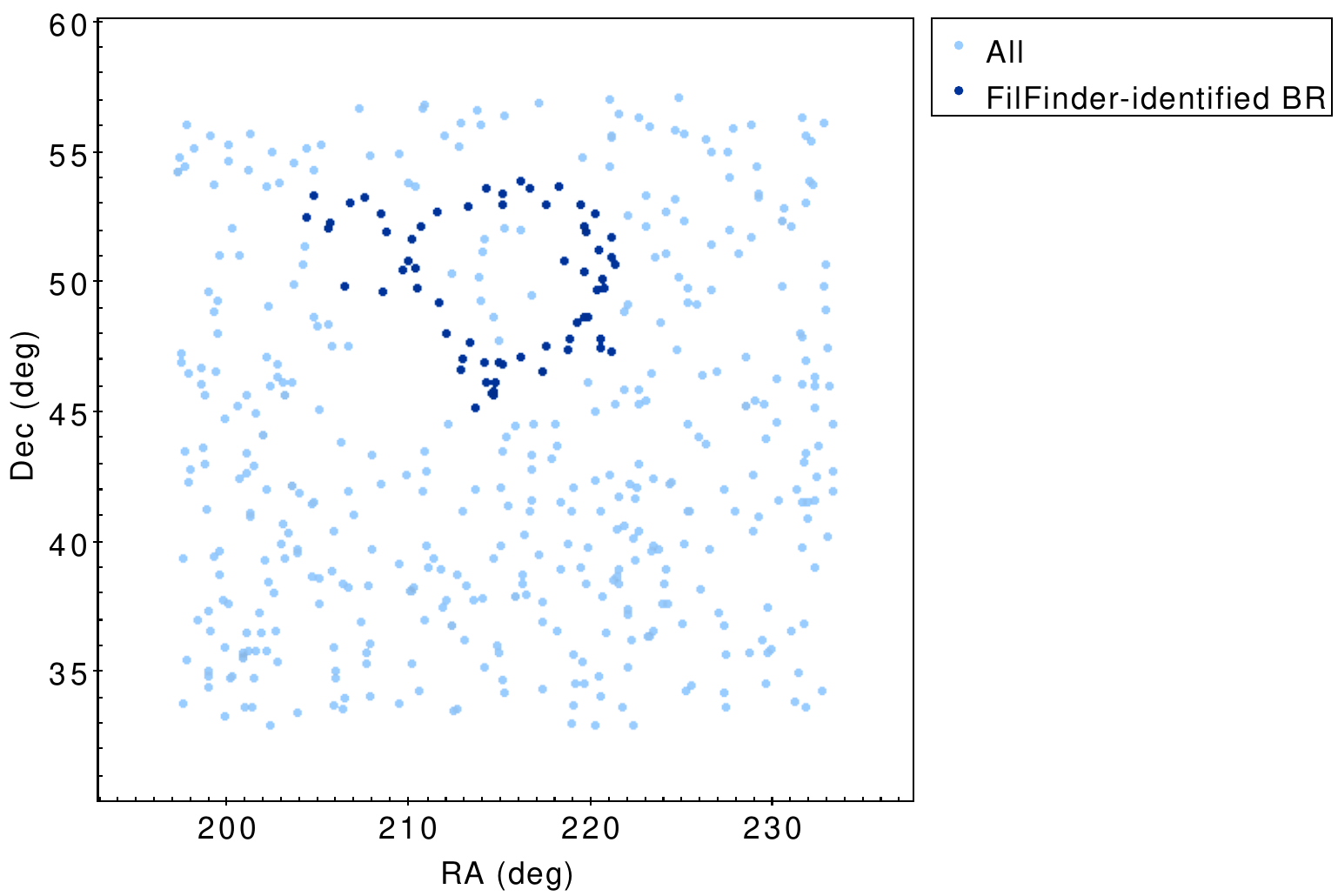}
\caption{\label{fig:topcat_filfinder_identified} The Mg~{\sc II} absorbers
  belonging to the whole BR field, represented by light blue points, and the
  FilFinder-identified BR absorbers represented by dark blue points.}
\end{figure}

Finally, we have seen the CHMS algorithm as well as the MST significance test
applied to the SLHC-identified and visually-identified absorbers, so now we
will apply both algorithms to the FilFinder-identified absorbers.  The
FilFinder work is discussed next, in Section~\ref{subsec:FilFinder}, and we
are referencing Figure~\ref{subfig:filfinder_AT_18_ST_2000} for the work
in this section.  The FilFinder algorithm is applied to 2D pixel images, so
the physical absorber points are irrelevant to the algorithm, but we can
estimate the absorbers identified by the filament by including those that are
attached to the filament in the figure.  Our estimation of the absorbers
connected by the FilFinder algorithm is shown in
Figure~\ref{fig:topcat_filfinder_identified}.
We can intuitively see that a volume containing all of the
FilFinder-identified absorbers will include many large, empty sub-volumes ---
e.g., the centre of the BR and the volume between the `extended filaments'
connected to the BR.  The BR volume was also overestimated with all previous
identifications of the BR, which can be confirmed with the comparison of, in
particular, the BR-all with the BR-only absorbers, and the CHMS with the MST
significances.  So, when the CHMS algorithm is applied to the
FilFinder-identified absorbers, it calculates a $2.5 \sigma$ significance.
In contrast, the MST-significance test estimates that the
FilFinder-identified absorbers have a significance of $3.6 \sigma$, which is
generally in agreement with all previous estimates of the BR significance
using the MST calculation and the Alpha Hull Poisson statistics.  So, again,
the BR volume is considerably overestimated, this time using the
FilFinder-identified absorbers, and the CHMS significance is therefore
underestimated as a consequence, which can be confirmed with the use of the
MST-significance test.

\subsubsection{Summary of significances}
We have shown our application of the CHMS and MST significance calculations
to four sets of uniquely-identified BR absorber members.
Table~\ref{tab:all_significances} summarises all of the above significance
calculations.
\begin{table}[]
\centering
\begin{tabular}{|c|c|c|c|c|}
\hline
& SLHC & \begin{tabular}[c]{@{}l@{}}Visual\\ BR-all\end{tabular} & \begin{tabular}[c]{@{}l@{}}Visual\\ BR-only\end{tabular} & FilFinder \\ \hline
CHMS & 3.6 & 5.2 & 3.3 & 2.5 \\ \hline
MST & 4.7 & 4.1 & 4.0 & 3.6 \\ \hline
\end{tabular}
\caption{\label{tab:all_significances} The CHMS and MST significances of the
  BR calculated for each set of identified absorbers (SLHC, visual BR-all,
  visual BR-only and FilFinder). }
\end{table}
The results are: the total mean significance of the BR (all results from
Table~\ref{tab:all_significances} considered) is ($3.88 \pm 0.83$)$\sigma$;
the mean CHMS-significance is ($3.65 \pm 1.13$)$\sigma$; and the mean
MST-significance is ($4.10 \pm 0.45$)$\sigma$.  From these results we can see
that the variation of the CHMS-significance is a considerable fraction of the
average, indicating its results are to be taken with caution.  On the other
hand, the MST-significance has much lower spread, indicating that for the
purpose of analysing the BR, the MST-significance test may be more
appropriate, given the difficulty of defining a volume around a ring-like
structure without incorporating over-estimations of the volume (as is the
case with the CHMS algorithm).

\subsection{Filament identification algorithm}
\label{subsec:FilFinder}
FilFinder is a filament identification algorithm created by \cite{Koch2015}.
It uses mathematical morphology to identify filaments ranging in size, shape
and brightness on a 2D pixel image. 
\begin{figure*}[h]
\centering
\begin{subfigure}[b]{0.475\textwidth}
\centering
\includegraphics[width=\textwidth]{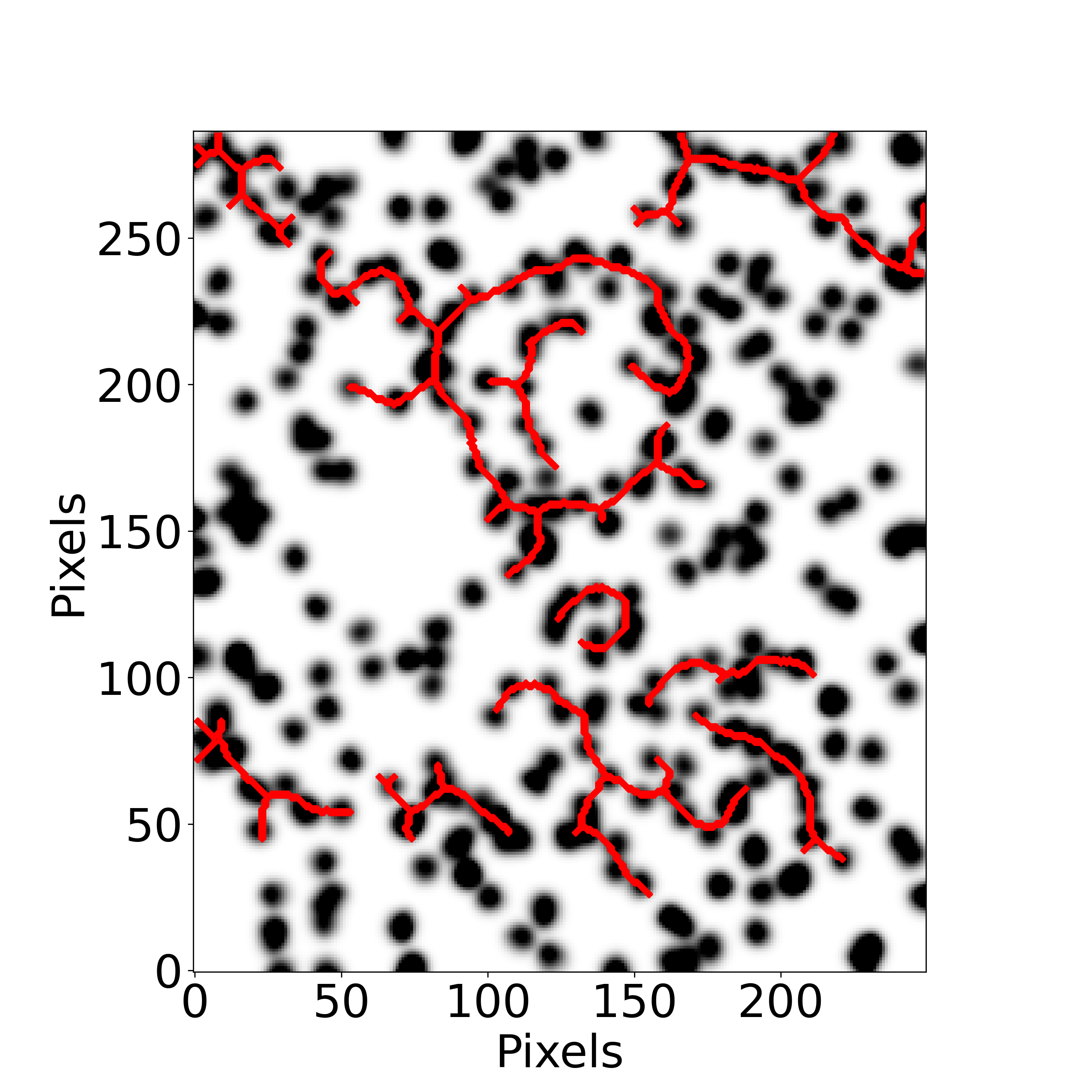}
\caption{\small }
\label{subfig:filfinder_new_standard}
\end{subfigure}
\hfill
\begin{subfigure}[b]{0.475\textwidth}
\centering
\includegraphics[width=\textwidth]{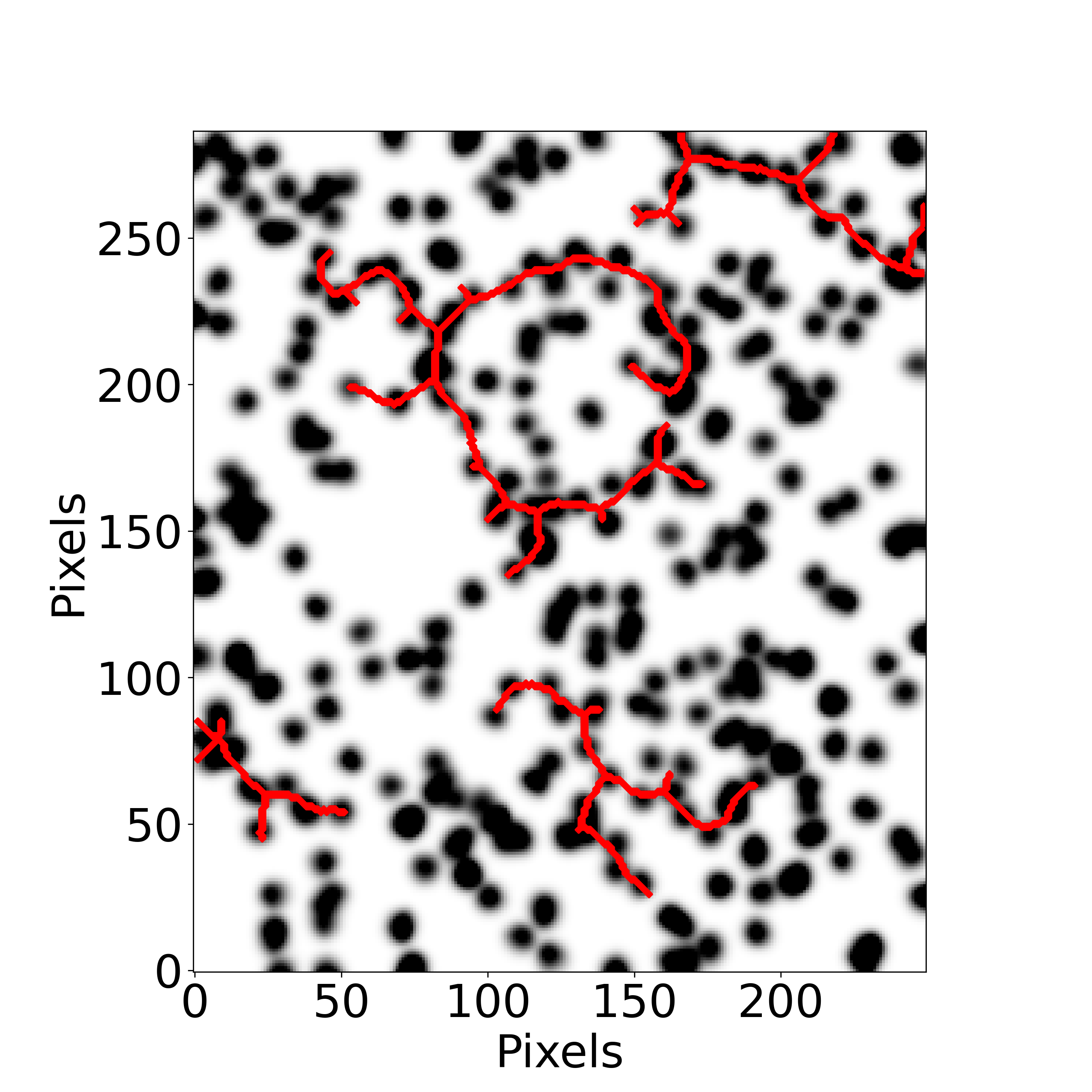}
\caption{\small }
\label{subfig:filfinder_AT_18_ST_800}
\end{subfigure}
\vskip\baselineskip
\begin{subfigure}[b]{0.475\textwidth}
\centering
\includegraphics[width=\textwidth]{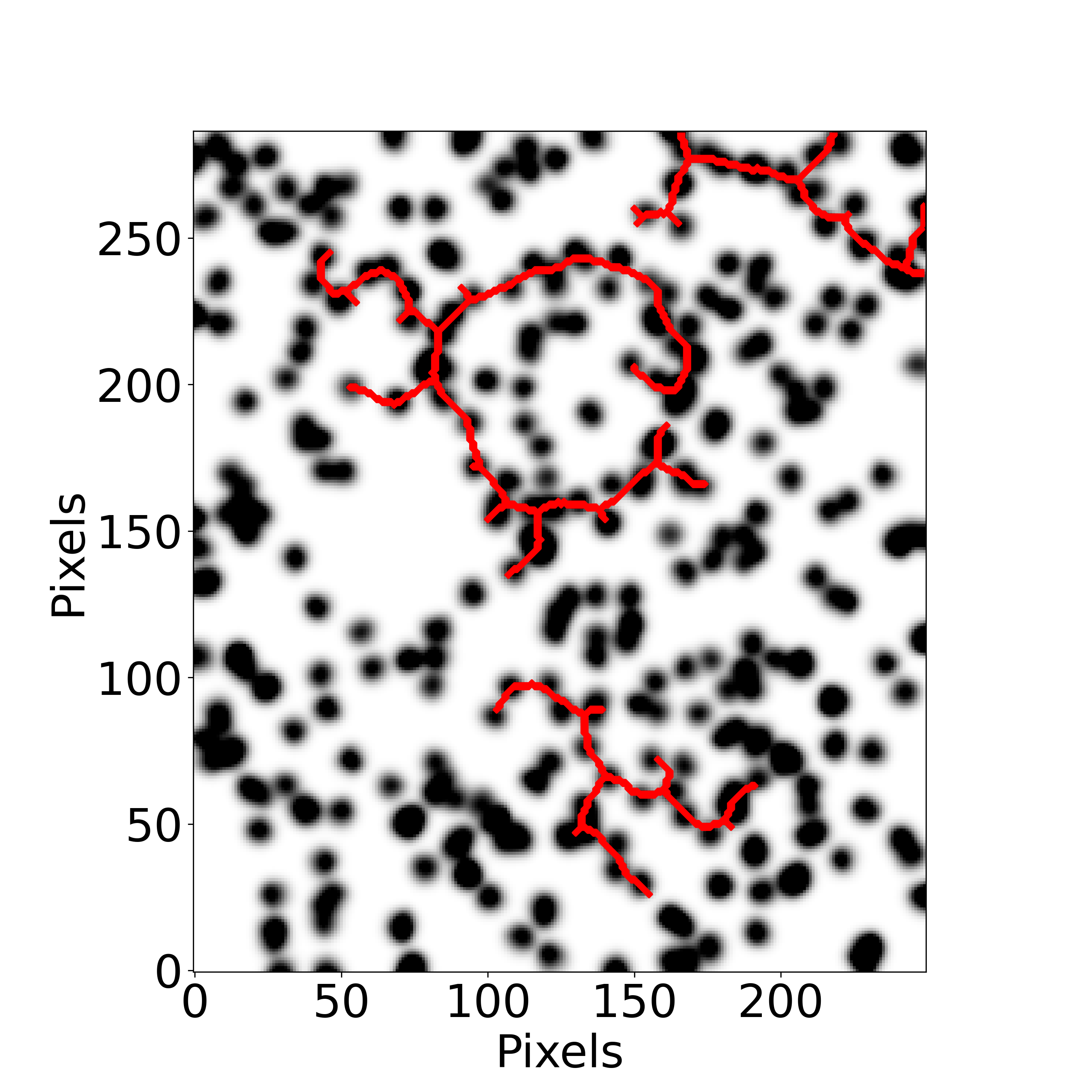}
\caption{\small }
\label{subfig:filfinder_AT_18_ST_1000}
\end{subfigure}
\hfill
\begin{subfigure}[b]{0.475\textwidth}
\includegraphics[width=\textwidth]{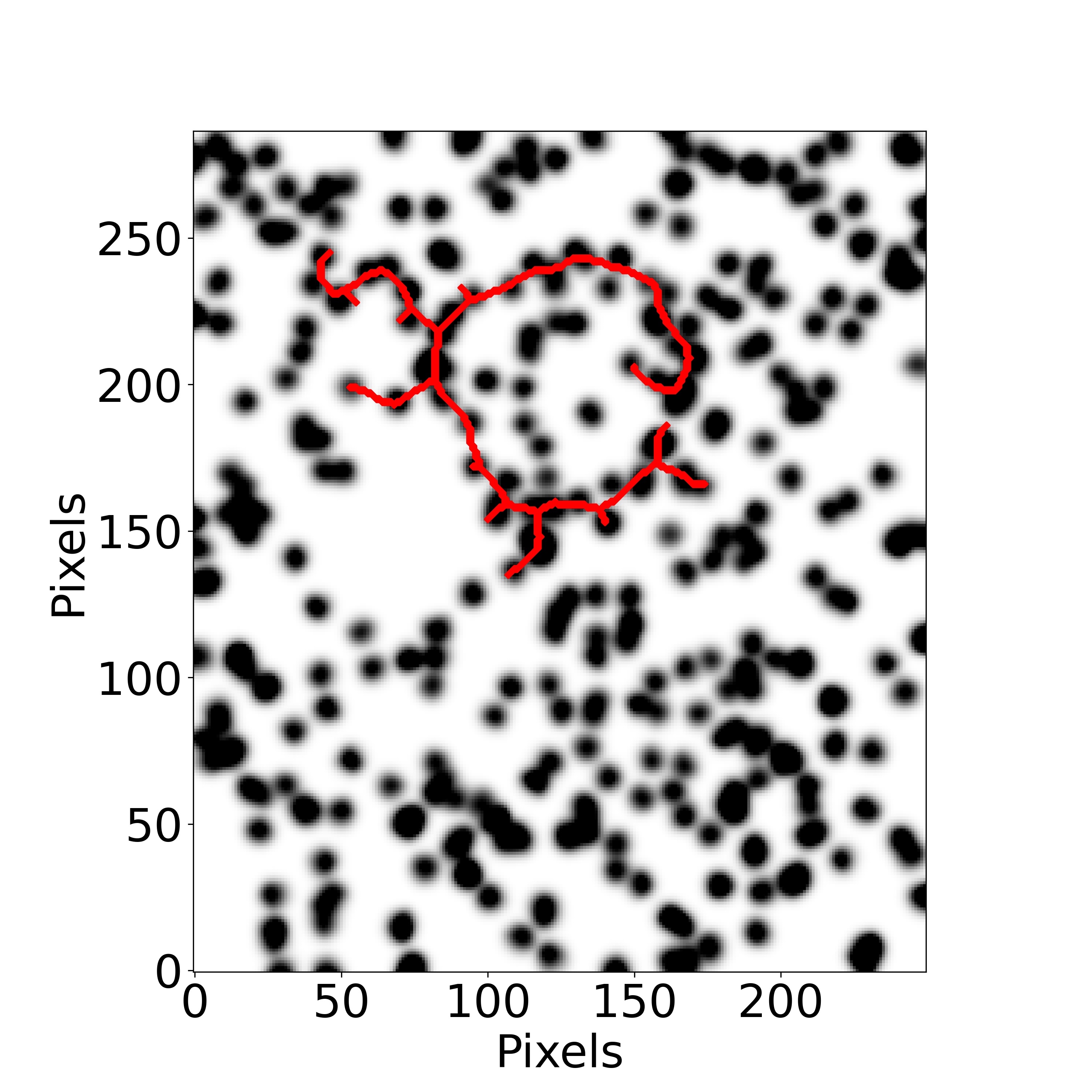}
\caption{\small }
\label{subfig:filfinder_AT_18_ST_2000}
\end{subfigure}
\caption{Applying the FilFinder algorithm to the standard BR field and
  increasing the size threshold incrementally to show the elimination process
  of small filaments. All axes are labelled in pixels, where $1$~pixel~$=
  4^2$~Mpc$^2$. (a) The standard parameters, for which adaptive threshold is
  $18$ pixels, smooth size is $12$ pixels, and size threshold is $576$
  pixels. For the remaining figures we incrementally increase the size
  threshold. (b) Size threshold is $800$ pixels. (c) Size threshold is $1000$
  pixels. (d) Size threshold is $2000$ pixels. }
\label{figs:filfinder_increasing_ST}
\end{figure*}
The algorithm was intended for use in
small ($10^{1}$ --- $10^{3}$~pc), gaseous areas, such as star formation
regions and the interstellar medium \cite{Mookerjea2023, Zhang2023,
  Meidt2023}.  
\begin{figure}[h]
\centering
\includegraphics[scale=0.6]{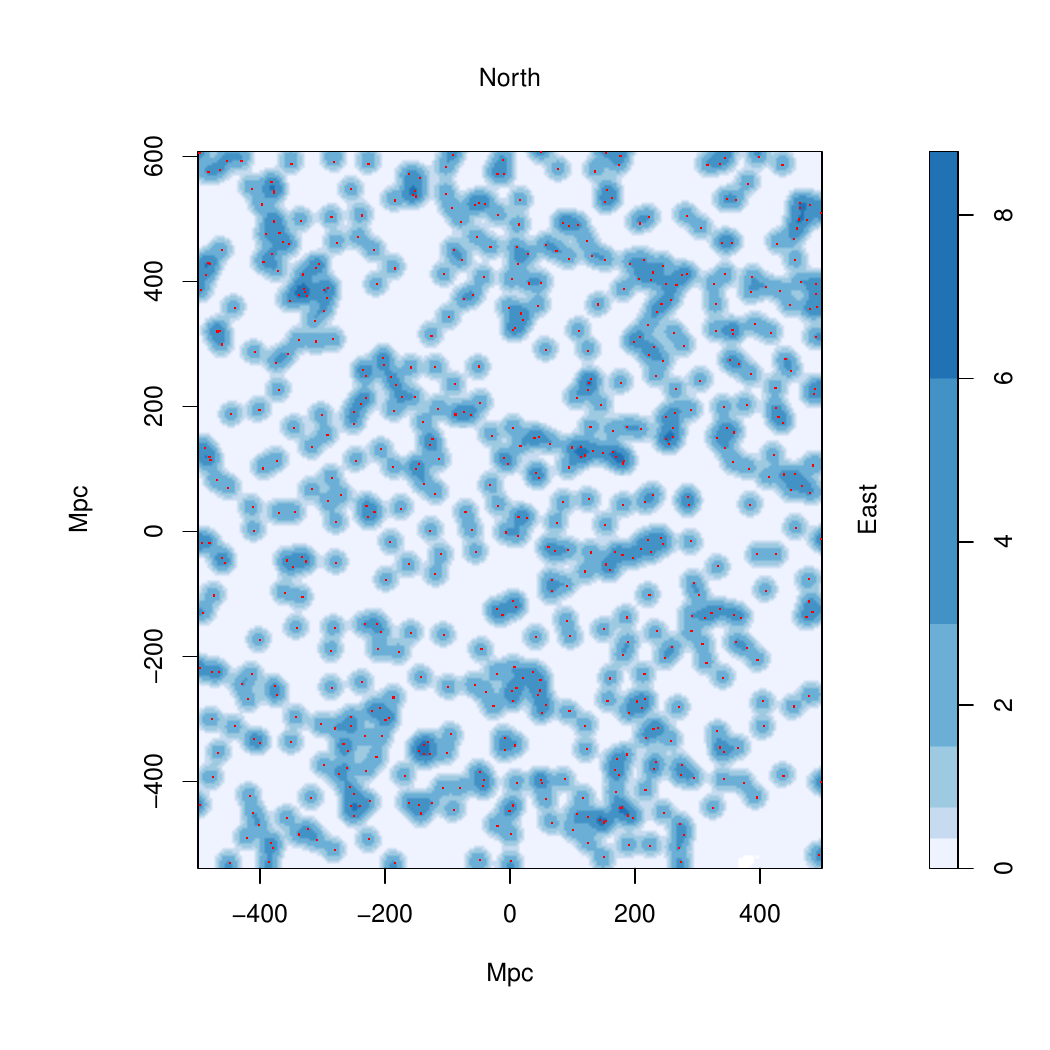}
\caption{\label{fig:BR_field_quasars} The tangent-plane distribution of
  quasars in the redshift slice $z=0.802 \pm 0.060$. The blue contours,
  increasing by a factor of two, represent the density distribution of the
  field quasars which have been smoothed using a Gaussian kernel of $\sigma =
  11$~Mpc. Magnitude limits have been applied to the quasars such that $i
  \leq 20.0$. The field-of-view corresponds to the small, pink area seen in
  Figure~\ref{fig:SDSS_control_fields}. There is no strikingly obvious
  structure by eye, but when the FilFinder algorithm is applied to this field
  it can detect a filamentary ring-like shape that coincides with the BR. }
\end{figure}
So, applying it to cosmological LSS is new, and we have had to
make adaptations in the parameter settings to suit the data.  We use the
following input parameter settings.  (1) Adaptive threshold --- the expected
width of a typical filament.  We choose a value (in number of pixels)
equivalent to a filament that is only one absorber wide.  Since the Mg~{\sc
  II} absorbers in the Mg~{\sc II} images span a diameter of approximately
$12$ pixels, the adaptive threshold is set to $12$.  (2) Smooth size ---
scale on which to smooth the data.  Our data is already smoothed and
flat-fielded, so it is possibly unnecessary to smooth beyond the size of a
Mg~{\sc II} absorber: i.e., we shall implement a smoothing size that does not
affect the already smoothed Mg~{\sc II} absorbers.  Therefore, we set smooth
size to a value of $12$ pixels.  (3) Size threshold --- the smallest area to
be considered a filament.  Using again the estimated value for the number of
pixels across a single Mg~{\sc II} absorber, we set this value at $4 \times
12^2 = 576$ pixels.  That is, the minimum area to be considered a filament is
made up of 4 Mg~{\sc II} absorbers (with the absorbers imagined as squares
rather than as circles, for simplicity).

We use the FilFinder package available in Python to objectively identify the
filaments present in the BR field.  For the above parameters we find that
there are no visually obvious consequences for relatively small changes
(e.g., within a few percent of the chosen value).  However, much larger
changes in the parameter choices render noticeably different, and
consequential, results.  For this reason we experiment with changing the
standard parameter values above to gain a better understanding of the
FilFinder algorithm when applied to cosmological data.  The FilFinder
algorithm is applied multiple times, with different parameter settings, on
the Mg~{\sc II} image containing the BR (Figure~\ref{fig:BR}), and the
results are discussed below.

\begin{figure*}[h]
\centering
\begin{subfigure}[b]{0.475\textwidth}
\centering
\includegraphics[width=\textwidth]{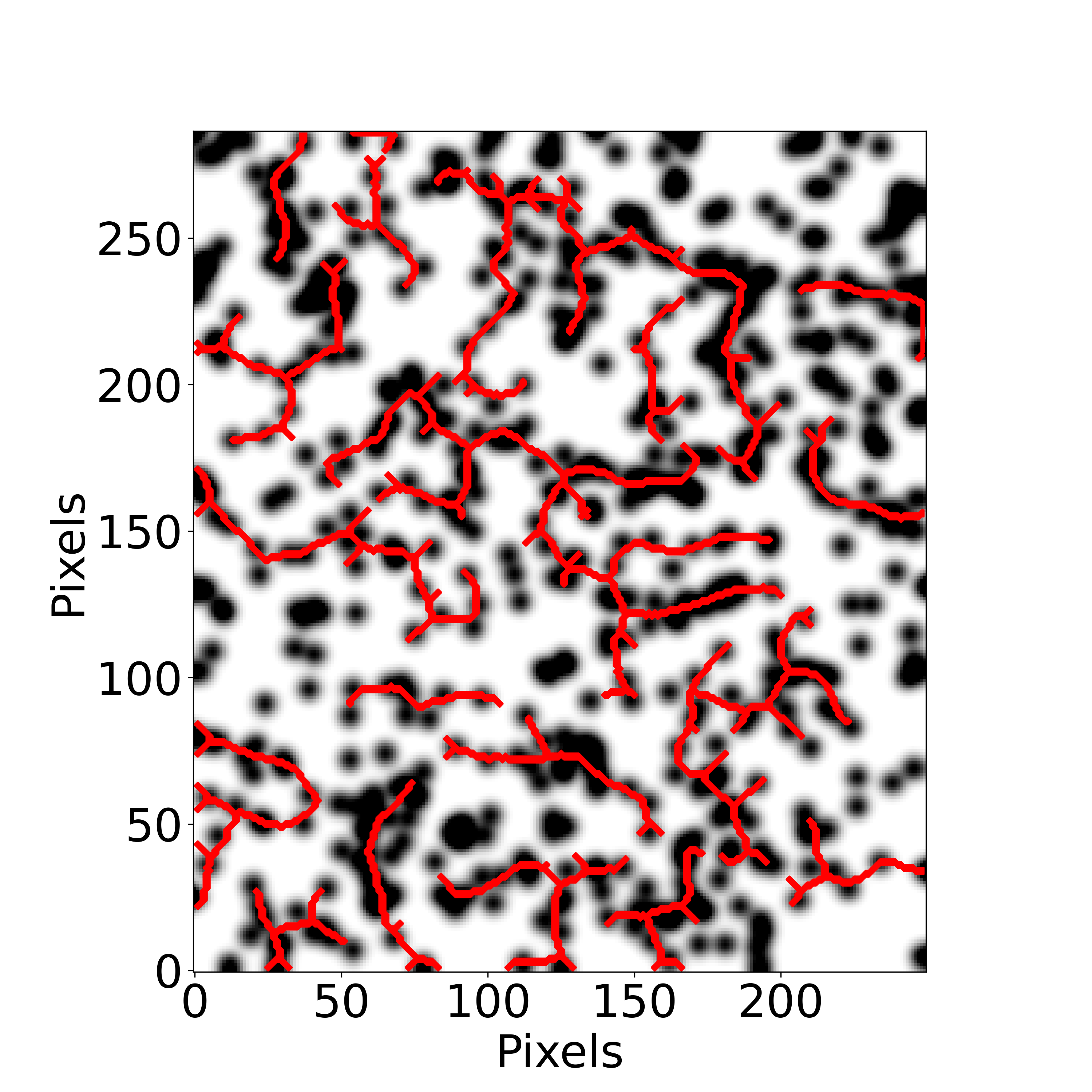}
\caption{}
\label{subfig:filfinder_qso_AT_18}
\end{subfigure}
\hfill
\begin{subfigure}[b]{0.475\textwidth}
\centering
\includegraphics[width=\textwidth]{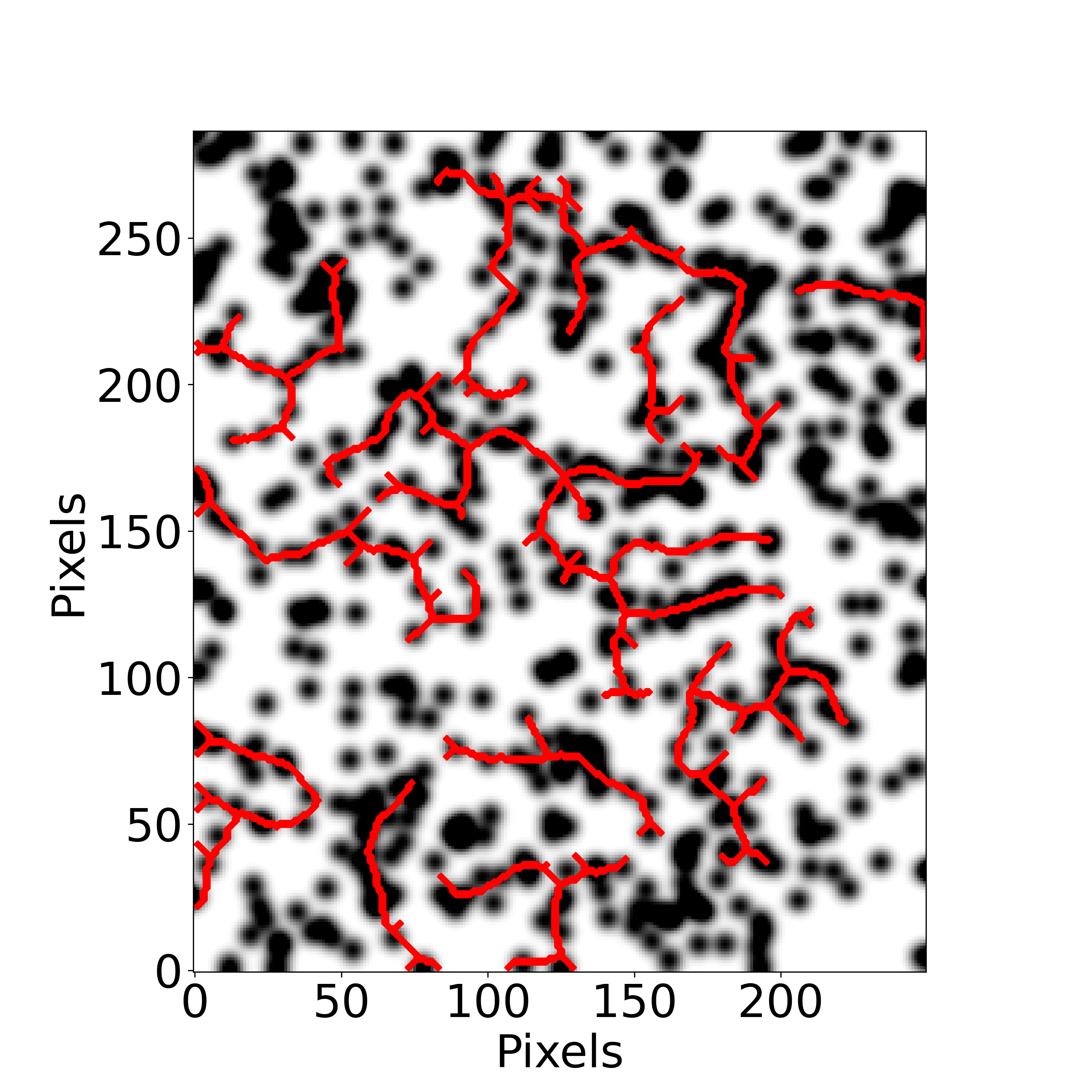}
\caption{}
\label{subfig:filfinder_qso_AT_18_ST_1000}
\end{subfigure}
\vskip\baselineskip
\begin{subfigure}[b]{0.475\textwidth}
\centering
\includegraphics[width=\textwidth]{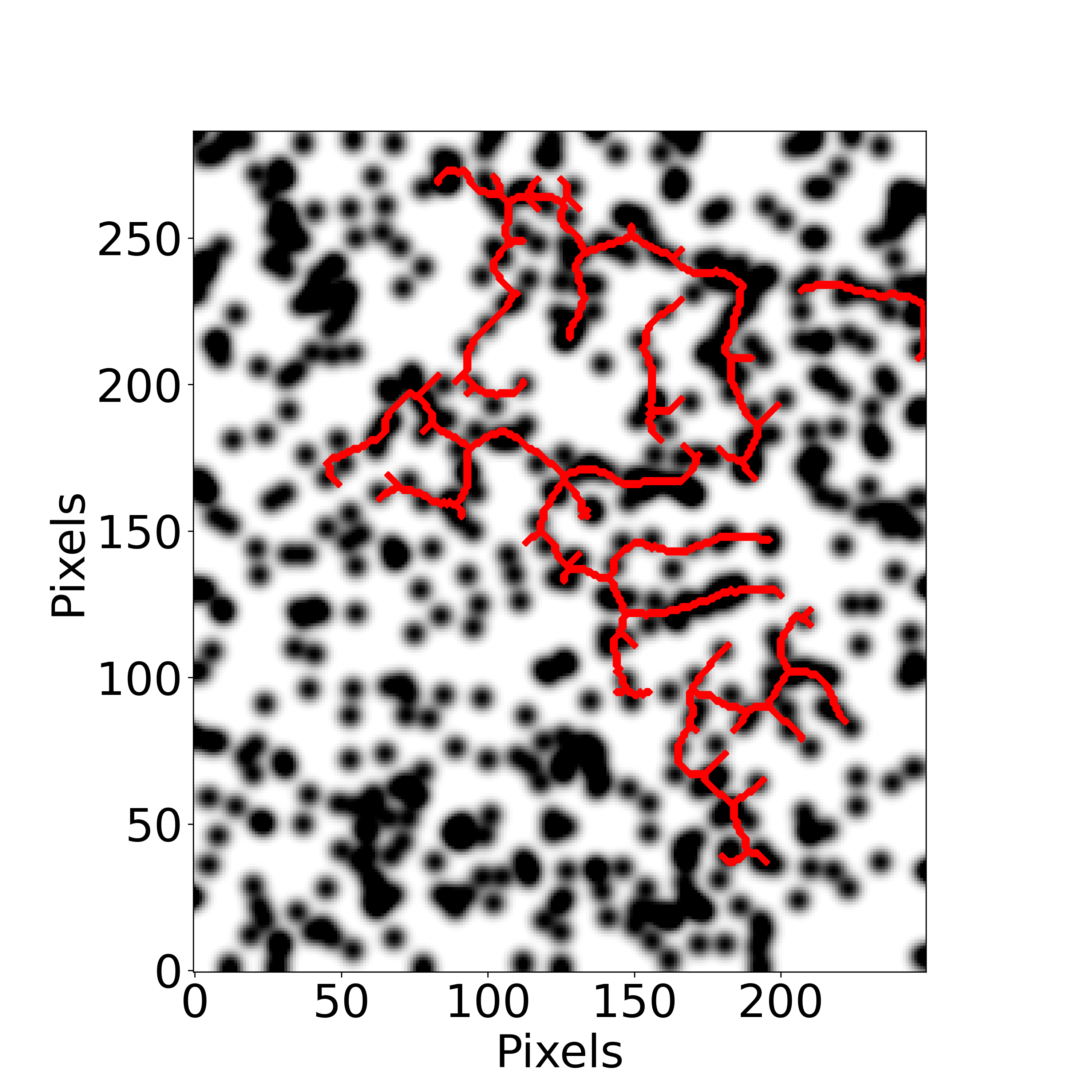}
\caption{}
\label{subfig:filfinder_qso_AT_18_ST_2000}
\end{subfigure}
\hfill
\begin{subfigure}[b]{0.475\textwidth}
\includegraphics[width=\textwidth]{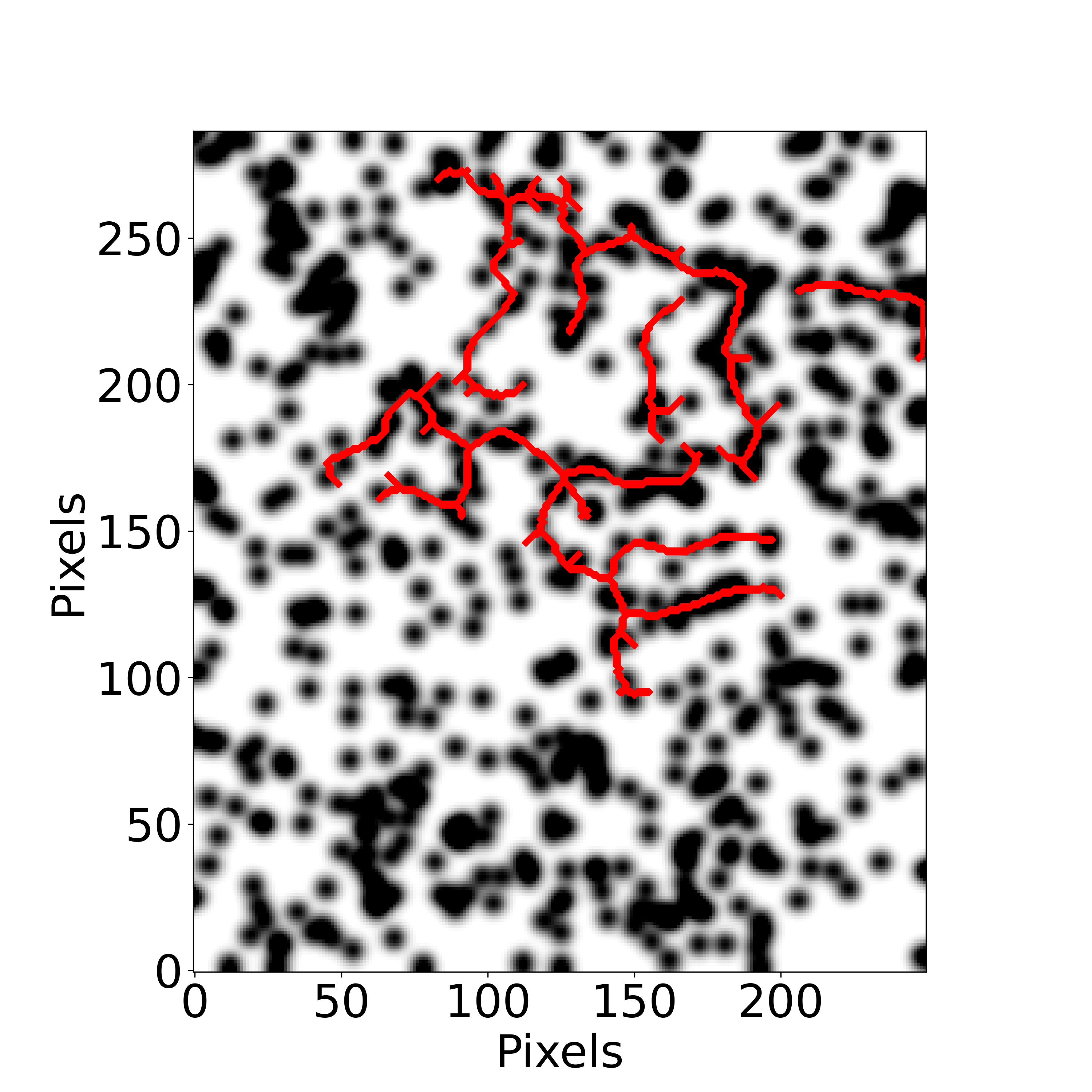}
\caption{}
\label{subfig:filfinder_qso_AT_18_ST_4000}
\end{subfigure}
\caption{Applying the FilFinder algorithm to the field quasars corresponding
  to the standard BR field and incrementally increasing the size threshold to
  show the elimination process of small filaments. All axes are labelled in
  pixels, where $1$~pixel~$= 4^2$~Mpc$^2$. (a) Using the standard FilFinder
  parameters where adaptive threshold is $18$ pixels, smooth size is $12$
  pixels, and size threshold is $576$ pixels. For the remaining figures we
  incrementally increase the size threshold. (b) Size threshold is $1000$
  pixels. (c) Size threshold is $2000$ pixels. (d) Size threshold is $4000$
  pixels. }
\label{figs:filfinder_qso}
\end{figure*}

\begin{figure}[tbp]
\centering
\includegraphics[scale=0.6]{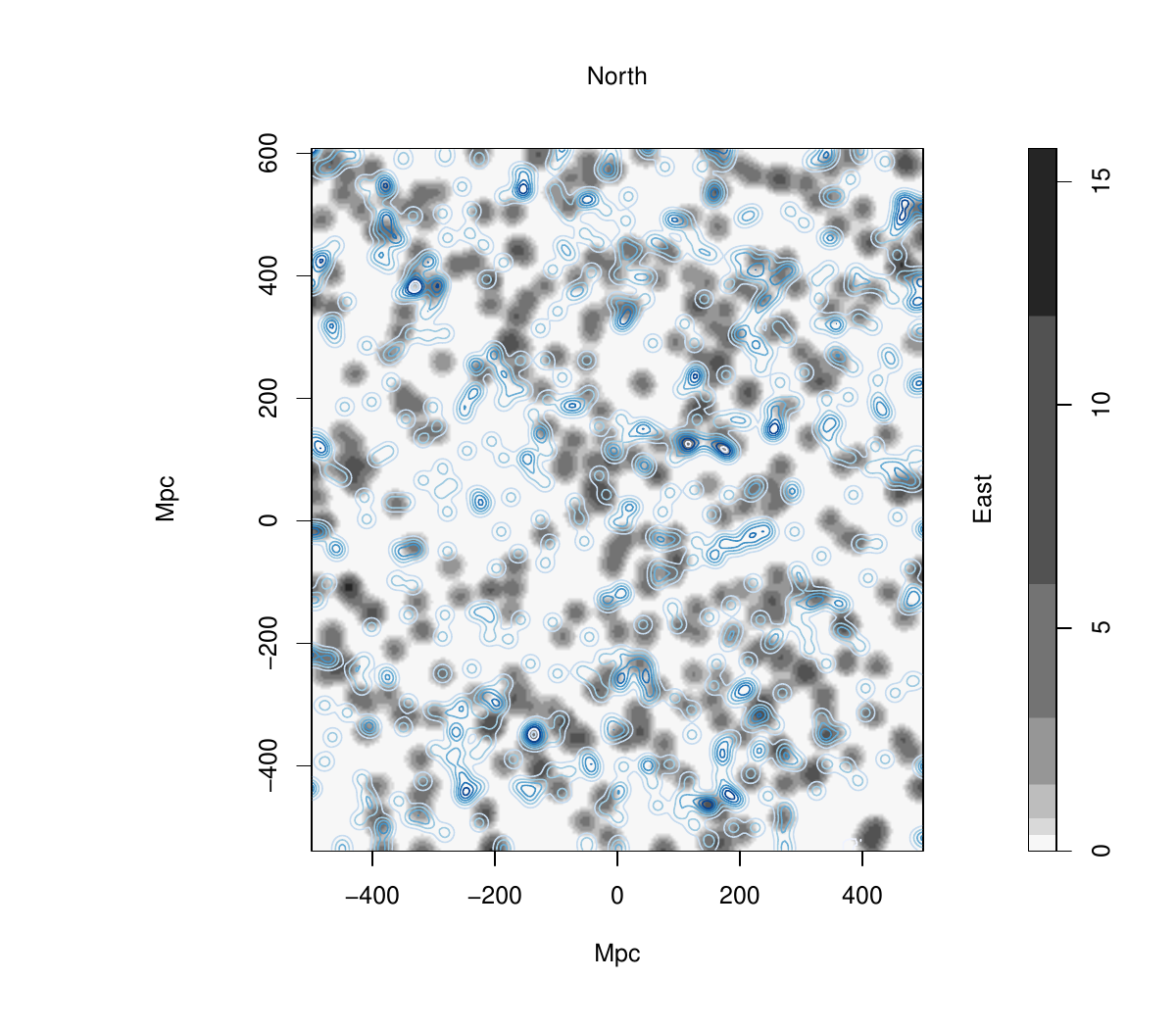}
\caption{\label{fig:BR_mgii_qso} The tangent-plane distribution of Mg~{\sc
    II} absorbers in the redshift slice $z=0.802 \pm 0.060$ superimposed with
  the tangent-plane distribution of quasar positions in the same redshift
  slice (not to be confused with the background probes). The grey contours,
  increasing by a factor of two, represent the density distribution of the
  absorbers which have been smoothed using a Gaussian kernel of $\sigma =
  11$~Mpc, and flat-fielded with respect to the distribution of background
  probes (quasars). The blue contours, increasing by a factor of two,
  represent the density distribution of the field quasars which have been
  smoothed using a Gaussian kernel of $\sigma = 11$~Mpc. In the Mg~{\sc II}
  image, S/N limits of $4, 2$ and $4$ were applied to the $\lambda_{2796},
  \lambda_{2803}$ Mg~{\sc II} lines and quasar continuum, respectively. In
  the quasar image, magnitude limits were applied such that $i \leq 20.0$. In
  general, we find that the blue contours follow the grey contours. }
\end{figure}

First, the adaptive threshold and smoothing size are related in a way that
increasing one has an almost indistinguishable effect from decreasing the
other, for reasonable values (e.g., < 40 pixels, since larger values flag a
warning within the FilFinder algorithm).  More specifically, the most
meaningful and interesting effects can be found when one focuses on the ratio
between the values.  This can be understood intuitively since smoothing size
is the scale on which to smooth the filaments and adaptive threshold is the
typical width of those filaments, the size of which is affected by the
smoothing size.  When the adaptive threshold and smoothing size ratio is
$1:2$ there is an undesirable effect of creating `blurred' masking borders
over the filaments; this is when the borders of the mask around individual
filaments overlap.  However, the opposite is true when the adaptive threshold
and smoothing size ratio is $2:1$, which has the effect of creating more
concise borders around the filaments, but it also drastically reduces the
number of filaments that can be detected.  With this in mind, we choose to
slightly increase the adaptive threshold from $12$, a single Mg~{\sc II}
absorber, to $18$, $1.5$ times a Mg~{\sc II} absorber, while keeping the
smoothing size at $12$ as usual, thus creating a ratio of $3:2$ for the
adaptive threshold and smooth size.

Secondly, the size threshold simply determines the minimum size to be
considered a filament.  Increasing the size threshold incrementally and
comparing results demonstrates the elimination process of small filaments.
So now using the updated parameter values, where the adaptive threshold is
set at $18$ pixels and the smoothing size is kept at $12$ pixels, we
incrementally increase the size threshold and show the filaments that survive
the elimination process (see Figure~\ref{figs:filfinder_increasing_ST}).  Immediately it becomes
clear that the BR is the largest, and most dense, filament in the field.  The
BR filament is the only filament left with the largest size threshold limit.
In fact, the BR is not eliminated until the size threshold exceeds $4200$
pixels.  Using the elimination process with the FilFinder algorithm has given
an impressive indication for the size and uniqueness of the BR compared with
the rest of the field.

Next, we apply the same method of analysis to the SDSS DR16Q quasars in the
same field and the same redshift slice as the Mg~{\sc II} absorbers
containing the BR for comparison with an independent data source.  We want to
reduce the noise of the high number density of quasars so we apply an
$i$-magnitude ($i$) limit of $i \leq 20.0$, such that only the intrinsically
very bright quasars are included.  The field size and redshift interval for
producing the quasar image is the same as that for the Mg~{\sc II} absorbers
from Figure~\ref{fig:BR}: that is, the quasars are chosen to be those that
lie in the same field as Mg~{\sc II} absorbers, not to be confused with the
background quasars responsible for the Mg~{\sc II} absorbers.  The quasar
image is seen in Figure~\ref{fig:BR_field_quasars}.
The FilFinder algorithm is applied to the quasar image in the same manner as
described above with the Mg~{\sc II} absorbers, with the following parameter
settings: adaptive threshold $= 18$, smoothing size $= 12$ and size threshold
$= 576$ (see Figure~\ref{subfig:filfinder_qso_AT_18}).  Given the much higher
density of the field quasars compared with the Mg~{\sc II} absorbers, even
after $i$~magnitude cuts have been made, we can see that there are generally
more filaments identified (Figure~\ref{subfig:filfinder_qso_AT_18}).
Incrementally increasing the size threshold will remove the small filaments
and leave only those able to survive the cuts --- so just the large filaments
of interest.  Figures \ref{subfig:filfinder_qso_AT_18} ---
\ref{subfig:filfinder_qso_AT_18_ST_4000} show the results from FilFinder applied to
the field quasars with increasing size threshold.

When the FilFinder algorithm is applied to the field quasars, and the size
threshold is incrementally increased, we are left with only a ring-like
filament that coincides with the BR.  This is particularly interesting as it
demonstrates from independent corroboration that the quasars also follow a
similar shape to the Mg~{\sc II} absorbers.  We can also visualise the
relationship of the Mg~{\sc II} absorbers with the bright field quasars by
superimposing the quasar image onto the Mg~{\sc II} image (see
Figure~\ref{fig:BR_mgii_qso}).
In Section~\ref{subsec:independent_data} we further investigate the
observational properties of both field quasars and DESI clusters in the BR
field.

\subsection{Cuzick and Edwards test}
\label{subsec:CE}
The Cuzick-Edwards (CE) test \cite{Cuzick1990} was created to assess the
clustering of cases in an inhomogeneous population.  It is a 2D,
case-control statistical method that deals with variations in spatial
populations.  This method has shown to be useful for the particular data that
we are working with; the Mg~{\sc II} are the cases to be assessed, which are
affected by the inhomogeneous spatial distribution of the background probes
(quasars).  This method assesses the statistical significance of clustering
by assessing the occurrence of cases within the $k$ nearest neighbours, but
it cannot, however, assess the significance of individual candidate
structures.  Therefore, it is a useful additional method to analyse the field
containing the BR.

The CE test was first applied to cosmological (and to our knowledge,
astrophysical) data when analysing the GA (section~$2.3.3$ in Lopez22);
before this, the method had mainly been used in the context of medical
research (e.g. for patterns in diseases across varying population sizes).  We
can now follow the steps and process presented in Lopez22 for the BR.  (We
are using the application $qnn.test$ in the R package \scalebox{0.7}{\sc
  SMACPOD} by \cite{French2023}.)

As mentioned previously in Lopez22, the power of the CE test is affected by
the control-case ratio, and a ratio between $4$ and $6$ is found to be
optimal \cite{Cuzick1990}.  For this reason, we continue to use a
control-case (i.e., probes to absorbers) ratio of $5:1$, which, for the BR
field, requires removing $\sim 90 \%$ of the probes.  The $\sim 10 \%$
remaining probes are randomly selected for each of the $200$
runs of the $qnn.test$, which itself is computed for $499$ simulations (the
default setting).  

\begin{figure*}
\centering
\begin{subfigure}[b]{0.475\textwidth}
\centering
\includegraphics[width=\textwidth]{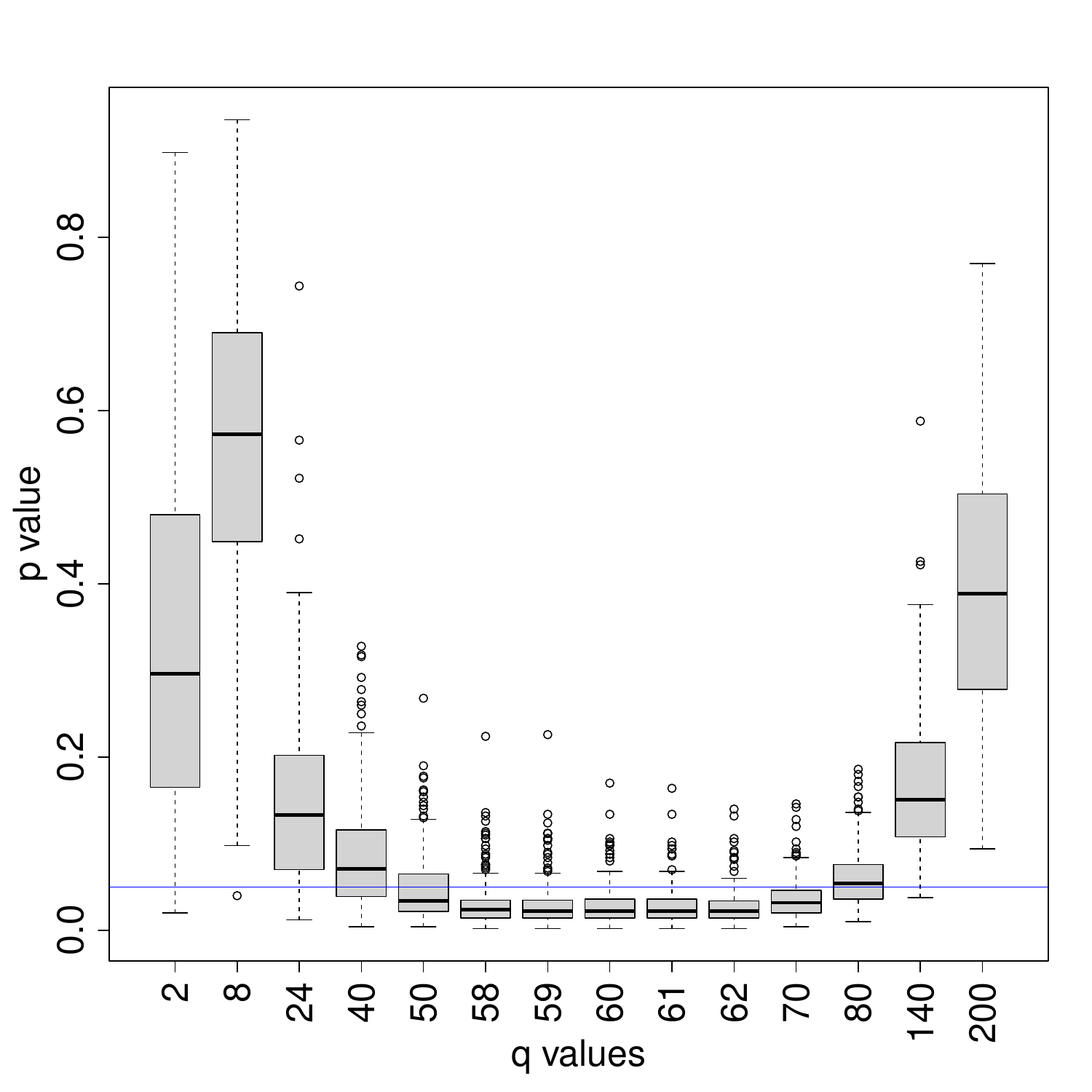}
\caption{}
\label{subfig:CE_BR_zoom2}
\end{subfigure}
\hfill
\begin{subfigure}[b]{0.475\textwidth}
\centering
\includegraphics[width=\textwidth]{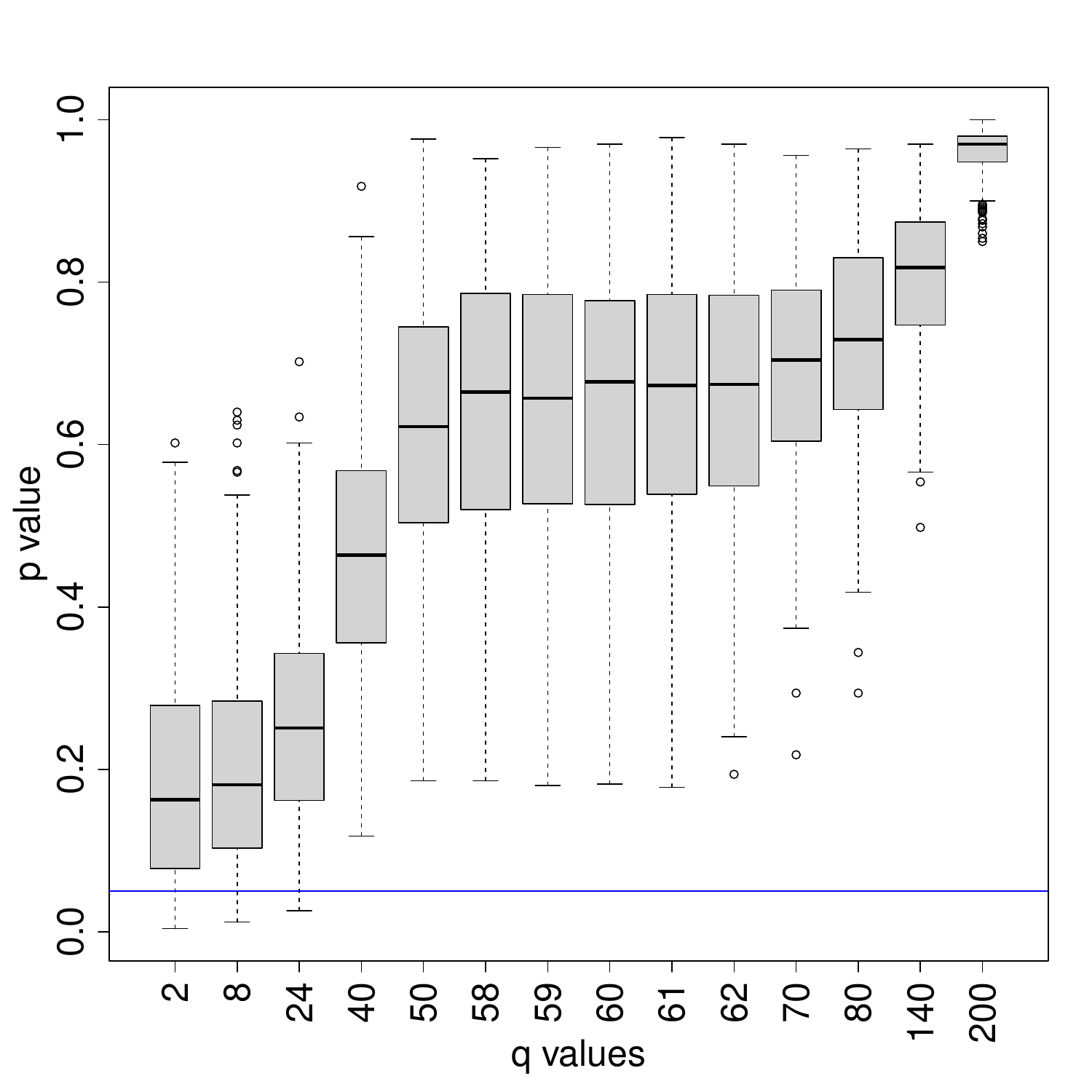}
\caption{}
\label{subfig:CE_random_zoom2}
\end{subfigure}
\caption{The Cuzick and Edwards (CE) test applied to the second `zoom' of the
  BR field and an unrelated field (a and b respectively) for comparison. The
  figures show a box-plot of the distribution of $p$-values over
    $200$ runs of $499$ simulations as a function of chosen $q$ ($k$)
  value. (a) The BR field shows tentative significant clustering
    ($p \leq 0.05$) between $q = 50$ and $q=70$, reaching a minimum of
    $p=0.022$ between $59$ and $q=62$ corresponding to a significance of $2.0
    \sigma$. Note, this is \emph{one} of the figures produced from the $10$
    repeated tests. (b) The unrelated field shows no significant clustering
  on any scale.}
\label{figs:CE_test}
\end{figure*}

We begin by assessing the BR field, and then we successively shrink the FOV
to `zoom in' on the BR, which is presumably the main feature contributing to
any clustering patterns detected in the field.  Unlike the GA, the BR is a
full ring shape, so the successive zooming is limited in both axes.
Therefore, we apply only two successive zooms to the BR, beyond which the BR
would extend beyond the FOV. We then apply the CE test on four
other fields that are unrelated and detached from the BR field, for
comparison with the BR field. The four unrelated fields are chosen to be at
the same redshift as the BR field ($z=0.802\pm0.060$), the same field size
(equivalent to the second `zoom' of the BR field) and the same declination
($\delta = 50^\circ$), so that the only change from the BR field to the
unrelated fields is to move the right ascension (RA) centre point --- the
unrelated fields are then outside the BR field but within version-1 of the
control area (see Figure~\ref{fig:SDSS_control_fields}).  For the BR field,
we repeat the CE test (with $200$ runs and $499$ simulations) $10$ times
altogether to investigate how the random selection of probes (controls)
could affect the CE statistics. The distribution of $p$-values calculated
from applying the CE test to the BR field indicates tentative significant
clustering ($p \leq 0.05$) for a range of chosen $q$ ($k$) values.  The
median $p$-value dropped to a minimum of $p=0.0238 \pm 0.0015$ over the
$10$ repetitions of the CE test at $q=61$, corresponding to a significance
of $2.0 \sigma$.  However, in the four other unrelated fields there was no
(with the exception of exactly two data points) significant (significant is
here defined as $p \leq 0.05$) clustering detected, suggesting that the
clustering seen in the BR field is special (see Figure~\ref{figs:CE_test})
--- as was similarly found with the GA previously.  In one of the unrelated
fields there were two points, at $q=2$ and $q=8$, where the median
$p$-value dropped below the $p=0.05$ threshold, which is here considered
tentative significant clustering.  However, since it was at a much lower
$q$~value, the CE test was likely here detecting very small-scale
clustering, and not related to the same type of clustering seen in the BR
(and GA) field.  Overall, the clustering seen in the BR field mimicked the
clustering that was also seen in the GA field, but here at a tentative
($p\leq 0.05$) significance level; the clustering
in the BR field was inconclusive at the $\sigma > 3.0$ significance level.

\section{Observational properties}
\label{sec:observational_properties}

\subsection{Corroboration with independent data}
\label{subsec:independent_data}
\begin{figure*}
\centering
\begin{subfigure}[b]{0.475\textwidth}
\centering
\includegraphics[width=\textwidth]{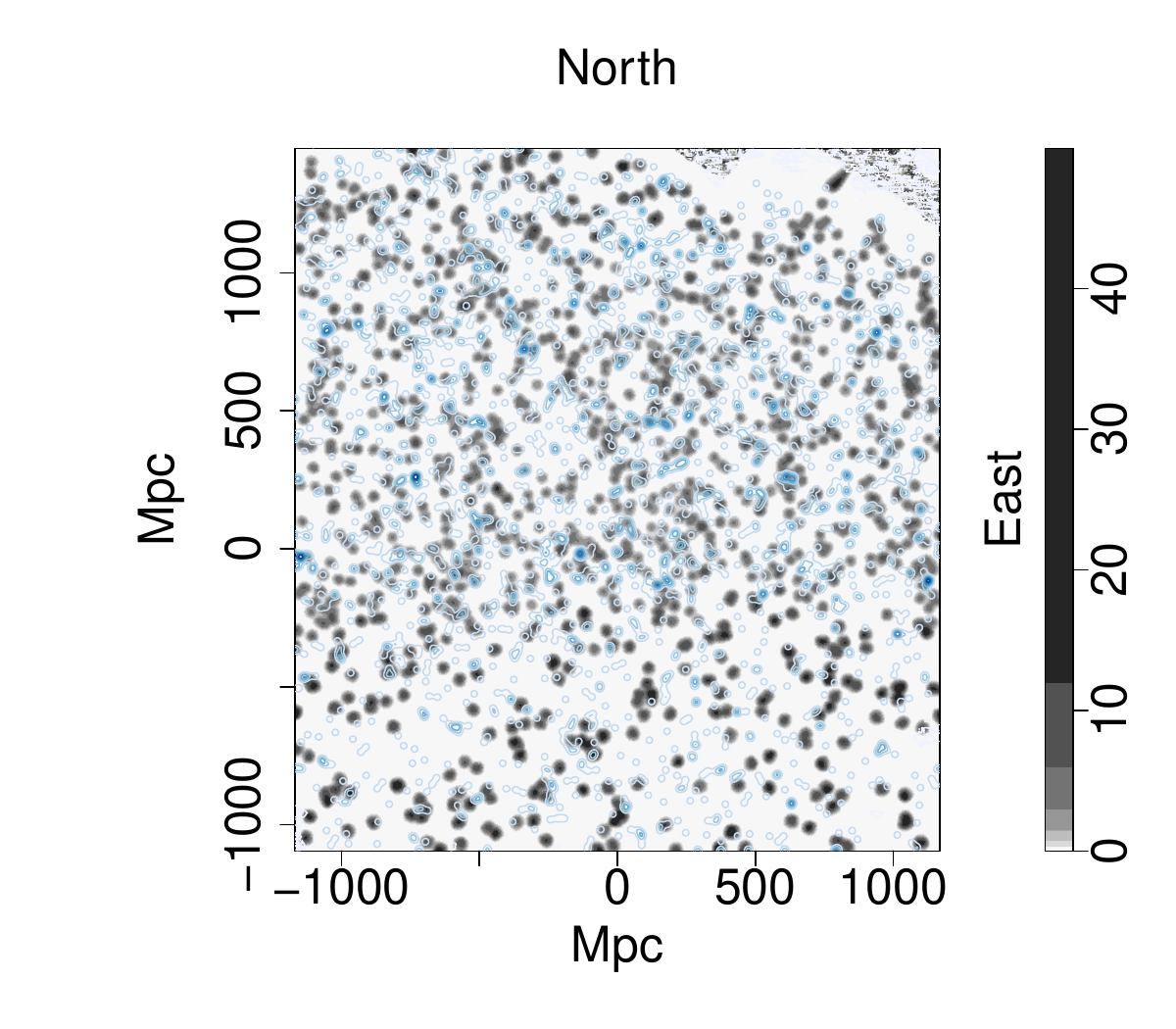}
\caption{}
\label{subfig:mgii_qso_i_20}
\end{subfigure}
\hfill
\begin{subfigure}[b]{0.475\textwidth}
\centering
\includegraphics[width=\textwidth]{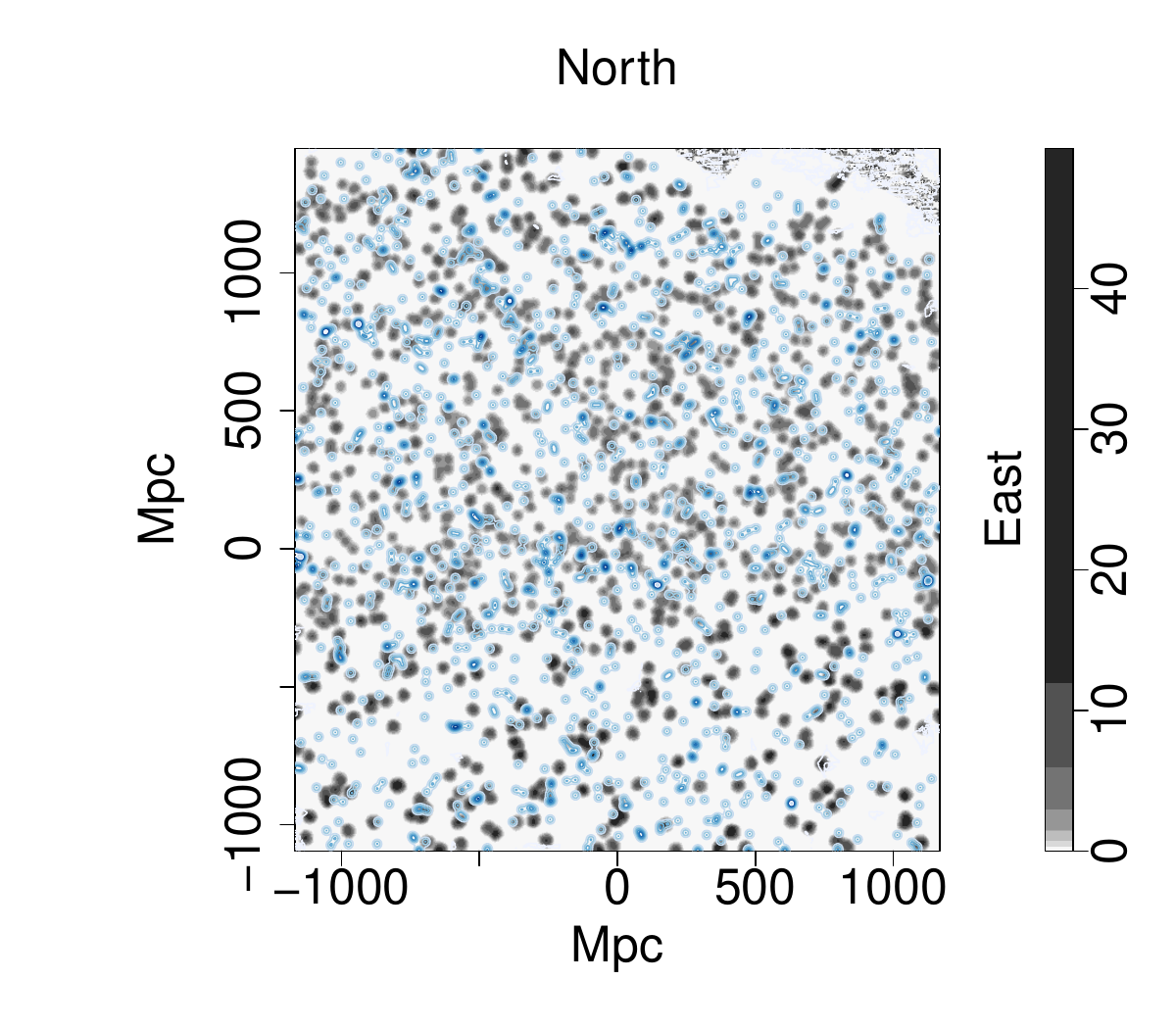}
\caption{}
\label{subfig:mgii_qso_i_19.5}
\end{subfigure}
\vskip\baselineskip
\begin{subfigure}[b]{0.475\textwidth}
\centering
\includegraphics[width=\textwidth]{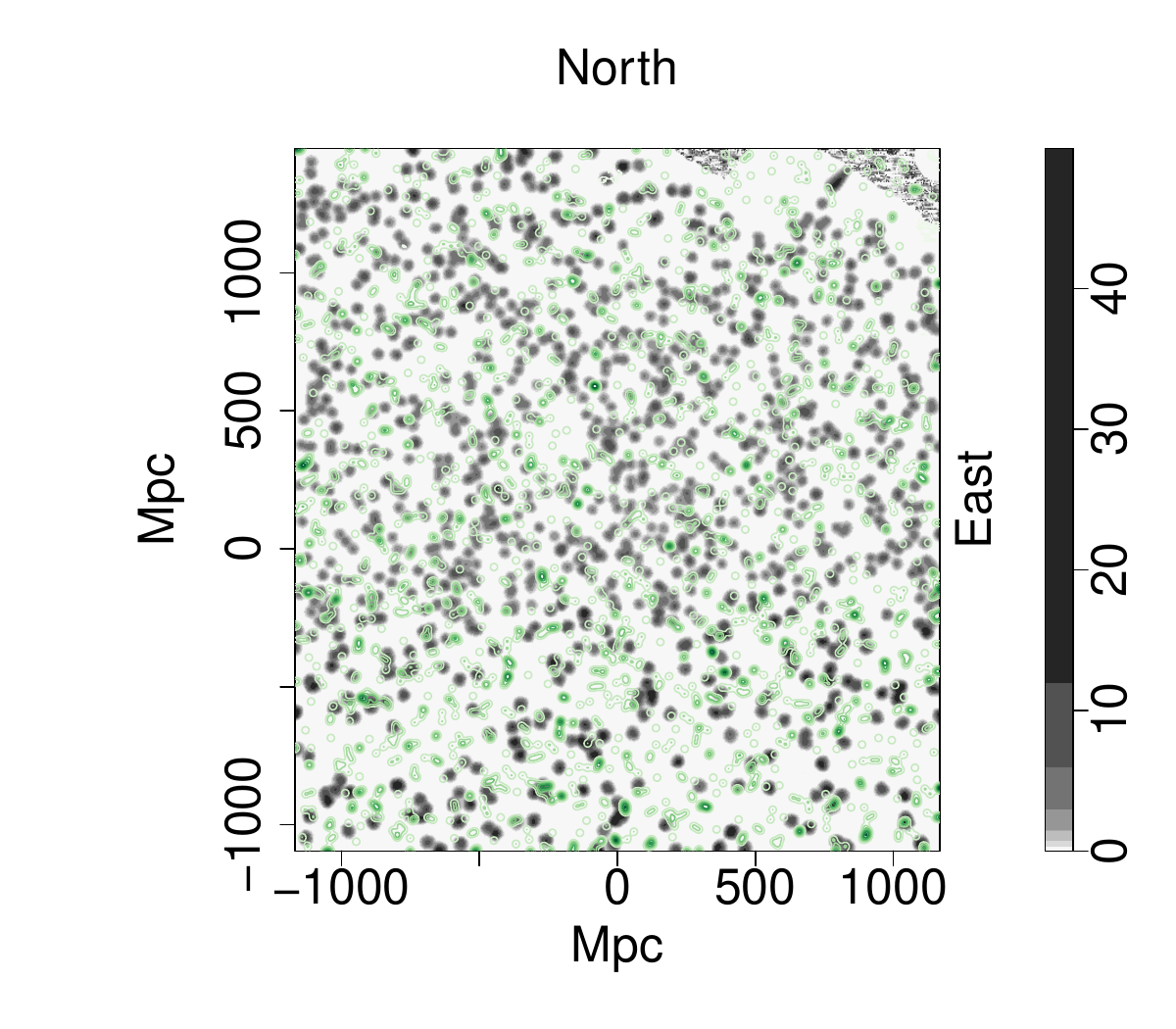}
\caption{}
\label{subfig:mgii_desi_R_leq_22_5}
\end{subfigure}
\hfill
\begin{subfigure}[b]{0.475\textwidth}
\includegraphics[width=\textwidth]{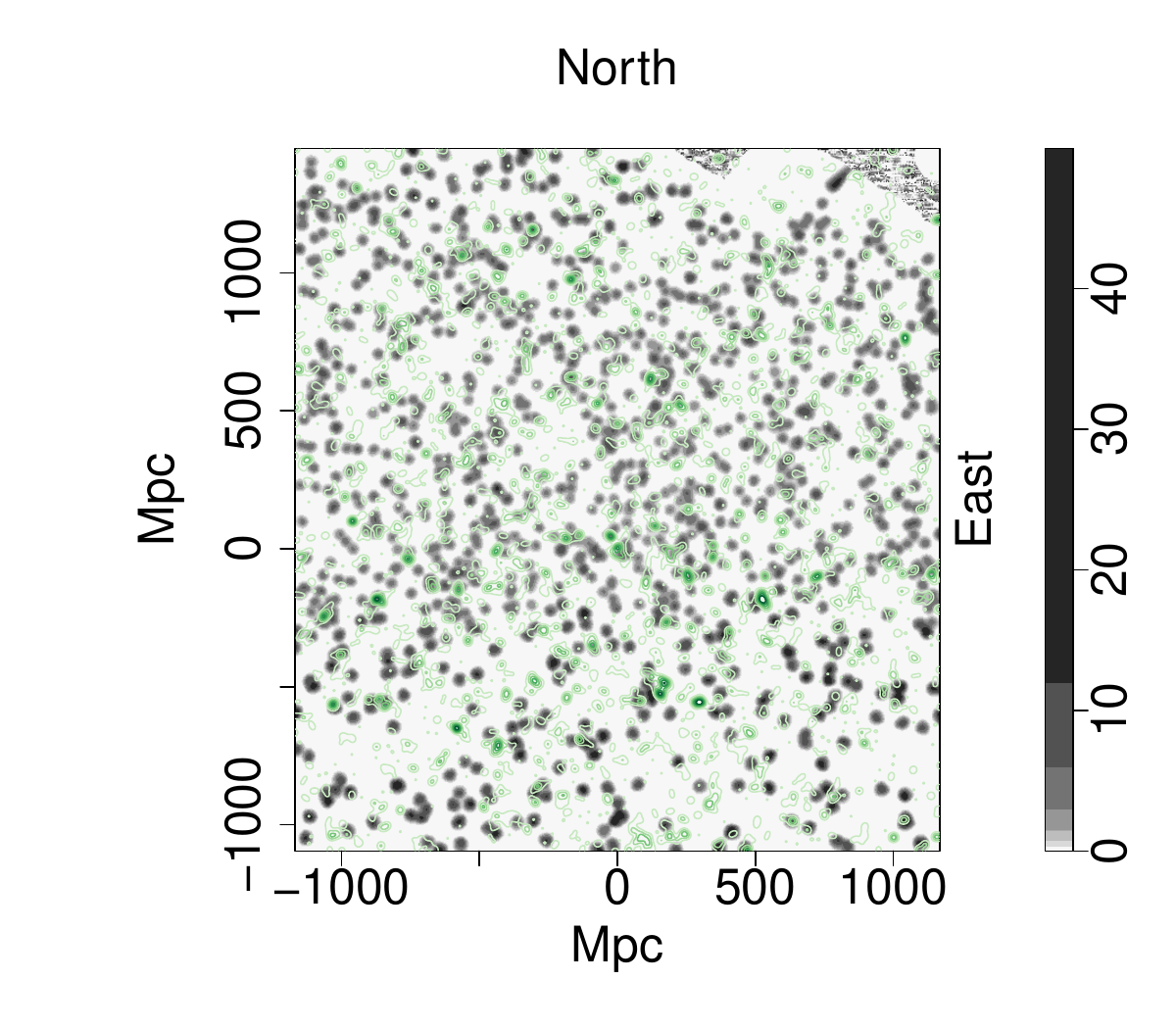}
\caption{}
\label{subfig:mgii_desi_R_geq_22.5}
\end{subfigure}
\caption{\small {Density distribution of the Mg~{\sc II} absorbers in the
    redshift slice $z=0.802 \pm 0.060$ in a large field-of-view represented
    by the grey contours which have been smoothed using a Gaussian kernel of
    $\sigma = 11$~Mpc and flat-fielded with respect to the background
    quasars. The additional contours (in blue and green) represent the
    superimposed density distribution of the field quasars (a and b) and the
    DESI clusters (c and d), respectively.
    (a) The field quasars, represented by the blue contours, are restricted
    to $i \leq 20.0$. The blue contours appear to have a plausible
    association with the grey contours. The $i$~magnitude limit is not very
    restrictive so the blue contours are visually noisy compared with the
    Mg~{\sc II} absorbers. Note, these quasars were found to have a FilFinder
    filament that linked closely to the shape of the BR, so clearly it is the
    visual impression that is difficult to determine due to the noise.
    (b) The field quasars, represented by the blue contours, are restricted
    to $i \leq 19.5$. With the reduced quasars (compared with a) a clearer
    trend of quasar and Mg~{\sc II} absorber association can be
    seen. Generally, the dense clumps of Mg~{\sc II} absorbers appear to have
    associated quasars rather than the thin Mg~{\sc II} filaments.
    (c) The DESI clusters, represented by the green contours, are restricted
    to $R \leq 22.5$. The green contours generally follow the grey contours,
    although the richness limit reduces the number of DESI clusters in the
    field.
    (d) The DESI clusters, represented by the green contours, are restricted
    to $R \geq 22.5$. Again, it can be seen that the green contours generally
    follow the grey contours. In particular, the lower half of the BR has a
    strong filament of DESI clusters following the filamentary shape of the
    Mg~{\sc II} absorbers.  Note, the larger field-of-view of the above
    figures crosses two of the SDSS borders where the quasar coverage drops
    sharply (see Figure~\ref{fig:SDSS_control_fields}).}}
\label{figs:independent_data}
\end{figure*}

Earlier, in Section~\ref{subsec:FilFinder}, we saw that the bright field
quasars have identified filaments similar to those of the Mg~{\sc II}
absorbers.  It was also shown that there is a plausible visual association of
the field quasars and Mg~{\sc II} absorbers from superimposing the contour
maps of each dataset (Figure~\ref{fig:BR_mgii_qso}).  We can continue to use
the contour maps from the quasars, and now also the DESI clusters
\cite{Zou2021}, looking at a larger FOV, and investigate the visual
association of the independent datasets with the Mg{\sc II} absorbers on a
larger scale (see Figure~\ref{figs:independent_data}).

\begin{figure}[h]
\centering 
\includegraphics[scale=0.6]{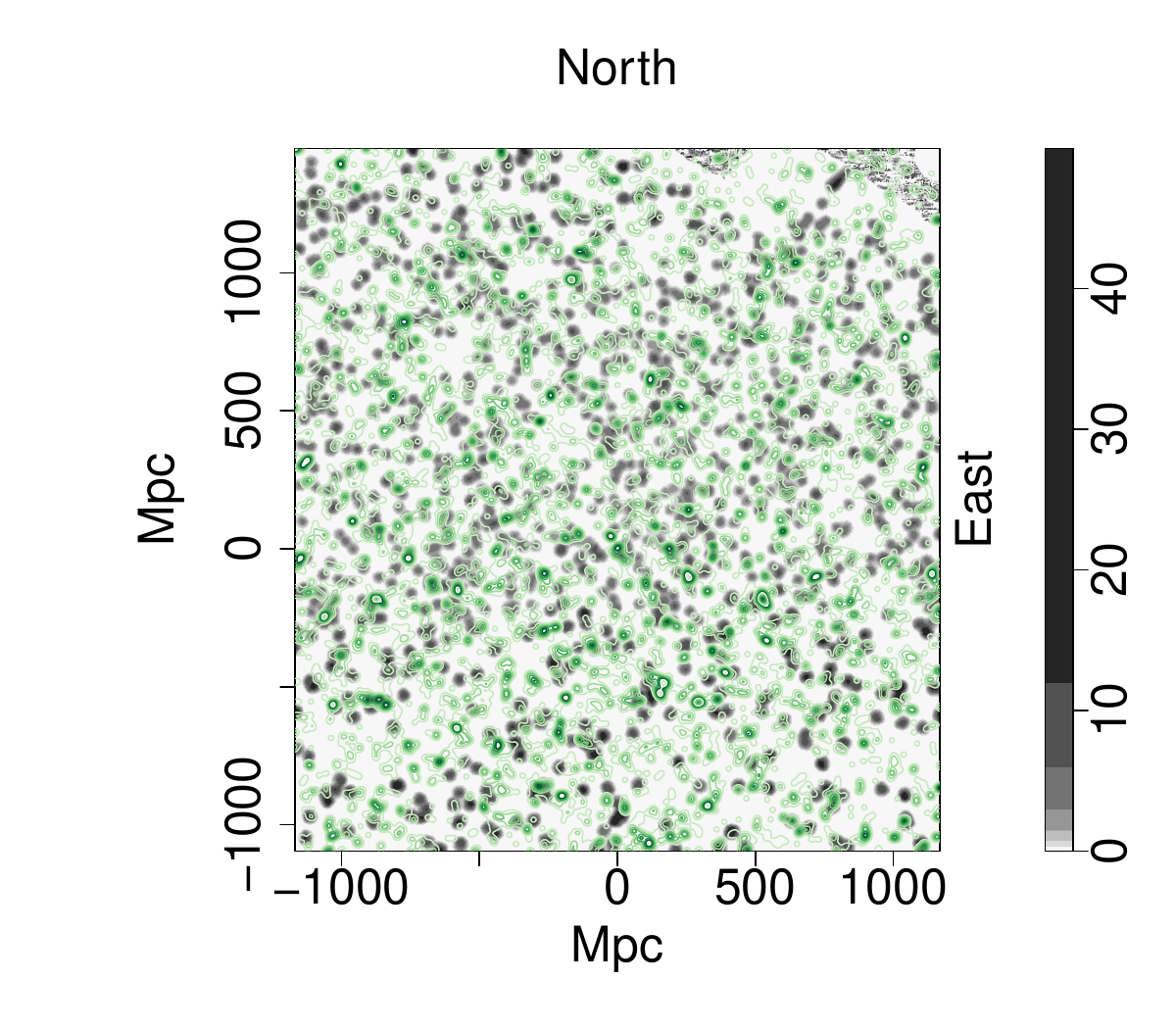}
\caption{\label{fig:mgii_desi_all} All of the DESI clusters with richnesses
  $0 < R \leq 300$, represented by the green contours. Here it can be seen
  that the green contours generally follow the grey contours quite well.
  Compare with Figures~\ref{subfig:mgii_desi_R_leq_22_5} and
  \ref{subfig:mgii_desi_R_geq_22.5}, which are for $R \leq 22.5$ and $R \geq
  22.5$. Note, the larger field-of-view of the figure crosses two of the SDSS
  borders where the quasar coverage drops sharply (see
  Figure~\ref{fig:SDSS_control_fields}).}
\end{figure}

The visual investigation of quasars and DESI clusters with Mg~{\sc II}
absorbers indicates that generally there is plausible association of both
independent datasets with the absorbers.  The association can be seen clearly
when comparing the `voids' in the data; where there are no absorbers there
tend also not to be very many quasars or DESI clusters.  The visually most
striking association is seen in Figure~\ref{subfig:mgii_desi_R_geq_22.5}
where the green contours of the DESI clusters seem to follow the same
filamentary trajectory as the Mg~{\sc II} absorbers in many cases.
Interestingly, however, in Figure~\ref{subfig:mgii_desi_R_leq_22_5}, similar
trajectory is also seen, but on other Mg~{\sc II} absorbers that were missed
in Figure~\ref{subfig:mgii_desi_R_geq_22.5}.  The two DESI cluster figures
(\ref{subfig:mgii_desi_R_leq_22_5} and \ref{subfig:mgii_desi_R_geq_22.5})
have richness limits applied of $R \leq 22.5$ and $R\geq 22.5$, respectively,
so the crossover of the two figures is minimal.  (The median
  richness limit for the DESI clusters is $22.5$ \cite{Zou2021}.)  We might
thus have a potentially useful technique of investigating the Mg~{\sc II}
absorbers and their association with low and high richness clusters, from
which we might learn more about the physical origin of the Mg~{\sc II}
absorbers.  We can also see in Figure~\ref{fig:mgii_desi_all}
the DESI clusters of all richnesses plotted in green contours over the BR
Mg~{\sc II} absorbers.  Comparison with Figures~\ref{subfig:mgii_desi_R_leq_22_5}
and \ref{subfig:mgii_desi_R_geq_22.5}, which are for $R \leq 22.5$ and $R \geq 22.5$,
indicates that Figure~\ref{fig:mgii_desi_all} shows the strongest association of the
DESI clusters with the Mg~{\sc II} absorbers.
Of course, the DESI clusters will have much larger redshift errors ($\Delta z
\sim 0.04$) than the Mg~{\sc II} absorbers and quasars, so the structures in
the DESI clusters will be blurred.  Given that the association seems quite
clear, then presumably smaller redshift errors would lead to an even clearer
association.

\begin{figure}[h]
\centering 
\includegraphics[scale=0.4]{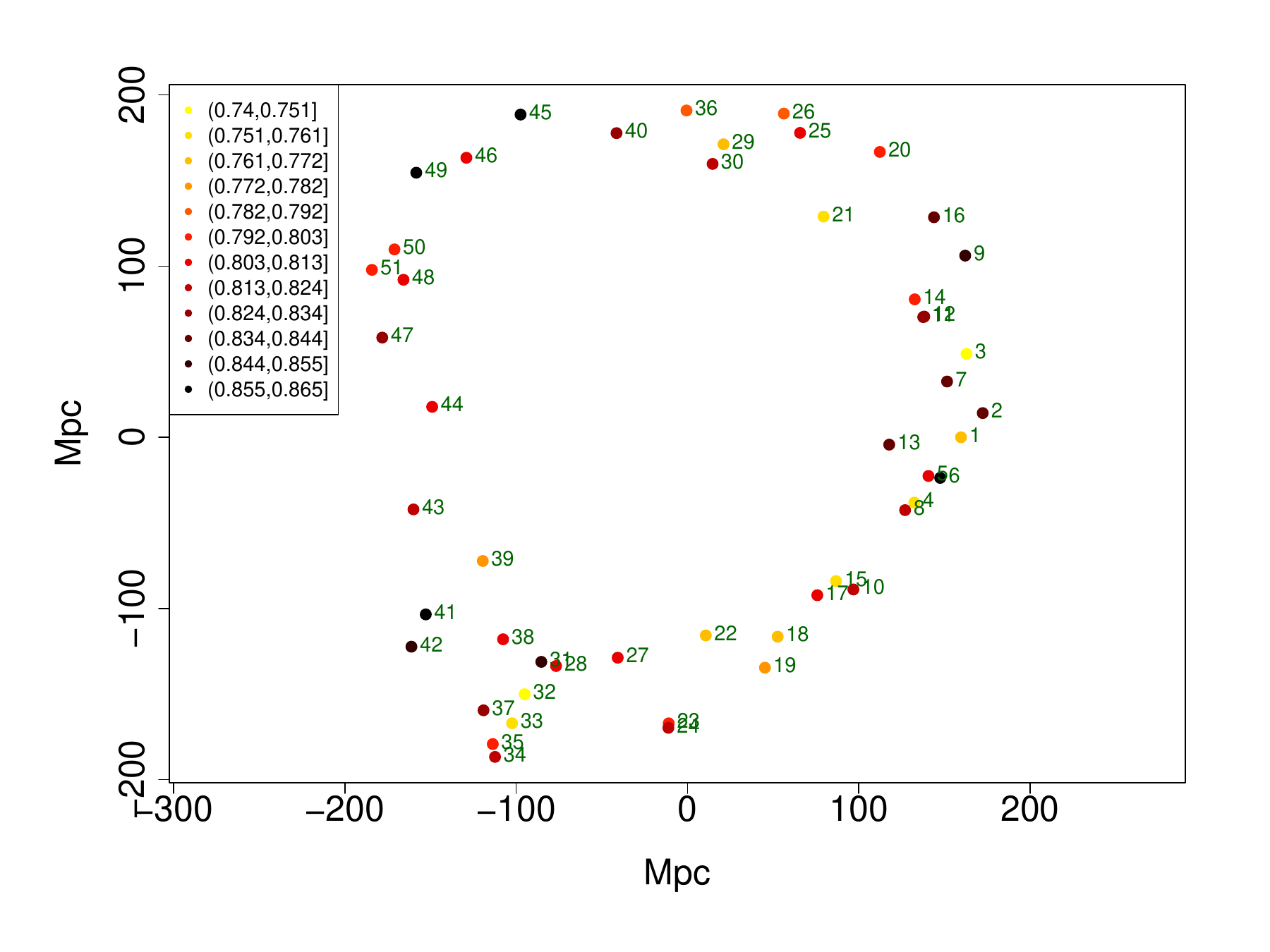}
\caption{\label{fig:BR_orig_proj} The visually-identified BR absorber members
  projected onto the plane perpendicular to the initial normal vector, $w_0$.
  The $x$-axis points towards the $u_0$ direction and the $y$-axis points
  towards the $v_0$ direction. By comparing the figure here with
  Figure~\ref{fig:BR_and_filament_topcat} one will notice the slight rotation
  of the BR projection, indicating the small misalignment of the
  tangent-plane coordinate system with the new $u_0, v_0, w_0$ system. The
  key in the top LHS of the figure indicates the redshifts of the absorbers
  associated by the colours, and the small numbers paired with each data
  point indicate their unique ID number.}
\end{figure}
\subsection{Viewing the BR from other angles}
\label{subsec:project_plane}
The BR and GA discoveries were made unexpectedly with the method of
intervening Mg~{\sc II} absorbers, and so are subject to observational bias
--- we have first detected the signal of a LSS by observing a curious shape
and/or visual overdensity in the initial Mg~{\sc II} images that we later
assess statistically.  Specifically, we are only observing these LSS
candidates from one line of sight (LOS), i.e., a 2D projection on the sky of
the 3D matter distribution.  Thus, from a different viewing angle, or LOS,
the LSS candidates could look entirely different, or furthermore, there could
be LSS candidates that have a seemingly average, or `uninteresting'
distribution of absorber members from our LOS that may be arcs, rings or
interesting filamentary shapes from a different viewing angle.
\begin{figure}[h]
\centering
\includegraphics[scale=0.4]{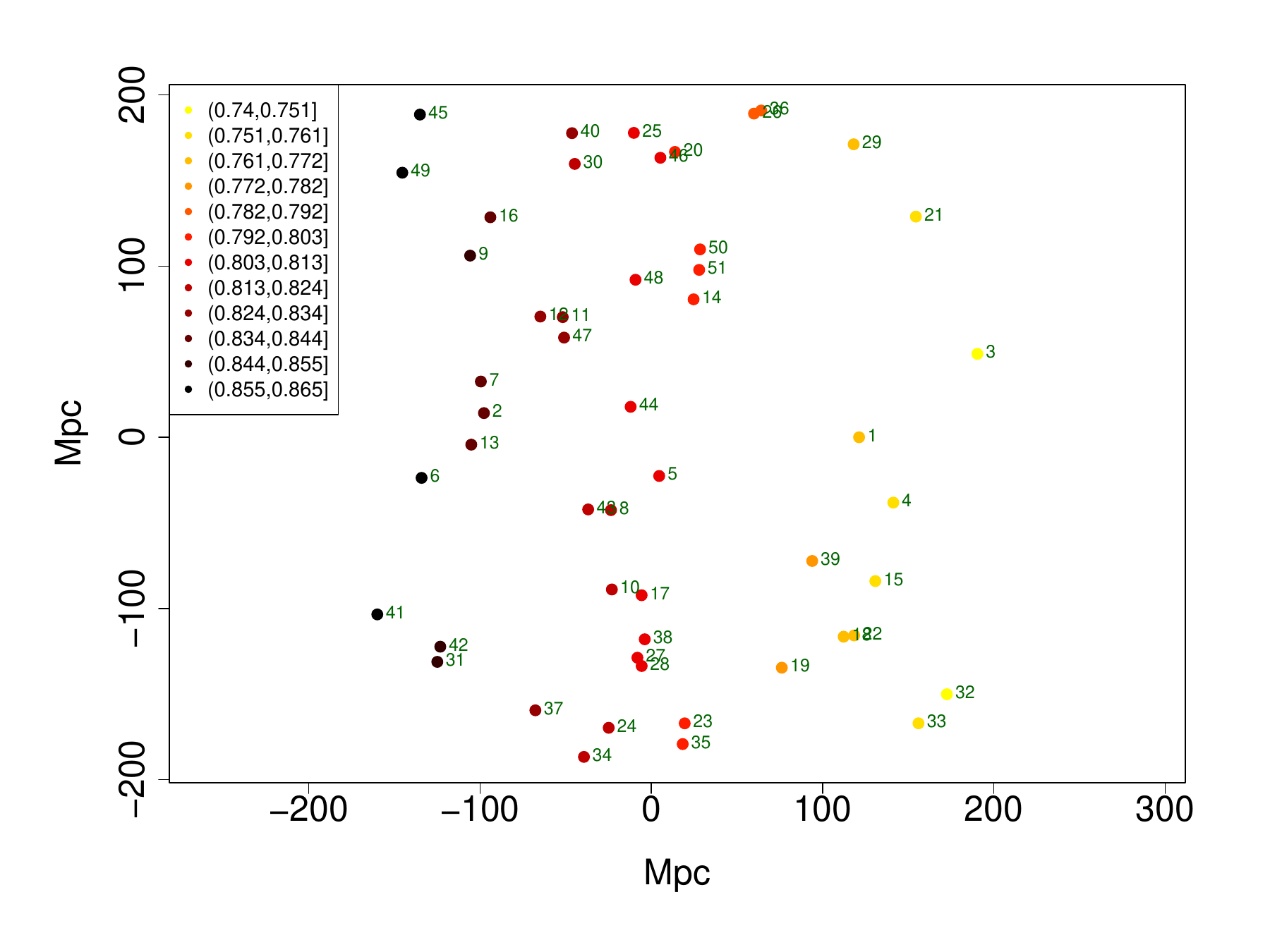}
\caption{\label{fig:BR_proj_u_hat_0} The BR absorber members projected onto
  the plane perpendicular to $u_0$. The new $u$ direction points towards the
  most easterly absorber, as usual. The colours represent the redshifts of
  the absorbers (see the key in the top LHS of the figure) and the small
  numbers paired with each data point indicate their unique ID number. }
\end{figure}
\begin{figure}[h]
\centering
\includegraphics[scale=0.4]{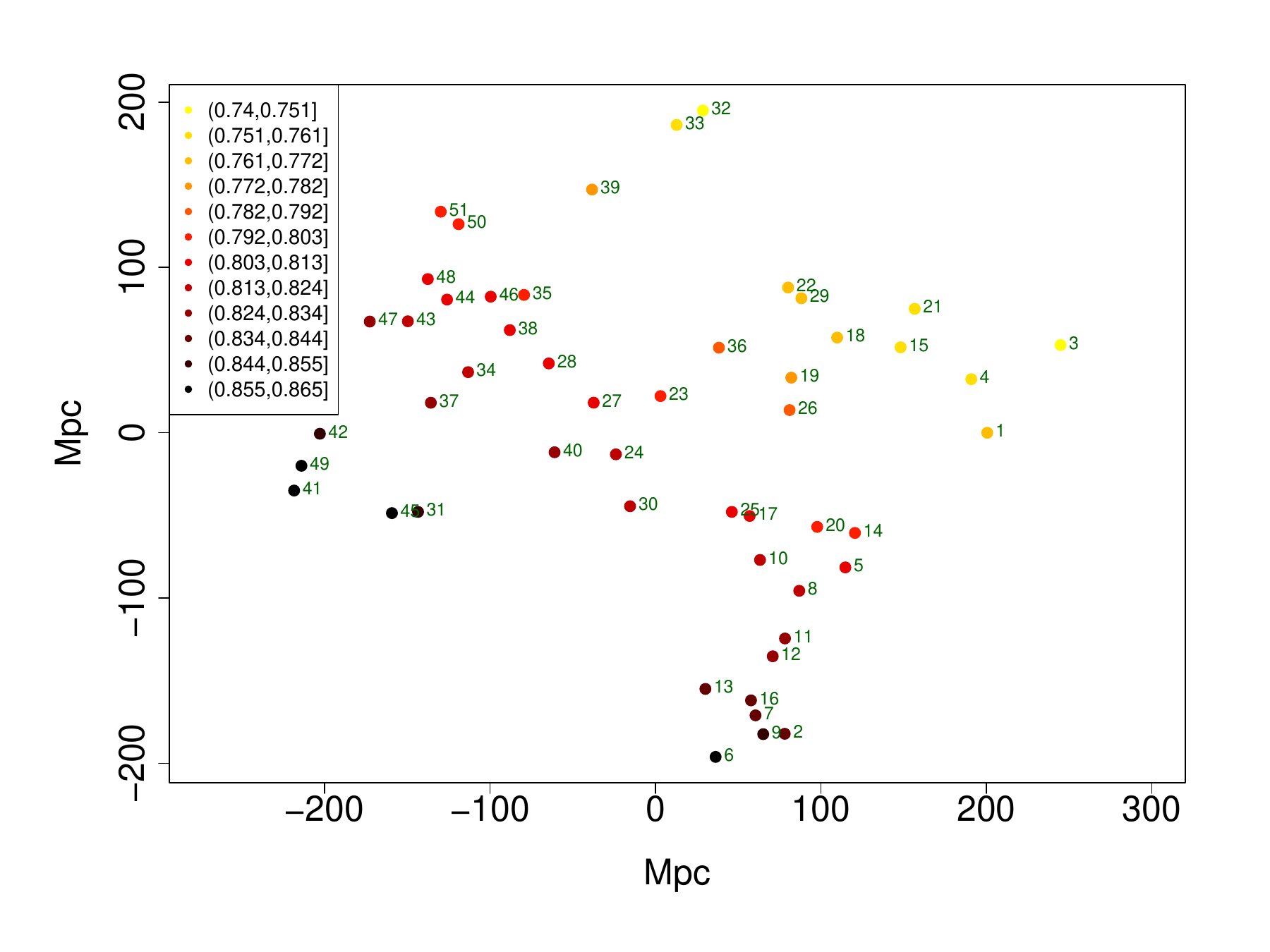}
\caption{\label{fig:BR_proj_v_hat_0} The BR absorber members projected onto
  the plane perpendicular to $v_0$. The new $u$ direction points towards the
  most easterly absorber, as usual. The colours represent the redshifts of
  the absorbers (see the key in the top LHS of the figure) and the small
  numbers paired with each data point indicate their unique ID number.}
\end{figure}
\begin{figure}[h]
\centering
\includegraphics[scale=0.4]{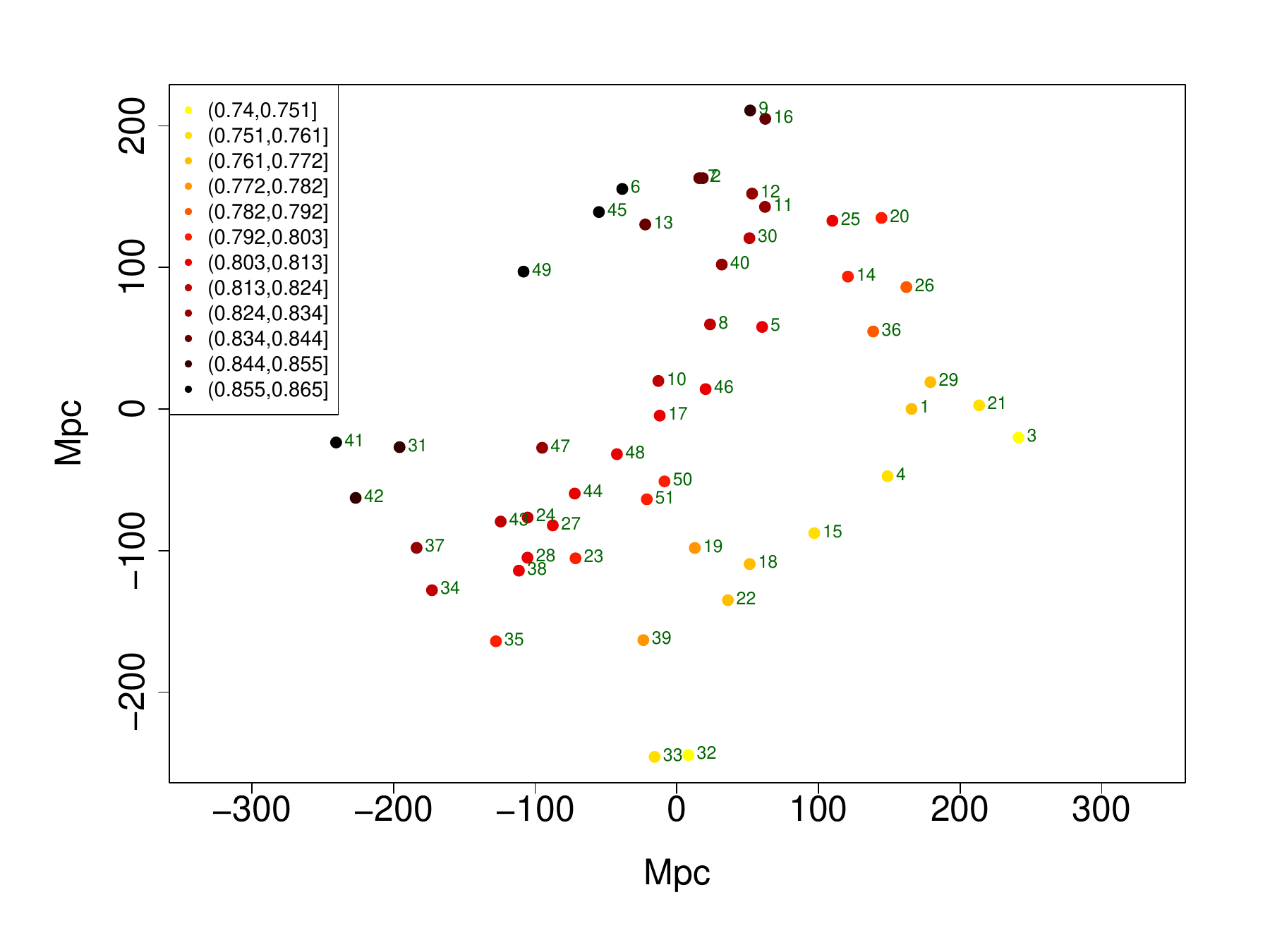}
\caption{\label{fig:BR_proj_u_hat_0_minus_v_hat_0} The BR absorber members
  projected onto the plane perpendicular to $u_0 - v_0$. The new $u$
  direction points towards the most easterly absorber, as usual. The colours
  represent the redshifts of the absorbers (see the key in the top LHS of
  the figure) and the small numbers paired with each data point indicate
  their unique ID number.}
\end{figure}

We investigate how different viewing angles change the perception of the BR,
which first requires redefining the coordinate system on which the BR-field
absorber members are projected, described here.  An initial, orthogonal,
3-vector coordinate system is defined such that $u_0, v_0, w_0$ are closely
linked to $x_{prop}, y_{prop}, z_{prop}$.  The initial normal vector ($w_0$)
is defined as the proper coordinate to the mean $x_{prop}, y_{prop},
z_{prop}$ of the absorber members (which can be thought of as the original
LOS).  Then, the plane perpendicular to the normal is rotated such that $u_0$
points towards the most easterly absorber.  Finally, all the absorber members
are projected onto the new plane (see Figure~\ref{fig:BR_orig_proj}).  Hence,
we call this method the `project-plane method'.  With the initial coordinate
system defined, we can define any new normal vector ($w$) as some combination
of the initial $u_0, v_0, w_0$.

In Figure~\ref{fig:BR_orig_proj} and other similar figures following, the
colours of the absorbers indicate the redshift of each absorber, with the
high-$z$ absorbers represented by a darker shade, and vice versa (see the key
on the top left of the figures).  The numbers associated with each absorber
are a simple ID system, labelled $1$ to $51$ in ascending order from the most
easterly absorber to the most westerly (i.e., in descending order of RA).
The ID system remains the same throughout all subsequent rotations of the
plane projections so that the numbers can be used to orientate oneself.
In this manner we are viewing the BR from different viewing angles, similar
to taking snapshots of the 3D structure from different LOSs.  Note, the
observational analysis presented on the BR here is based on the visually
selected BR absorber members, which are the blue points in
Figure~\ref{fig:BR_and_filament_topcat}.
  
We then redefine the normal with combinations of the original, orthogonal,
3-vector system to give different projected planes.  The newly-defined
vectors $u, v, w$ are similarly orthogonal.  Figures
\ref{fig:BR_proj_u_hat_0} --- \ref{fig:BR_proj_u_hat_0_minus_v_hat_0} each
have a different viewing angles of the BR, and the observational analysis of
each plane projection follows.

In Figure~\ref{fig:BR_proj_u_hat_0} the new normal is set to $u_0$, which can
be thought of as the side-on view of the BR.  To emphasise this point, the
redshift colours are banded vertically indicating that the $u$-axis has
essentially become the redshift $z$-axis.  Of interest in this figure is the
curious backwards `S' shape which is in both the dark-coloured high-$z$
absorbers and the light-coloured low-$z$ absorbers.  The impression is that
the BR has three distinct redshift bands, and the absorbers in each band
create a noticeable backwards `S' shape, but more so in the nearest and
farthest redshift bins (the lightest and darkest coloured absorbers).  It is
not entirely clear what the backwards `S' shape might indicate, or if it
warrants further investigation.  Initially, it could be thought that the
absorbers belonging to the three distinct redshift bands may also be
associated particularly with some of the SLHC identified groups from
Figure~\ref{fig:SLHC_CHMS_BR_SN_4_2_4}, but this is not the case.  However,
instead we find that almost all (one exception, starred in the list below) of
the nearest redshift absorbers (those forming the backwards `S' shape with
the light-coloured absorbers) belong to one half of the BR closest to the
RHS.  This can be confirmed by cross-correlating the number IDs in
Figure~\ref{fig:BR_orig_proj} and Figure~\ref{fig:BR_proj_u_hat_0}).  (IDs of
absorbers contained in the near-redshift band, i.e., the light-coloured
backwards `S' shape: $36, 26, 29, 21, 3, 1, 4, 39*, 15, 22, 18, 19, 32,
33$. $39$ has been starred as the one absorber belonging in the near-redshift
band that does not sit with the rest of the absorbers on one half of the BR.)
So the near-redshift absorbers form an arc of a circle closest to the RHS of
the BR.  Later we will see that this further implies a somewhat spiral or
`cork-screw' interpretation of the BR, that is aligned face-on with our LOS.

In Figure~\ref{fig:BR_proj_v_hat_0} the new normal is set to $v_0$, which can
be thought of as viewing the BR from the bottom, looking up.  Now we can see
a thin, central arc which is formed from the most central redshift absorbers,
with clumps of absorbers at lower and higher redshift either side of this
arc.  If the BR had been discovered from this particular angle, had we been
located at a different position in the Universe, then we may have named this
structure the `Big Arc'.  This highlights the observational bias mentioned
earlier: our viewing angle only allows one LOS in which to discover
interesting structures and filaments when using the Mg~{\sc II} image method.
However, it also presents the opportunity to re-examine LSS candidates from
multiple viewing angles.
 
In Figure~\ref{fig:BR_proj_u_hat_0_minus_v_hat_0} the new normal is set to
$u_0 - v_0$, which can be thought of as looking at $45^\circ$ through the BR
towards the south-east direction.  The BR now very clearly resembles that of
a coil, and the inner filament is very thin and filamentary.  The high-$z$
and low-$z$ absorbers form ovoid shapes connected by the inner filament,
although the high-$z$ absorbers appear more broken and sparse compared with
the rest of the coil.
 
All three viewing angles, plus the original face-on view of the BR, indicate
that the BR structure is a coil shape with a thin, flat, dense, central
component.  The flatness of the central component is particularly interesting
given the recent work of finding flat patterns in cosmic structure presented
in \cite{Peebles2023}.  It is not yet clear how a structure like this will
have formed; however, one possibility could be cosmic strings, which have
been suggested to explain other recent discoveries and data \cite{Ahmed2023,
  Cyr2023, Ellis2023, Gouttenoire2023, Jiao2023, Peebles2023, Sanyal2022,
  Wang2023}.

Previously, the diameter of the BR was noted as similar to the expected size
of a BAO.  However, the coil shape seems likely to be inconsistent with an
origin in BAOs.

\begin{figure}[h]
\centering
\includegraphics[scale=0.35]{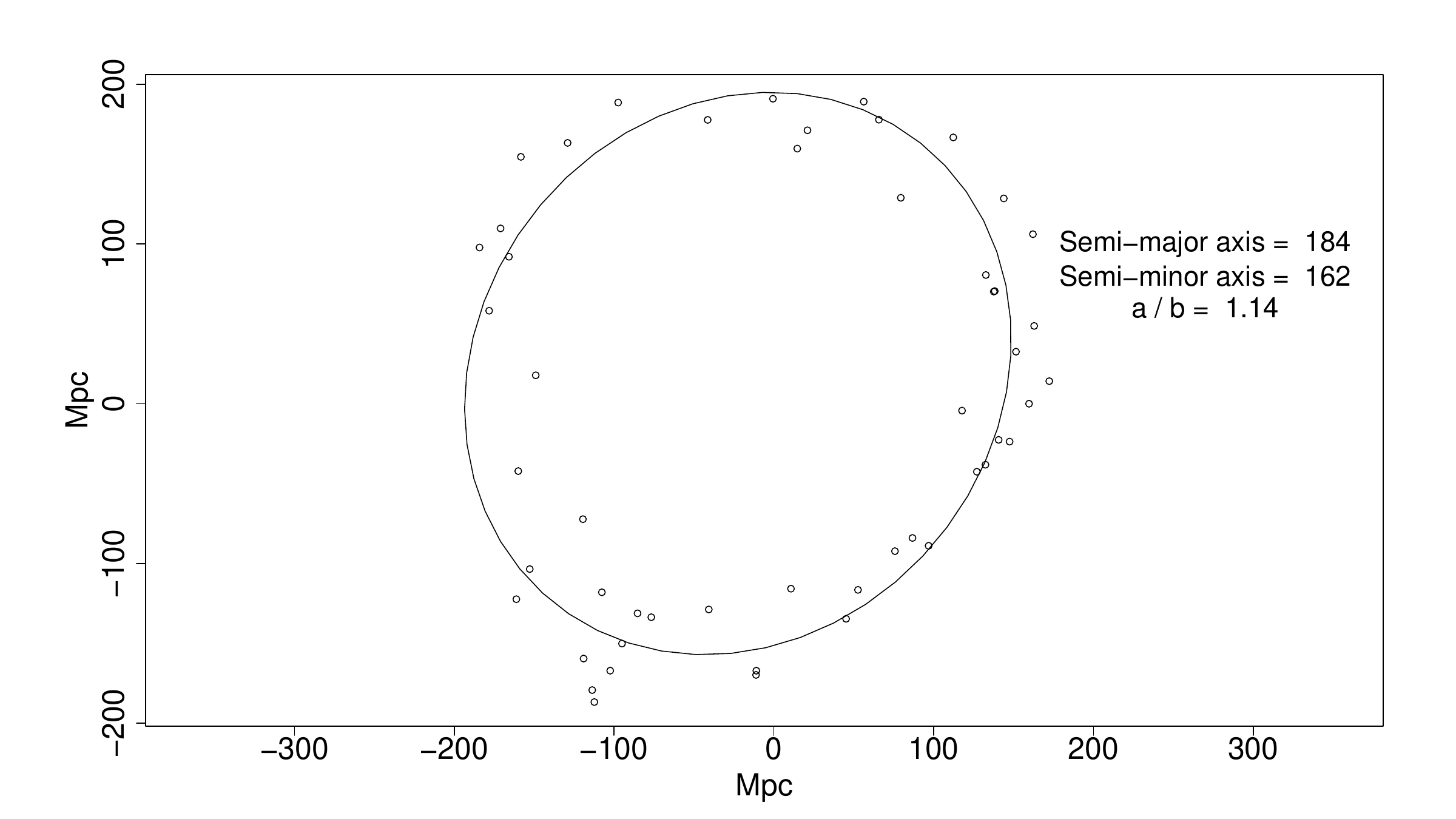}
\caption{\label{fig:BR_ellipse} The visually-identified BR absorber members
  projected onto the plane perpendicular to $w_0$ with an added fitted
  ellipse.}
\end{figure}
\begin{figure}[h]
\centering 
\includegraphics[scale=0.4]{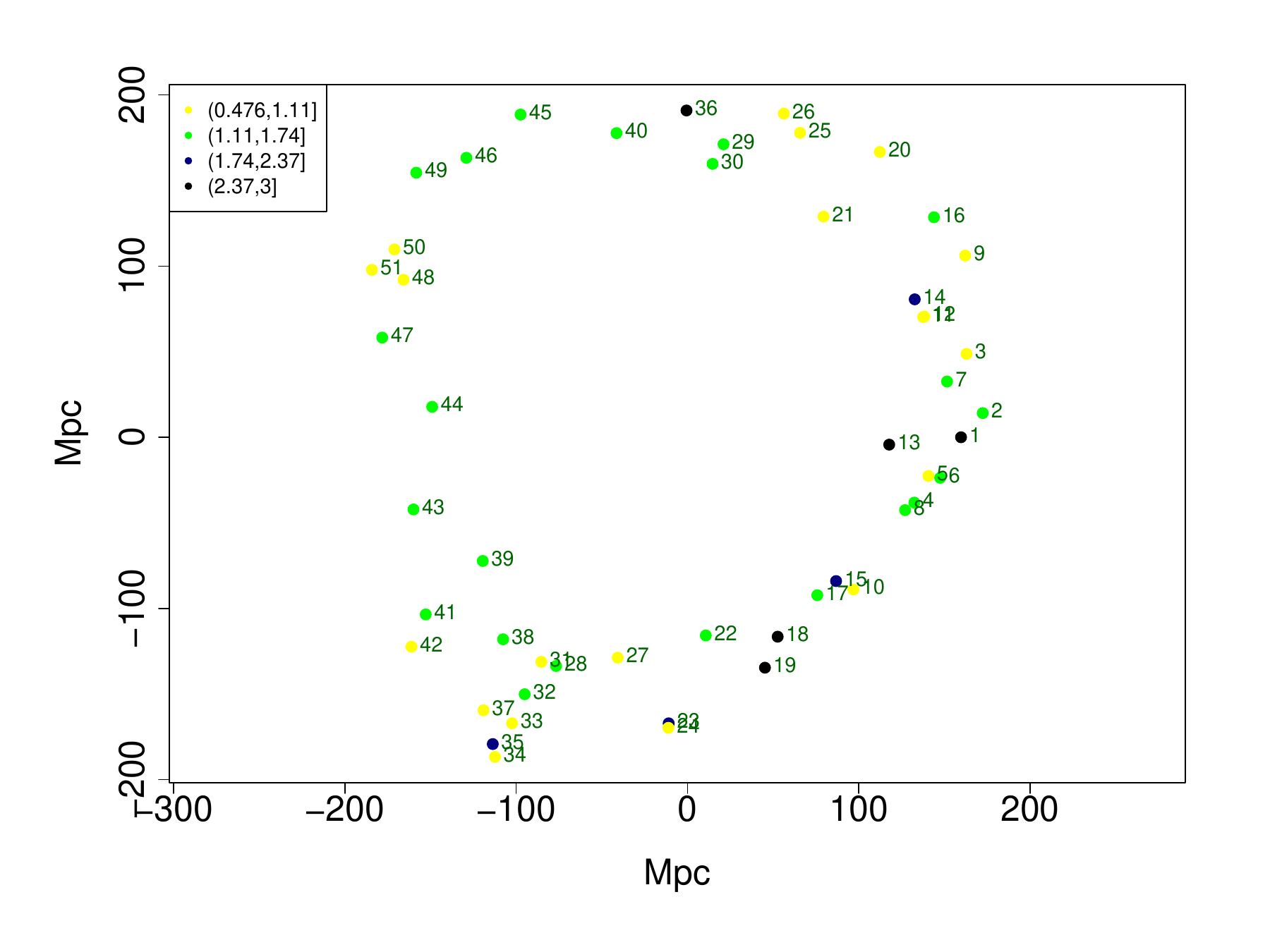}
\caption{\label{fig:BR_orig_proj_EW} The visually-identified BR absorber
  members projected onto the plane perpendicular to the initial normal
  vector, $w_0$. The $x$-axis points towards the $u_0$ direction and the
  $y$-axis points towards the $v_0$ direction.  The key in the top LHS of
  the figure indicates the rest-frame equivalent widths (in \AA) of the
  $\lambda_{2796}$ component of the absorbers associated by the colours, and
  the small numbers paired with each data point indicate their unique ID
  number. The equivalent widths are taken from the Mg~{\sc II} database of
  \cite{Anand2021}.}
\end{figure}
\begin{figure}[h]
\centering 
\includegraphics[scale=0.4]{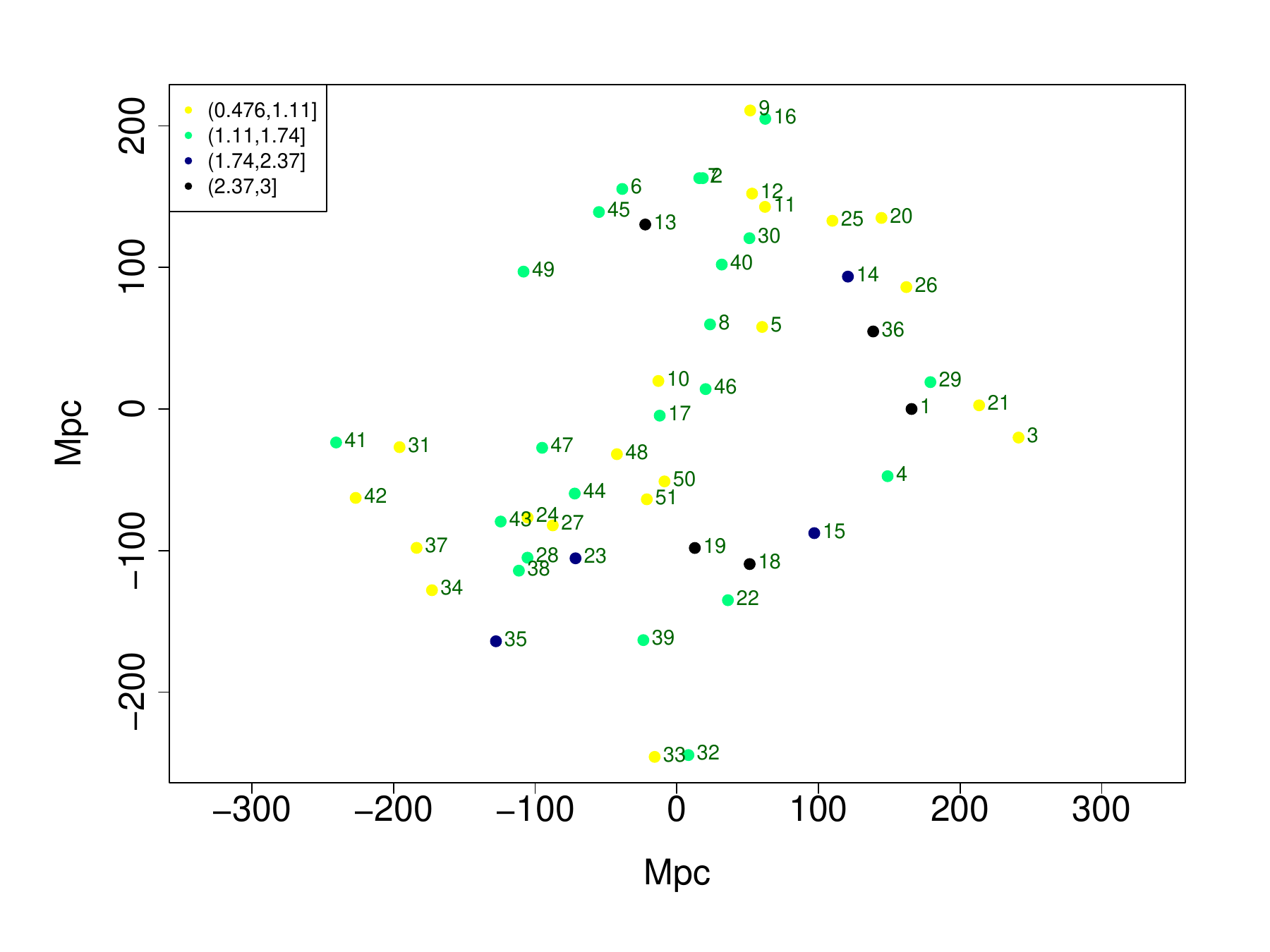}
\caption{\label{fig:BR_proj_EW_u_hat_0_minus_v_hat_0} The BR absorber members
  projected onto the plane perpendicular to $u_0 - v_0$. The new $u$
  direction points towards the most easterly absorber, as usual. The key in
  the top LHS of the figure indicates the rest-frame equivalent widths (in
  \AA) of the $\lambda_{2796}$ component of the absorbers associated by the
  colours, and the small numbers paired with each data point indicate their
  unique ID number. The equivalent widths are taken from the Mg~{\sc II}
  database of \cite{Anand2021}.}
\end{figure}

Finally, an ellipse can be fitted to the BR (Figure~\ref{fig:BR_ellipse}).
The ellipse calculates the semi-major and semi-minor axes as $184$~Mpc and
$162$~Mpc, respectively.  Given that the signature of an individual BAO has a
characteristic size of $150$~Mpc, we again argue against the possibility of
the BR occurring from a BAO.  Given the ellipse calculations of the
semi-major and semi-minor axes we can use Ramanujan's approximative perimeter
to approximate the circumference of the ellipse,
\begin{equation*}
p_R = \pi \Bigg\{ (a+b) + \dfrac{3(a-b)^2}{10(a+b)+ \sqrt{a^2 + 14ab + b^2}} \Bigg\}
\end{equation*}
\noindent where $a$ and $b$ are the semi-major and semi-minor axes.
The circumference of the ellipse is then $\sim 1.1$~Gpc.

\subsection{Equivalent widths}
\label{subsec:equivalent_widths}
In Figure~\ref{fig:BR_orig_proj} we showed the visually-identified BR
absorber members projected onto a plane normal to the vector $w_0$ (which
corresponds to the LOS), with the points colour-coded according to
redshift. Here, in Figure~\ref{fig:BR_orig_proj_EW}, we show a corresponding
figure but with the points now colour-coded according to the rest-frame
equivalent widths (in \AA) of the $\lambda_{2796}$ component of the Mg~{\sc
  II} doublet. The equivalent widths and their uncertainties, both taken
from the Mg~{\sc II} database of \cite{Anand2021}, are tabulated in
Table~\ref{tab:EW_sigmaEW}.

In Figure~\ref{fig:BR_orig_proj_EW} there is no obvious pattern for the lower
equivalent widths, but we note that the higher equivalent widths tend to
be concentrated on the RHS of the plot, especially the points that are
numbered $14, 1, 13, 15, 18, 19, 23$ (but not $36, 35$).

The earlier Figure~\ref{fig:BR_proj_u_hat_0_minus_v_hat_0}, for the plane
perpendicular to $u_0 - v_0$, drew attention to ovoid shapes at high-$z$ and
low-$z$. Here, Figure~\ref{fig:BR_proj_EW_u_hat_0_minus_v_hat_0} corresponds
to that figure, but again with colour-coding according to the equivalent
widths. It shows that the points $14, 36, 1, 15, 18, 19$ (and possibly $23,
35$ but not $13$) are actually concentrated in the low-$z$ ovoid
shape. Possibly a preferential alignment with respect to the LOS of the host
galaxies in this ovoid shape is enhancing the column densities.

\begin{table}[h]
\centering
\begin{tabular}{|c|c|c|c|c|c|}
\hline
{\bf ID} & EW (\AA) & $\sigma_{EW}$ (\AA) & {\bf ID} & EW (\AA) & $\sigma_{EW}$ (\AA) \\ \hline
\textbf{ 1} & 2.99 & 0.17 & \textbf{27} & 0.96 & 0.12 \\ \hline
\textbf{ 2} & 1.46 & 0.20 & \textbf{28} & 1.15 & 0.27 \\ \hline
\textbf{ 3} & 0.94 & 0.12 & \textbf{29} & 1.63 & 0.19 \\ \hline
\textbf{ 4} & 1.71 & 0.43 & \textbf{30} & 1.35 & 0.30 \\ \hline
\textbf{ 5} & 1.05 & 0.23 & \textbf{31} & 0.76 & 0.09 \\ \hline
\textbf{ 6} & 1.29 & 0.26 & \textbf{32} & 1.22 & 0.30 \\ \hline
\textbf{ 7} & 1.13 & 0.28 & \textbf{33} & 0.93 & 0.12 \\ \hline
\textbf{ 8} & 1.31 & 0.22 & \textbf{34} & 0.99 & 0.18 \\ \hline
\textbf{ 9} & 0.51 & 0.13 & \textbf{35} & 2.07 & 0.43 \\ \hline
\textbf{10} & 0.72 & 0.15 & \textbf{36} & 2.49 & 0.35 \\ \hline
\textbf{11} & 1.08 & 0.06 & \textbf{37} & 0.51 & 0.05 \\ \hline
\textbf{12} & 0.82 & 0.05 & \textbf{38} & 1.33 & 0.16 \\ \hline
\textbf{13} & 2.70 & 0.46 & \textbf{39} & 1.32 & 0.29 \\ \hline
\textbf{14} & 2.29 & 0.32 & \textbf{40} & 1.20 & 0.22 \\ \hline
\textbf{15} & 1.75 & 0.25 & \textbf{41} & 1.73 & 0.33 \\ \hline
\textbf{16} & 1.45 & 0.32 & \textbf{42} & 0.76 & 0.05 \\ \hline
\textbf{17} & 1.55 & 0.36 & \textbf{43} & 1.28 & 0.25 \\ \hline
\textbf{18} & 2.62 & 0.08 & \textbf{44} & 1.50 & 0.17 \\ \hline
\textbf{19} & 2.58 & 0.21 & \textbf{45} & 1.35 & 0.12 \\ \hline
\textbf{20} & 1.07 & 0.05 & \textbf{46} & 1.55 & 0.25 \\ \hline
\textbf{21} & 0.86 & 0.05 & \textbf{47} & 1.50 & 0.36 \\ \hline
\textbf{22} & 1.66 & 0.12 & \textbf{48} & 0.91 & 0.21 \\ \hline
\textbf{23} & 2.08 & 0.20 & \textbf{49} & 1.56 & 0.29 \\ \hline
\textbf{24} & 1.05 & 0.25 & \textbf{50} & 0.48 & 0.11 \\ \hline
\textbf{25} & 0.79 & 0.09 & \textbf{51} & 1.08 & 0.11 \\ \hline
\textbf{26} & 0.58 & 0.09 &             &      &      \\ \hline
\end{tabular}
\caption{\label{tab:EW_sigmaEW} The table is for the 51 visually-identified
  BR absorber members. The identification numbers (ID) correspond to those in
  Figure~\ref{fig:BR_orig_proj} and the other similar figures that follow
  it. The rest-frame equivalent widths (EW) and their associated
  uncertainties ($\sigma_{EW}$, taken from the Mg~{\sc II} database of
  \cite{Anand2021}, are for the $\lambda_{2796}$ component of the Mg~{\sc II}
  doublet; the units are \AA. Note that the EW for point $12$ is likely to be
  enhanced by the superimposed $\lambda_{2803}$ component from point $11$.}
\end{table}

\section{Discussion and conclusions} 

In this paper we have presented a `Big Ring on the Sky', which is the second
ultra-large large-scale structure (uLSS) detected in the Mg~{\sc II}
catalogues The BR is detected most prominently in the redshift slice $z=0.802
\pm 0.060$.  The redshift slice containing the BR is the exact same redshift
slice containing the previously detected GA, and both structures on the sky
are separated by only $\sim 12^\circ$, meaning that these two intriguing
structures are in the same cosmological neighbourhood.
  
We applied several inspection tests and statistical tests to the BR to
support it as an ultra-large LSS.  A summary of the results is as follows.
(1) Each of the visually-identified BR+filament absorbers was confirmed
visually in the corresponding spectra, establishing that $100 \% $ of the
absorbers are real detections, and not false-positives.  (2) The SLHC
algorithm identified $46$ out of $59$ of the visually-identified BR+filament
absorbers across $5$ individual, overlapping or adjacent, groups.  The SLHC
algorithm also detected a statistically-significant arc in the redshift
slices centred at $z=0.862 \pm 0.060$ and $z=0.922 \pm 0.060$.  The arc
appears to be an extension of the bottom arc of the BR, given the agreement
in the on-sky position and that the redshift slices overlap by $50\%$.  (3)
We calculated the CHMS and MST significance of the SLHC-identified absorbers,
the visually-identified BR absorbers (two versions), and the
FilFinder-identified absorbers.  The CHMS method was found to have a much
larger variation in its reported significances due to the nature of the
method: the CHMS calculates the volume of the unique structure containing all
the absorbers, so in the example of a ring-like structure, there is a large
volume in the centre of the ring with far fewer absorbers than the annulus.
The CHMS calculates a ($3.65 \pm 1.13$)$\sigma$ significance.  By using the
visually-identified absorbers, and keeping the BR innards, the CHMS
calculates a (likely) upper limit of $5.2 \sigma$ significance.  The MST
significance on the other hand was more consistent, as this method relies
only on the mean MST edge-length, and this method calculates a significance
of ($4.10 \pm 0.45$)$\sigma$.  (4) The FilFinder algorithm was applied to the
Mg~{\sc II} image of the BR field to identify filaments objectively. By
incrementally increasing the size-threshold, the algorithm left only one
identified filament --- a ring --- and that ring corresponded to the
visually-identified BR. In this way, the BR was established as: (i) a ring,
independently of visual perception; and (ii) the most-connected and largest
filament in the image.  We also applied the FilFinder algorithm to the field
quasars (not to be confused with the background probes) and this found that
there was a large, connected, ring-like filament that mostly aligns with the
BR.  (5) We applied the CE test to the field containing the BR and then
`zoomed' in on the BR to assess the contribution of the BR to the spatial
clustering in the field.  Tentative significant clustering in the field was
detected in the second zoom of the BR, at a $p$-value $p = 0.022$
  (corresponding to a significance of $2.0 \sigma$). We compared this with
four other unrelated fields at the same redshift as the BR and found no
significant spatial clustering.  However, the CE test results for the BR
field are still inconclusive at the $\sigma > 3.0$ significance level,
suggesting that the clustering in the BR field is not statistically
significant, unlike what was seen with the GA field.  Note, this test
assesses the clustering in the \emph{field} and not the statistical
significance of any individual candidate structure.  (6) By superimposing
contour maps of the field quasars and DESI clusters on the Mg~{\sc II}
absorber images we were able to show plausible association between the
datasets, thus providing independent corroboration.  We found in particular
that the DESI clusters that were subject to richness limits were mapping
different Mg~{\sc II} absorbers, suggesting that there could be scope for
using DESI clusters as a way to investigate the physical environment around
the Mg~{\sc II} absorbers.  (7) Using a project-plane method we could
investigate the 3D distribution of the BR.  Viewing the BR on different
angles showed that there are $3$ distinct redshift bands, where the central
redshift band contains the majority of the absorbers in a ring shape in a
thin, flat region.  The near-redshift band is contained almost entirely on
the RHS of the BR and even appears to angle backwards into the central
redshift bin of the BR creating a spiral-like shape.  The farthest-redshift
band had a similar coiling effect, but appeared to be distributed randomly
throughout the BR (it made a broken, ring shape).  We also found, with the
project-plane method, curious backwards `S' shapes when viewing the BR side
on.  The S shapes are most apparent in the near and far redshift bands but
can also be seen in the central redshift band.
  
The data and analysis show that the BR is of particular interest for LSS
studies in cosmology.  In cosmology, we assume statistical homogeneity in the
Cosmological Principle (CP) as the foundations of the $\Lambda$CDM model.  We
have shown that the BR is real and statistically significant, adding it to a
growing list of LSS candidates that are in tension with the CP.  The growing
list of LSSs also indicates that the chance of all of these structures being
a matter of statistical noise is decreasing.  In addition, the list of LSS in
tension with the CP is also generally difficult to understand as there are
gaps in the knowledge for explaining the formation of these ultra-large
structures.  Perhaps we could speculate that the existence of these
structures requires an extension to the standard model, perhaps in the form
of cosmic strings.

The BR and GA together, given their sizes and morphologies, are presumably
telling us something intriguing, and quite possibly important, about the
Universe, but at the moment we can only speculate what that might be.

\acknowledgments
We acknowledge the use of the public R software (v4.1.2)
\footnote{https://www.R-project.org/}.  Our data has depended on the
publicly-available Sloan Digital Sky Survey quasar catalogue and the
corresponding Mg~{\sc II}
catalogues\footnote{https://wwwmpa.mpa-garching.mpg.de/SDSS/MgII/}.  AML was
supported by a UCLan/JHI PhD studentship.

We thank the referee for thoughtful comments and useful suggestions.

\end{document}